\documentclass[reprint,
 superscriptaddress,
 longbibliography,
 nofootinbib,
  amsmath,amssymb,
 prx,
 ]{revtex4-2}

\usepackage{siunitx}
\DeclareSIUnit{\torr}{torr}

\usepackage{amsthm}
\usepackage{xcolor}
\usepackage{siunitx}

\definecolor{mygreen}{rgb}{0,0.5,0}
\definecolor{mygrey}{rgb}{0.5,0.5,0.5}
\definecolor{myred}{rgb}{0.75,0,0}
\definecolor{myblue}{rgb}{0,0,0.75}
\definecolor{mymagenta}{cmyk}{0,1,0,0.12}
\definecolor{mycyan}{cmyk}{1,0,0,0.12}
\definecolor{myorange}{rgb}{1.,0.5,0}
\definecolor{myviolet}{rgb}{0.6,0.15,0.6}
\definecolor{mybrown}{cmyk}{0,0.50,1,0.41}

\usepackage[normalem]{ulem}

\usepackage{graphicx}
\usepackage{dcolumn}
\usepackage{bm}
\usepackage{lipsum}
\usepackage{float}

\usepackage{bbold}

\newcommand{\FisherInfo}{\mathcal{F}}

\usepackage{cancel}
\usepackage{color,graphicx}
\usepackage{hyperref}

\begin{document}

\newtheorem{theorem}{Theorem}
\newtheorem{lemma}[theorem]{Lemma}
\newtheorem{corollary}{Corollary}[theorem]
\newtheorem{prop}{Proposition}

\newcommand{\vecb}[1]{\mbox{\boldmath$#1$}}
\newcommand{\red}[1]{\mbox{\textcolor{red}{$#1$}}}
\preprint{APS/123-QED}
\newcommand\expect{\mathbb{E}}
\newcommand\Tr{\textrm{Tr}}
\newcommand{\order}{\mathcal{O}}

\newcommand{\expe}[1]{{\color{orange} #1}}
\newcommand{\jcc}[1]{{\color{magenta}JCC: #1}}
\newcommand{\jccr}[2]{{\color{magenta} \sout{#1}}{\color{magenta}#2}}

\newcommand{\ers}[1]{{\color{mygreen} ERS: #1}}
\newcommand{\ersr}[2]{{\color{mygreen} \sout{#1}}{\color{mygreen}#2}}

\newcommand{\mss}[1]{{\color{red} MS: #1}}
\newcommand{\msr}[2]{{\color{red} \sout{#1}}{\color{red} MS: #2}}

\newcommand{\GG}[1]{{\color{teal}GG: #1}}
\newcommand{\GGr}[2]{{\color{teal} \sout{#1}}{\color{teal}#2}}

\newcommand{\mwm}[1]{{\color{cyan}MWM: #1}}
\newcommand{\mwmr}[2]{{\color{cyan} \sout{#1}}{\color{cyan}#2}}

\title{Sequential analysis in a continuous spin-noise quantum sensor}

\author{ E. Roda-Salichs}
\email{elisabet.roda@uab.cat}
\affiliation{Física Teòrica: Informació i Fenòmens Quàntics, Department de Física, Universitat Autònoma de Barcelona, 08193 Bellaterra (Barcelona), Spain}
\author{ G. Gasbarri}
\altaffiliation[Present address: ]{Naturwissenschaftlich–Technische Fakultät, Universität Siegen, Walter-Flex-Straße 3, 57068 Siegen, Germany}
\affiliation{Física Teòrica: Informació i Fenòmens Quàntics, Department de Física, Universitat Autònoma de Barcelona, 08193 Bellaterra (Barcelona), Spain}
\author{ A. Alou}
\altaffiliation[Also at ]{Port d’Informació Científica (PIC), Campus UAB, C. Albareda s/n, 08193 Bellaterra}
\affiliation{Física Teòrica: Informació i Fenòmens Quàntics, Department de Física, Universitat Autònoma de Barcelona, 08193 Bellaterra (Barcelona), Spain}
\affiliation{Centro de Investigaciones Energéticas, Medioambientales y Tecnológicas (CIEMAT), Avda Complutense 22, 28040 Madrid, Spain}
\author{M. Skotiniotis}
\altaffiliation[Present address: ]{Quantum Thermodynamics and Computation Group, Department of Electromagnetism and Condensed Matter, Universidad de Granada, 18071 Granada, Spain}
\altaffiliation[and at ]{Instituto Carlos I de Física Teórica y Computacional, Universidad de Granada, 18071 Granada, Spain\looseness=-1}
\affiliation{Física Teòrica: Informació i Fenòmens Quàntics, Department de Física, Universitat Autònoma de Barcelona, 08193 Bellaterra (Barcelona), Spain}
\author{A. Sierant}
\affiliation{ICFO - Institut de Ciències Fotòniques, The Barcelona Institute of Science and Technology, 08860 Castelldefels (Barcelona), Spain}
\author{D. Méndez-Avalos}
\affiliation{ICFO - Institut de Ciències Fotòniques, The Barcelona Institute of Science and Technology, 08860 Castelldefels (Barcelona), Spain}
\author{ M. W. Mitchell}
\email{morgan.mitchell@icfo.eu}
\affiliation{ICFO - Institut de Ciències Fotòniques, The Barcelona Institute
of Science and Technology, 08860 Castelldefels (Barcelona), Spain}
\affiliation{ICREA - Institució Catalana de Recerca i Estudis Avançats, 08010 Barcelona, Spain}
\author{ J. Calsamiglia}
\email{john.calsamiglia@uab.cat}
\affiliation{Física Teòrica: Informació i Fenòmens Quàntics, Department de Física, Universitat Autònoma de Barcelona, 08193 Bellaterra (Barcelona), Spain}


\begin{abstract}
Many control and detection applications require real-time analysis of signals from sensors, in order to  quickly and accurately act upon events revealed by the sensors. Such signal analysis benefits from statistical models of signal and sensor behavior. This creates a need for data analysis methods that are simultaneously model-based,  computationally efficient and causal, in the sense that they employ only sensor data available prior to a specific point in time. In this work, we implement sequential data analysis techniques on a spin-noise-based quantum sensor,  
to perform two key tasks: hypothesis testing and quickest change-point detection. These online protocols allow us to detect weak magnetic fields by adaptively collecting 
measurement data until a predefined confidence threshold is reached.
We demonstrate these methods in a realistic experimental setting and derive performance bounds for the achievable precision and response time.  
Our approach has potential utility  when detecting small perturbations to the magnetic field, in both applied and fundamental contexts including biomagnetism, geophysical surveys, detection of concealed materials, searches for dark matter candidates and exotic spin interactions. Our results demonstrate that sequential techniques enable faster and more sensitive detection, making them a powerful tool for quantum sensing. 
\end{abstract}

\maketitle

\section{\label{sec:Introduction}Introduction}

Continuously monitored quantum systems, such as atomic magnetometers and optomechanical sensors, offer a promising avenue for ultraprecise quantum sensing \cite{wiseman2009quantum, 
kitching2011atomic, serafini2012feedback, Aspelmeyer2014Cavity,rogers2014hybrid,yu2016cavity, millen2020optomechanics,kumar2021cavity,li2021cavity, 
barzanjeh2022optomechanics,angOptomechanicalParameterEstimation2013}. These platforms provide opportunities for real-time statistical inference in applications ranging from biomedical diagnostics to materials characterization~\cite{budker2007optical,rogers2014hybrid,barzanjeh2022optomechanics}. Their ability to operate in real time makes them particularly suitable for detecting and tracking stochastic time-varying signals. As such, they play a key role in applications 
requiring online and efficient detection of abrupt transitions or sudden changes in dynamics.

Despite remarkable theoretical and experimental progress~\cite{wiseman1993quantum,kitching2006chip, forstner2012sensitivity, Tsang2012Continuous,jacobs2014quantum, molmer_hypothesis_2015, Albarelli_2017, kiilerich_hypothesis_2018, ralph2018dynamical, Jimenez2018Signal,liu2020quantum, mitchell2020colloquium, Am2021, marchese2023optomechanics, wang2023beating,blais21QuantumCircuit, k7nk-lrwd}, many aspects of real-time quantum sensing protocols remain largely unexplored. 

In most standard approaches to inference tasks, particularly in hypothesis testing, signal information is analyzed only after data collection or after multiple experimental repetitions, thereby limiting the possibility of real-time decision-making. To address these limitations, recent work has begun to investigate how sequential decision-making strategies can be integrated into quantum systems under continuous observation.

A few seminal works have laid the foundation for quantum sequential analysis in settings where identically prepared copies of a quantum system are supplied on demand. In~\cite{vargas2021quantum, li_optimal_2022}, the quantum analogue of Wald’s \cite{Wald1945Sequential} problem of \emph{sequential hypothesis testing} was developed. In this setting, one aims to determine whether a quantum source is emitting states $\rho_0$ or $\rho_1$, using measurements on successive, identically prepared systems, and stopping the process once sufficient evidence has been accumulated to make a reliable decision within a given confidence level. Similarly, Lorden's \cite{lorden1971procedures} classical problem of \emph{quickest change-point detection}, which seeks to identify as promptly as possible the moment when the statistical properties of a sequence change, was extended to the quantum domain in~\cite{fanizza_ultimate_2023}. There, the system emits quantum states initially prepared as $\rho_0$, which at an unknown time switch to $\rho_1$, and the task is to detect this change with minimal delay while keeping false alarms under control. These works have established fundamental bounds on performance in such independent and identically distributed (IID) quantum scenarios, showing that the relevant asymptotic figure of merit is the quantum relative entropy $D(\rho_0 \| \rho_1)$, just as the Kullback–Leibler divergence plays a central role in classical statistics. More recently, quantum-enhanced strategies for quickest change-point detection have been applied in the context of quantum communication~\cite{reichert2023a, gongQuantumEnhancedChangeDetection2025, guhaQuantumenhancedQuickestChange2025}, indicating their relevance for real-time decision-making across a broad range of quantum technologies.

This paper considers a distinct quantum sensing paradigm in which, instead of performing a sequence of measurements on an increasing number of independently prepared copies, a continuous sequence of measurements is carried out on the \emph{same} quantum system. The goal is to discriminate between alternative internal dynamics governing the system. This approach, explored theoretically in the context of binary hypothesis testing using sequential strategies~\cite{gasbarri_sequential_2024}, offers notable advantages: it enables real-time analysis of individual trajectories without requiring ensemble averages, and it can significantly reduce measurement time by allowing data collection to stop as soon as a hypothesis is identified with the desired level of confidence.

In this work, we report what is, to our knowledge, the first experimental implementation of sequential analysis methodologies in continuously monitored quantum sensors. We demonstrate this framework using two foundational sequential analysis primitives, sequential hypothesis testing and quickest change-point detection, applied to atomic spin-noise magnetometry. Atomic magnetometry is among the most mature and widely adopted in quantum technologies, with demonstrated relevance in diverse fields \cite{ mitchell2020colloquium}. 
The sensing platform of optically-pumped and -probed atomic vapors is applied in medicine \cite{BotoNI2017, BotoN2018}, materials science \cite{KnappePatent2024}, geophysics \cite{OelsnerPRAppl2022}, space science \cite{MrozowskiSR2024} and fundamental physics \cite{HunterS2013, SafronovaRMP2018}. The statistical properties of spin-noise dynamics has been studied extensively  \cite{SinitsynRPP2016, ZapasskiiAOP2013, CrookerN2004}, including quantum-enhanced spectroscopy with squeezed light \cite{LuciveroPRA2016, LuciveroPRA2017} and measurement-induced spin squeezing \cite{KongNC2020}, and found to have excellent agreement with gaussian statistical models \cite{JimenezMartinezPRL2018, MouloudakisPRA2023}, 
including in high-density regimes dominated by spin-exchange interactions \cite{MouloudakisPRA2022, MouloudakisPRA2024}.

This work is structured as follows: In Sec.~\ref{sec:seqstrategies} we
present hypothesis testing, both deterministic and sequential strategies, and quickest change-point detection scenarios. Here, and throughout this work, we use the term \textit{deterministic} to refer to standard approaches that rely on a pre-determined, fixed number of samples, in contrast to sequential strategies, where the number of samples is stochastic, as it depends on previous measurement outcomes. In Sec.~\ref{sec:experiment} we explain the physical model, the dynamics behind the spin-noise quantum sensor and the probability distribution that the observations follow. We then proceed in Sec.~\ref{sec:experimentaldata} to show the experimental data and how we tackle the problem of 
low-frequency spurious noise in the system. With this at hand, in Sec.~\ref{sec:experimentalimplementation} we present the results of sequential analysis in the spin-noise system. Finally, in Sec.~\ref{sec:ultimatelimits} we derive the optimal asymptotic performance limits for the studied inference strategies.

\section{\label{sec:seqstrategies}
Sequential analysis: Basic concepts and methods} 

To introduce the class of sequential strategies considered in this work, we begin with sequences of $n$ observations denoted by $\vecb{R}_n = (R_1, \ldots, R_n)^T$, where $n$ may grow as new data becomes available. In hypothesis testing we are given two known models labeled $h=0$ and $h=1$, with the assumption that the sequence $\vecb{R}_n$ is generated by one of them. The goal is to sequentially evaluate the plausibility of these hypotheses as more observations are collected, terminating when the evidence reaches a predetermined confidence level.

The change-point detection scenario, on the other hand, assumes that the observations initially follow model $h=0$, but may transition to model $h=1$ at some unknown point in the sequence. Here, the task becomes monitoring the stream of observations to detect such a change as quick as possible while controlling  the rate of false alarm.

As we will shortly see, in both sequential hypothesis testing and {quickest} change-point detection, the central quantity that must be computed and monitored is the log-likelihood ratio (LLR) statistic.
\begin{equation}
L_{R,n}:=\log\frac{p_1\left(\vecb{R}_n\right)}{p_0\left(\vecb{R}_n\right)}.
\label{llr}
\end{equation}
Here $p_i\left(\vecb{R}_n\right):=p\left(\vecb{R}_n| h_i\right)$, for $i=0,1$, denotes the probability density function of $\vecb{R}_n$ under $h_i$. Tracking this single scalar statistic is sufficient to implement our optimal stopping rules and inference decisions.

\subsection{\label{sec:hyptesting} Binary hypothesis testing}

In binary hypothesis testing, the goal is to decide which of two mutually exclusive hypotheses, labeled $h=0$ and $h=1$, correctly describes a system or process based on observed data.  Each hypothesis corresponds to a distinct probability distribution for the observations. Since decisions are made from finite noisy data, errors are unavoidable. These errors are quantified by two standard metrics: the type-I error (false positive), $\alpha_1$, which occurs when $h=0$ is rejected even though it is true, and the type-II error (false negative), $\alpha_0$, which occurs when $h=0$ is accepted while the true hypothesis is $h=1$. Broadly, the goal of hypothesis testing is to design decision strategies that minimize these errors using the least amount of resources, such as the minimum number of measurements or samples.

\subsubsection{\label{sec:hyptestingdet} Hypothesis testing with deterministic time} 

In the standard approach, a sequence of measurement data of predetermined length $n$ is collected, after which a definitive decision regarding the true hypothesis is made. Such decision faces an inherent trade-off between the two error probabilities $\alpha_0$  and $\alpha_1$, which generally cannot be minimized simultaneously. Different testing paradigms emerge depending on the optimization criteria. An asymmetric test strategy minimizes one error probability while constraining the other below a specified threshold, which is particularly relevant when one error type carries greater consequences than the other. In contrast, symmetric testing naturally appears in Bayesian frameworks where prior probabilities $\{\pi_0,\pi_1\}$ are assigned to the different hypotheses. Here, the optimal strategy is the one that minimizes the average error probability
\begin{equation}
  P_{\text{error}} = \pi_0 \alpha_1 + \pi_1 \alpha_0.
  \label{perr}
\end{equation}

In order to perform the optimal decision in these problems, it is sufficient to compute the LLR statistic $L_n(\vecb{R}_n)$ from Eq.~\eqref{llr}, choose an appropriate threshold $a$, and carry out the \textit{likelihood ratio test} \cite{cover2006}:

\begin{itemize}
	    \item Accept $h=1$  if $L_n> a$
	    \item Accept $h=0$  if $L_n\leq a$
     \end{itemize}
which renders the error probabilities, 
\begin{equation}
  \begin{aligned}
        \alpha_0 &= P_1\left(L_n < a\right) = \int_{-\infty}^a p_1(\ell)\,d\ell, \\
        \alpha_1 &= P_0\left(L_n \geq a\right) = \int_a^{\infty} p_0(\ell)\,d\ell,
  \end{aligned}
\end{equation}
where $p_h(\ell)$ is the probability density function of the test statistic $L_n$ under hypothesis $h = 0,1$. 
The threshold $a$ controls the tradeoff between type-I and type-II errors: decreasing $a$ reduces $\alpha_1$ at the cost of increasing $\alpha_0$, and vice versa. 
For equal prior probabilities ($\pi_0 = \pi_1$), the optimal decision rule is the \textit{maximum likelihood estimator}, which corresponds to setting $a = 0$.

For a large number $n$
of independent and identically distributed (IID) observations, the error probabilities decay exponentially with $n$, with rates characterized by fundamental information-theoretic quantities~\cite{cover2006}. In particular, the Chernoff information
\begin{equation}
\label{eq:Cheriid}
C(p_0, p_1) := -\min_{0 \leq s \leq 1} \log \int p_0(x)^s p_1(x)^{1-s}  dx
\end{equation}
determines the optimal error exponent for symmetric hypothesis testing, satisfying
\begin{equation}
-\lim_{n\to\infty}\frac{1}{n}\log P_\mathrm{err} = C(p_0,p_1).
\label{eq:errorrate}
\end{equation}
In asymmetric hypothesis testing the Kullback-Leibler divergence (relative entropy)
\begin{equation}
\label{eq:KLiid}
D(p_0\| p_1) := \int p_0(x) \log \frac{p_0(x)}{p_1(x)}  dx,
\end{equation}
governs the optimal type-II error exponent. Specifically, when constraining the type-I error probability $\alpha_1<\epsilon$
 for some constant $\epsilon<1/2$, we have
\begin{equation}
-\lim_{n\to\infty}\frac{1}{n}\log \alpha_0 = D(p_0 \| p_1).
\end{equation}

This represents the fastest achievable error rate, as guaranteed by the converse of Stein's lemma \cite{cover2006}: any attempt to achieve a better error exponent would necessarily drive the type-I error probability to 1, resulting in complete loss of discrimination power.

Equivalent asymptotic expressions can also be derived for (non-IID) Gaussian noise processes, like those arising in our setup (see~\cite{kazakos_spectral_1980,Tsang2012Continuous} and Sec.~\ref{sec:ultimatelimits}).

\subsubsection{\label{sec:hyptestingseq} Sequential hypothesis testing}

In the context of continuously monitored sensors,  rather than fixing the duration of the experiment and then assessing the error probability, it is more natural to perform inference \textit{on the fly}, continuing measurements until a desired error threshold is reached. In this setting, the inference strategy involves not only a decision rule but also a stopping time ---a criterion that dictates whether to continue or stop based on the observations. Note that since the observations are stochastic, the stopping time is itself a random variable.

In the Bayesian framework, such a stopping rule directly follows from the operational requirement that one may stop only when, based on the current measurement record, the probability of correctly identifying hypothesis $h=0$ is at least $s_0 = 1 - \epsilon_0$, and that of $h=1$ is at least $s_1 = 1 - \epsilon_1$. i.e. 
\begin{equation}
p(h_k|\vecb{R}_n)=\frac{p(\vecb{R}_n|h_k) \pi_k }{ p(\vecb{R}_n|h_0) \pi_0+ p(\vecb{R}_n|h_1)\pi_1} \geq 1-\epsilon_k,
 \label{ineq}
\end{equation}
where $k=0,1$ and we have made use of  Bayes' rule. The above inequalities can be conveniently expressed in terms of the  log-likelihood ratio:

\begin{itemize}

	    \item Stop and accept $h=1$ 
     if $L_n  \leq -a_0:=-\log\left(\frac{\pi_0}{\pi_1}\frac{1-\epsilon_0}{\epsilon_0}\right)$
	    \item Stop and accept $h=0$ 
   $ L_n  \geq a_1:=-\log\left(\frac{\pi_1}{\pi_0}\frac{1-\epsilon_1}{\epsilon_1}\right)$
	    \item Continue if 
     $ -a_0 <  L_n< a_1.$
\end{itemize}
This protocol is known as the \textit{Sequential Probability Ratio Test} (SPRT) \cite{Wald1945Sequential} and its stopping time is given by 
\begin{equation}
   N:=\inf\{n : L_{n}(\vecb{R}_n)\notin (-a_0,a_1)\} \,.
\end{equation}
Since the decision to stop depends only on data observed up to time $n \leq N$, it defines a valid stopping time in the sense of stochastic decision theory. Notice that for $\epsilon_0=\epsilon_1=\epsilon$,  the error probability from Eq.~\eqref{perr} satisfies $P_{\textrm{err}}\leq \epsilon$.

Wald \cite{Wald1945Sequential,wald2004sequential} showed that the thresholds ${a_0,a_1}$ in the SPRT algorithm are also in one-to-one correspondence with the type-I and type-II errors, whose definitions do not involve Bayesian priors.
Specifically, for a process that stops either at $L_N = -a_0$ or at $L_N = a_1$, 
\begin{equation}
    \alpha_0=\frac{1-\mathrm{e}^{-a_1}}{\mathrm{e}^{a_0}-\mathrm{e}^{-a_1}},\quad \alpha_1=\frac{1-\mathrm{e}^{-a_0}}{\mathrm{e}^{a_1}-\mathrm{e}^{-a_0}}.
\end{equation}
 Wald and Wolfowitz \cite{wald_optimum_1948} showed that, among all strategies constrained by given type-I and type-II error bounds, the SPRT minimizes the expected stopping time under each hypothesis. Analogously to the deterministic error rates, Eq.~\eqref{eq:Cheriid} and Eq.~\eqref{eq:KLiid}, for sequential strategies an asymptotic relation connects the expected number of observations to the error bounds: 
\begin{equation}
 \expect_{0}[{N}]\sim -\frac{\log\alpha_0}{D(p_{0}\|p_{1})}  \mbox{ and } \expect_{1}[{N}]\sim -\frac{\log\alpha_1}{D(p_{1}\|p_{0})} 
\end{equation}
in the limit of small error bounds $\alpha_0,\alpha_1\ll 1$. Clearly, sequential strategies achieve the same error probabilities with far fewer observations \textit{on average}. In particular, the SPRT achieves the fastest possible exponential decay of both error probabilities, each governed by the corresponding Kullback-Leibler divergence. This is in contrast to fixed-length (deterministic) strategies that can attain such a decay rate for one error, but only at the expense of keeping the other error probability finite.

In non-IID and quantum scenarios, recent works \cite{gasbarri_sequential_2024,QuantumVargas2021,li_optimal_2022}, have also demonstrated the advantages of the SPRT, further extending its relevance beyond classical settings.

\subsection{\label{sec:changepoint} Quickest change-point detection}

Detecting abrupt changes in continuously monitored quantum systems is a fundamental task, as it captures the general challenge of identifying sudden shifts in the underlying dynamics while balancing detection speed and reliability.  Beyond its conceptual importance, this problem has direct practical relevance in contexts such as early detection of fires or hazardous substances, or other environmental and medical monitoring, where timely responses to unexpected events are crucial.

In this subsection, we review the problem of quickest change-point detection, which aims at identifying a sudden change in the underlying quantum dynamics while minimizing both false alarms and detection delay.  
The cumulative sum (CUSUM) algorithm first proposed by Page \cite{page1954continuous} is one of the most widely used methods for quickest change-point detection. Its popularity stems from two main features: its computational efficiency, which allows real-time implementation, and optimality under certain mild conditions. In the IID scenario  Lorden \cite{lorden1971procedures} showed that CUSUM minimizes the worst mean detection delay among the strategies with a given bound on the mean false alarm time.
Lorden's   asymptotic optimality proof was  later  shown to hold also for finite times \cite{moustakides1986optimal,
ritov1990decision}.  The CUSUM algorithm was extended to non-iid settings and  its asymptotic optimality was also proven under some reasonable conditions by Lai \cite{737522}.

In its extended version the quickest change-point problem assumes the statistical model obeys different distributions before and after the (unknown) change-point time $\nu$, i.e. the
 probability density function of \( R_{n} \), conditional on $R_{1}\ldots R_{n-1}$  is given by
\begin{align}
p_0(R_n| R_{1}\ldots{R}_{n-1})& \quad \text{for } n < \nu \quad (\text{before change})\\
p_{1}(R_n|R_{1}\ldots R_{n-1})&  \quad \text{for } n \geq \nu \quad (\text{post-change}),
\end{align}

The CUSUM algorithms can be described in terms of the likelihood ratio between the change-point occurring at time $1\leq k \leq n$ and no change occurring in that time.
\begin{align}
L_{n}^{(k)}:&=\log \frac{p_{1}(R_{1:n})}{p_{0}(R_{1:n})}\label{eq:CUSUMstat}\\
&= \log \frac{\prod_{i=1}^{k-1} p_{0} (R_i | R_{1:i-1})  \prod_{i=k}^{n} p_{1} (R_i | R_{1:i-1})}{\prod_{i=1}^{n} p_{0} (R_i | R_{1:i-1})}\nonumber\\
&= \log \prod_{i=k}^{n} \frac{p_{1}(R_i | R_{1:i-1})}{p_{0} (R_i | R_{1:i-1})}
= \sum_{i=k}^{n} \Delta L_{i},
\end{align}
where the likelihood ratio increment at step $i$ is defined as
\begin{equation}
 \Delta L_{i} = \log \frac{p_{1} (R_i | R_{1:i-1})}{p_{0} (R_i | R_{1:i-1})},
\end{equation}
which is positive whenever the event at $i$ is more likely to be caused under the changed hypothesis  than under the default one.

The crucial statistic is  the cumulative sum (CUSUM):
\begin{equation}
\label{eq:Mn}
M_n = \max_{1 \leq k \leq n} \sum_{i=k}^{n} L_{i}\, .
\end{equation}

The CUSUM algorithm outputs  change-point  detection whenever this quantity exceeds a given threshold $a$  and the corresponding stopping, or detection, time is given by 
\begin{equation}
    T = \inf \{n : M_n \geq a\}\, .
    \label{eq:Tdetection}
\end{equation}

The threshold $a$ is chosen so that the mean false alarm time is guaranteed to exceed a given positive number.
\begin{equation}
T_{\text{FA}}:=\expect_0 (T) \geq \log a\, .
 \label{eq:Tfalse}
\end{equation}

Under such condition on the false alarm time, Lai  showed that the stopping time \( T \) minimizes  the worst-{worst case} mean delay~\cite{737522,tartakovsky2014sequential} 
    \begin{align}
    \bar\tau^{\star} :=\sup_{\nu\geq 0} \mathrm{ess} \textrm{ } \mathrm{ sup}\; \expect_{\nu}(T-\nu | T > \nu, R_{1:\nu}), 
    \end{align}
This figure of merit considers the worst delay over all possible locations of the change-point and over (almost) all measurement outcomes before the change. A less strict way of assessing the performance of the protocol is the maximal conditional average delay of detection:
\begin{equation}
    \bar\tau :=\sup_{\nu\geq 0} \expect_{\nu}(T-\nu | T > \nu)\leq \bar\tau^{\star}. 
\end{equation}

We note, that as a valuable byproduct, whenever the CUSUM algorithms signals a detection, i.e. $M_n>a$, the value of $k$ which maximizes Eq.~\eqref{eq:Mn}  provides a very good guess of where the change actually occurred. Indeed, from the definition in Eq.~\eqref{eq:CUSUMstat} it is clear that it corresponds the most likely hypothesis $1\leq k\leq n$. That is, the CUSUM algorithm effectively checks at every time step if the current measurement record  is a given factor more likely to be given due to a change at a previous time than due to the default hypothesis. 
    
In the IID setting one can give a closed form formula that quantities the trade-off between the false alarm time and the mean detection time:
\begin{equation}
   \bar\tau \sim \frac{\log T_{\mathrm{FA}}}{D(p_{0}\| p_{1})}\, .
   \label{delaytimes}
 \end{equation}

As we will see in Sec.~\ref{sec:sequentialstrategiesexp}  and Sec.~\ref{sec:ultimatelimits}, in our setting 
a similar relation holds where the regularized expectation value of the likelihood ratio plays the role of the relative entropy $D(p_{0}\| p_{1})$.

The same notation for $T, T_{\mathrm{FA}}, \nu, \bar \tau^*, \bar \tau$ will be used for both index values in a time series and actual times (discrete or continuous); The intended meaning will be clear from the context, or explicitly specified when needed.

\section{\label{sec:experiment}Spin-noise quantum sensor}

This section outlines the experimental setup of the quantum sensor (Sec.~\ref{sec:expdesc}) and the corresponding probability distribution of the measured signal (Subsec.~\ref{sec:probdist}). These elements provide the foundation for calculating the $L_n$ statistic in order to apply the sequential protocols described in Sec.~\ref{sec:seqstrategies}.  However, since computing $L_n$  directly from this distribution in real time is impractical, in Sec.~\ref{sec:kalmanfilter} we present a recursive method to obtain it using the so-called Kalman filter (KF). Finally, in Sec.~\ref{subsec:processasfilter} to enable an analytical and asymptotic analysis of the log-likelihood ratio in further sections, we examine the signal in the frequency domain and treat the dynamics of the system as a linear time invariant filter (LTI). 

The spin-noise system consists of an ensemble of $^{87}$Rb atoms, whose joint state is well described by the collective spin operators $\vecb{J}_t$, with components $J_i(t)$ for $i = x, y, z$. These operators capture the total spin of the ensemble at time $t$ and govern its dynamical evolution. Under the action of an external magnetic field $\vecb{B}=B \vecb{e_x}$, the atomic spin $\vecb{J}_t$ undergoes a precession around the direction of $\vecb{B}$, at the so-called Larmor frequency $\omega_L$, which is proportional to the magnitude of $\vecb{B}$.  This precession can be continuously measured via the magneto-optical Faraday effect, using an off-resonant light beam propagating along the $z$ axis and linearly polarized in the $x$ direction. As the spin precesses, the plane of polarization of the probe beam rotates proportionally to $J_z$, and this rotation is detected by a polarimeter that outputs a photocurrent signal carrying information about the spin dynamics, superimposed with optical shot noise.

\begin{figure}[t]
    \centering   \includegraphics[width=0.5\textwidth]{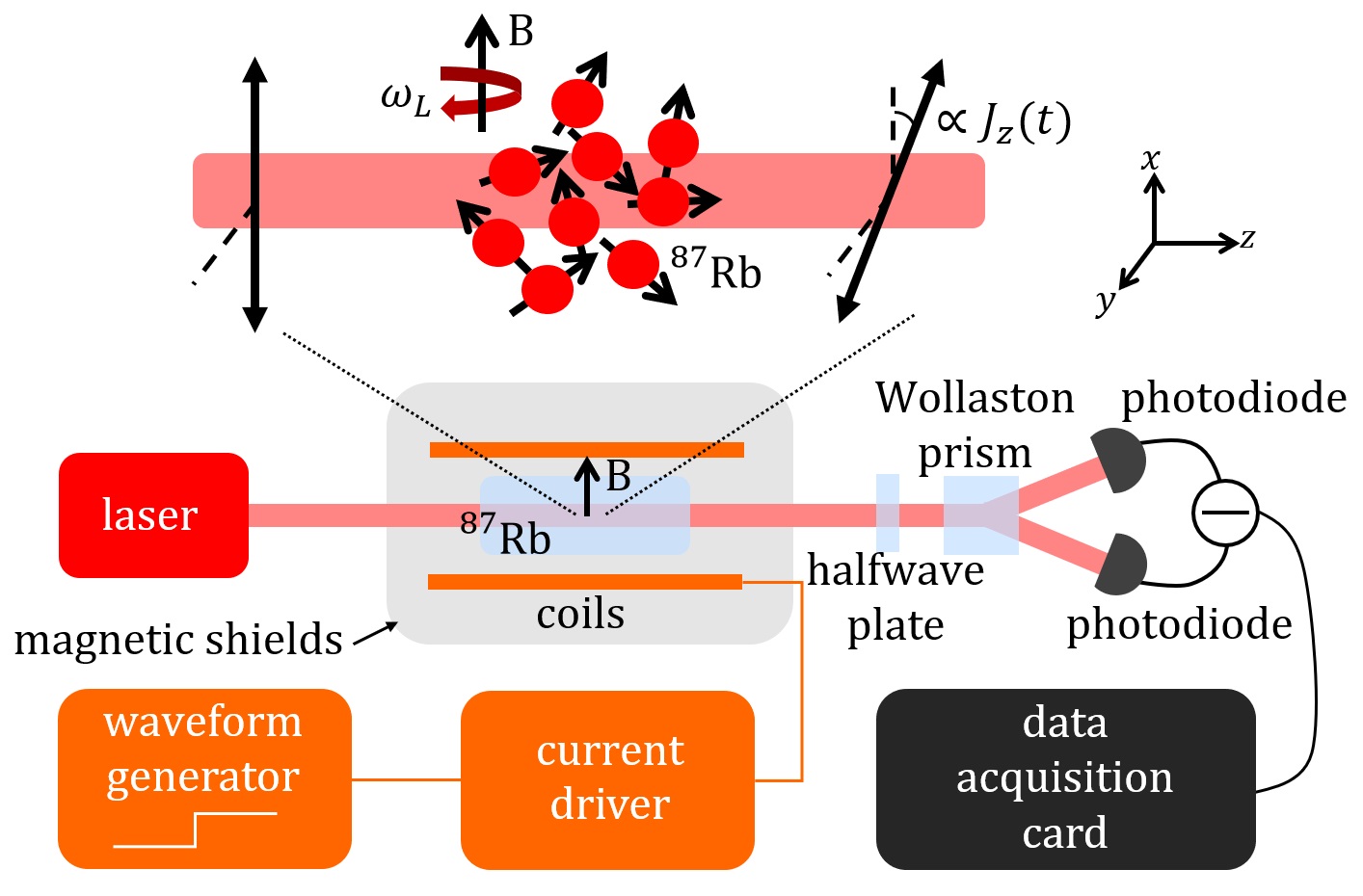}    
 \caption{Scheme of the atomic sensor: An ensemble of $^{87}$Rb atoms precesses at the Larmor frequency $\omega_L$ under the action of an external magnetic field $\mathbf{B}$. A probe laser beam monitors this precession via the Faraday effect, generating a photocurrent proportional to $J_z$, the $z$ component of the collective spin, with added shot noise. }
    \label{expscheme}
\end{figure}

The experimental set-up is shown schematically in Figure \ref{expscheme}. 

\subsection{\label{sec:expdesc} Experimental 
details}

Isotopically enriched $^{87}\mathrm{Rb}$ atoms are contained in a \SI{3}{\centi\meter}-long vapor cell filled with \SI{100}{\torr} of N\textsubscript{2} buffer gas (pressure specified at room temperature). The cell is enclosed in four layers of mu-metal shielding to suppress environmental magnetic fields. The atoms are heated inside a ceramic oven to \SI{102}{\celsius} using intermittent Joule heating, yielding an atomic number density of \SI{8.2e12}{atoms\per\centi\meter\cubed}. Induction coils driven by a waveform generator (Keysight 33500B) followed by a low-noise current source (TwinLeaf CSUA300) generate dc magnetic fields along the $x$-axis with strengths  around $B \approx \SI{7.22} {\micro\tesla}$, corresponding to a Larmor frequency of $\omega_{L} \approx 2\pi \times \SI{50.5}{\kilo\hertz}.$ The bias magnetic field step was implemented via a pulse function from the waveform generator and synchronized with the data acquisition system. 

The polarization rotation of a linearly polarized probe beam (\SI{830}{\micro\watt}) is measured with a shot-noise-limited polarimeter consisting of a half-wave plate, a Wollaston prism, and a balanced photodetector (Thorlabs PDB450A). The probe is derived from a tunable diode laser (Toptica DL100) operating at \SI{795}{\nano\meter}, frequency-stabilized via a fiber interferometer \cite{Kong2015} to be \SI{27}{\giga\hertz} blue-detuned from the D\textsubscript{1} line. The detector output is digitized with a National Instruments PCI-4462 data acquisition card with 200kSa/s sampling rate. Data were collected in 200 repeated samples, each acquisition lasting 2 seconds.

\subsection{\label{sec:probdist} Physical model and underlying probability distribution of the signal}

The stochastic model that describes the dynamics of the components of spin and the measured photocurrent $I$ is described in detail in \cite{Jimenez2018Signal}. Relying on the Itô formulation for stochastic differential equations (SDEs)  \cite{jacobs2010stochastic}, the dynamics of the spin components in the transverse $yz$-plane, $\boldsymbol{J}_t:=(J_y(t),J_z(t))^T$, 
are described by the Ornstein–Uhlenbeck process: 
\begin{equation}
d\boldsymbol{J}_t= \hat A \cdot\boldsymbol{J}_{t} dt+ \hat N \boldsymbol{ d W}  \textrm{ with }  \hat A = \begin{pmatrix}
           -\gamma & \omega_L \\
            -\omega_L &-\gamma 
           \end{pmatrix}, 
\label{syquation}
\end{equation}
 $\hat N=\sqrt{Q} \mathbb{1}_2$ and  $\boldsymbol{dW}=\left(dW_x, dW_y\right)^T$ with $dW_{x}$ and $dW_{y}$ two independent Wiener noises. 
The matrix $\hat A$ encapsulates coherent Larmor precession at frequency $\omega_L$  and transverse relaxation at rate  $\gamma$. Both $\gamma$ and the noise term with strength $\hat N$ originate from different noise processes including atomic collisions, optical depolarization, and transit-time broadening. Together, these parameters define a linear Gaussian model for the spin dynamics under continuous monitoring.

The photocurrent $I$ generated by the optical detection process  is sampled at finite time intervals  $t_k= k\Delta$, where $k \in \mathbb{N}$  and $\Delta$ denotes the sampling period.  The photocurrent captures information about the collective spin while being affected by optical shot-noise $\xi(t_k)$, which is assumed to be Gaussian and statistically independent of $\vecb{J}_t$. The sensor output is thus described by the discrete-time stochastic equation:
\begin{equation}
     I_k=\hat C\boldsymbol{J}_{t_k}+\xi(t_k)=g_DJ_z(t_k)+\xi(t_k),
     \label{signal}
\end{equation}
where $\hat C=g_D(0,1)$ defines the measurement coupling, with $g_D$  the transduction constant. The measurement noise $\xi(t_k)$ is zero-mean Gaussian white noise  $\xi(t_k) \sim \mathcal{N}(0,G_k)$, with variance  $G_k=S_{\mathrm{ph}}/\Delta$ $\forall k \in \mathbb{N}$, where $S_\mathrm{ph}$ is the spectral density of the photon shot noise.

In the steady state, the probability distribution of $n$ photocurrent observations, $\boldsymbol{I}_n=\left(I_1,..,I_{n}\right)^T$, follows a multivariate normal distribution $\boldsymbol{I}_n\sim \mathcal{N} (\boldsymbol{\mu}_n,\hat T_n )$, with mean $\boldsymbol{\mu}_n=\vecb{0}$ and covariance matrix $\hat T_n$  with elements (see Appendix \ref{sec:probdiststeadystate}):
\begin{align}\label{eq:CovarianceME}
T_{i,j}&=K_{i-j} \mbox{ with }\\
K_k&:=g_D^2 \frac{Q}{2\gamma}\mathrm{e}^{-\gamma |k|\Delta} \cos\left(k \omega_L\Delta\right)+\delta_{ k,0}\frac{S_{\mathrm{ph}}}{\Delta}.\nonumber
\end{align}
Explicitly:
\begin{equation}
    p(I_1,...,I_n)=\frac{\exp{\left(-\frac{1}{2}\boldsymbol{I}_n^T \hat T_n^{-1}\boldsymbol{I}_n\right)}}{\sqrt{(2\pi)^n \det \hat T_n}}\, .
    \label{probgaussian}
\end{equation}
From  Eq.~\eqref{probgaussian} one can see that the log-likelihood ratio $L_{I,n}$ in Eq.~\eqref{llr} is given by
\begin{equation}
L_{I,n}=\frac{1}{2}\log\left(\frac{\det\hat T_{0,n}}{\det \hat  T_{1,n}}\right)+\frac{1}{2}\boldsymbol{I}_n^T \left(\hat T_{0,n}^{-1} -  \hat T_{1,n}^{-1}\right)\boldsymbol{I}_n,
\label{LIToeplitz}
\end{equation}
where $\hat T_{h,n}$ stands for the covariance matrix under hypothesis $h$ after $n$ measurements. Notice that, since the process is stationary, each descending diagonal (from left to right) of the covariance matrix is constant, i.e. $\hat T_n$ is a Toeplitz matrix. We will exploit this structure, as the spectral properties of Toeplitz matrices and their associated functionals are well studied (see Appendix \ref{sec:Toeplitz}). 
For instance, the mean of $L_{I,n}$ can be immediately read-off Eq.~\eqref{LIToeplitz}, by noting that
$\expect_h[\boldsymbol{I}_n\boldsymbol{I}_n^T]=\hat T_{h,n}$. Similarly, the characteristic function of $L_{I,n}$ under hypothesis $h$, which completely determines its probability density function after $n$ observations, can be computed as:
\begin{equation}
    \phi_{L_I}(u):=\expect_h[\mathrm{e}^{i u L_I}]=\frac{\eta^{\tfrac{iu}{2}}}{\sqrt{\det(\mathbb{1}-iu \hat D \hat T_{h,n})}},
\end{equation}
where we defined $\hat D:=\hat T_{0,n}^{-1} -  \hat T_{1,n}^{-1}$ and $\eta=\frac{\det\hat T_{0,n}}{\det\hat  T_{1,n}}$. From here we can easily  obtain the $m$-th cumulant $\kappa_{m,n}$ of the LLR at time step $n$: 
\begin{align}
 \kappa_{m,n}^h&=i^{-m} \frac{d^m \log\phi(u)}{du^m}|_{u=0}\nonumber \\
&= \frac{1}{2}(m-1)! \;\textrm{Tr}\left(  (\hat D \hat T_{h,n})^m\right)+\delta_{m,1} \frac{1}{2}\log\eta\, .
 \label{ctoeplitz}
\end{align}
In the limit of large $n$, for Toeplitz matrices arising from short-memory stationary processes, such as those defined in Eq.~\eqref{LIToeplitz}, the operator traces appearing in Eq.~\eqref{ctoeplitz} can be computed (see  Appendix \ref{sec:Toeplitz}):
    \begin{align}
     \kappa_{m,t}^h&=\frac{t}{2\pi}(m-1)!\int_0^\infty \left(\frac{\bar S_{I,h}(\omega)}{\bar S_{I,0}(\omega)}-\frac{\bar S_{I,h}(\omega)}{\bar S_{I,1}(\omega)} \right)^m d\omega +\nonumber\\ 
    &+\frac{t}{2 \pi}\delta_{m,1}\int_0^\infty \log\left(\frac{\bar S_{I,0}(\omega)}{\bar S_{I,1}(\omega)} \right)d\omega +  \mathcal{O}\left(1\right)
    \label{cumulants}
\end{align}
where $t=n\Delta$ and $\bar S_{I,h}(\omega)$ is the Power Spectral Density (PSD) of the signal, given by the Fourier transform of the autocorrelation
function, i.e. 
\begin{align}
\label{lorentzian}
       &\bar S_{I,h}(\omega):=\int_{-\infty}^\infty K_{h,t} e^{-i\omega t}dt=\\
       &= S_{\mathrm{at}}\left(\frac{\gamma^2}{\gamma^2+ (\omega-\omega_L)^2}+\frac{\gamma^2}{\gamma^2+ (\omega+\omega_L)^2}\right)+S_{\mathrm{ph}}\nonumber
\end{align}
with $S_{\mathrm{at}}:=\frac{g_D^2Q}{2\gamma^2}$.   Here, and in the following, the sub-leading $\mathcal{O}(\Delta)$ terms from the continuum approximation are omitted for clarity.
Also, with a slight abuse of notation,  we write $K_t$ for the continuous-time analogue of the discrete $K_k$ of Eq.~\eqref{eq:CovarianceME} in the limit $\Delta \to 0$, and similarly generalize this notation to other quantities.

Equation~\eqref{cumulants} shows that the mean, variance, and higher-order cumulants of the log-likelihood ratio $L_{I}$ grow linearly with time (up to subleading constant corrections), indicating that the distribution of $L_I$ becomes increasingly peaked and concentrates around its mean.

\subsection{\label{sec:kalmanfilter} Hybrid Kalman filter for probability distribution of the signal}

The Kalman Filter (KF) provides a recursive Bayesian algorithm to infer the hidden state of linear Gaussian systems. It is widely used to estimate time-varying signals, as it yields the optimal estimator minimizing the mean-squared error (MSE) \cite{kalman_new_1960, bucy1961new}, e.g. in \cite{Jimenez2018Signal}. In our case, we use the KF not to estimate the hidden signal itself, but to recursively compute the probability distribution in Eq.~\eqref{probgaussian}.

The main idea is that the probability of the full measurement record $p(I_1, \dots, I_n)$ can be built iteratively via the chain rule. The recursion starts from a prior $p(\boldsymbol{J}_{t_0})$, which we take as the stationary distribution of the hidden process under the hypothesis being tested.  The state is then updated recursively through prediction and update steps as new observations become available.

At step $k=1$, after observing $I_1$, we update the prior via Bayes' rule:
\begin{equation}
p(\boldsymbol{J}_{t_1} | I_1) = \frac{p(I_1 | \boldsymbol{J}_{t_1}) \, p(\boldsymbol{J}_{t_1})}{\int p(I_1 | \boldsymbol{J}_{t_1}) \, p(\boldsymbol{J}_{t_1}) \, d\boldsymbol{J}_{t_1}}.
\end{equation}

For $k > 1$, the prediction step propagates the posterior via the state-space transition model:
\begin{align}
p(\boldsymbol{J}_{t_k} | I_{1:k-1}) 
&= \int p(\boldsymbol{J}_{t_k} | \boldsymbol{J}_{t_{k-1}})\, 
p(\boldsymbol{J}_{t_{k-1}} | I_{1:k-1})\, d\boldsymbol{J}_{t_{k-1}},\nonumber
\end{align}
followed by the update:
\begin{align}
p(\boldsymbol{J}_{t_k} | I_{1:k}) 
&= \frac{p(I_k | \boldsymbol{J}_{t_k})\, p(\boldsymbol{J}_{t_k} | I_{1:k-1})}{p(I_k | I_{1:k-1})},
\end{align}
with normalization:
\begin{align}
p(I_k | I_{1:k-1}) 
&= \int p(I_k | \boldsymbol{J}_{t_k})\, 
p(\boldsymbol{J}_{t_k} | I_{1:k-1})\, d\boldsymbol{J}_{t_k}.
\end{align}
where we have used the shorthand notation $I_{1:k} \equiv (I_1, \ldots, I_k)$.

Since both the dynamical model and the measurement model are Gaussian, all these integrals can be computed analytically, yielding Gaussian distributions at each step \cite{gurajala2021derivation}. In particular, $p(I_k | I_{1:k-1}) = \mathcal{N}({\mu}_k, {\sigma}_k^2)$
where the mean \({\mu}_k\) and variance \({\sigma}_k^2\) are explicit symbolic functions of the model parameters and the previous data, computed recursively through the Kalman filter equations.

By iterating these steps, and using the chain rule we obtain the full likelihood of the observed data as
\begin{align}
p(I_{1:n}) = \prod_{k=1}^n p(I_k | I_{1:{k-1}})
= \prod_{k=1}^n \frac{\mathrm{e}^{-\tfrac{(I_k - {\mu}_k)^2}{2 \sigma_k^2}}}{\sqrt{2\pi \sigma_k^2}} .
\label{probkalman}
\end{align}

When comparing two hypotheses \(h=0,1\), where each hypothesis specifies different system parameters and therefore distinct Kalman estimators \(({\mu}_{k,h}, {\sigma}_{k,h}^2)\), the log-likelihood ratio  reads:
\begin{equation}
\begin{aligned}
L_{I,n} = \sum_{k=1}^n  &\left(\log \frac{{\sigma}_{k,0}}{{\sigma}_{k,1}} + \frac{1}{2 {\sigma}_{k,0}^2} (I_k - {\mu}_{k,0})^2\right. \\
&\quad \left.
- \frac{1}{2 {\sigma}_{k,1}^2} (I_k - {\mu}_{k,1})^2\right).
\label{lrkalman}
\end{aligned}
\end{equation}

Notice that Eq.~\eqref{lrkalman} corresponds to the discrete-time version of the statistic derived in \cite{gasbarri_sequential_2024}. To account for continuous dynamics with discretely sampled measurements, we employ the continuous-discrete (hybrid) Kalman filter. The explicit algorithm is detailed in Appendix~\ref{sec:Kalmanfilter}.

\subsection{\label{subsec:processasfilter} Process as a filter and probability distribution in the asymptotic regime 
}
As mentioned in the beginning of Sec.~\ref{sec:experiment}, the dynamics of observations comprised in Eqs.\eqref{syquation}-\eqref{signal}, can be understood as a linear time-invariant (LTI) filter that inputs Wiener noise and outputs the detected photocurrent.

A complete characterization of such continuous-time systems is provided by the transfer matrix function, which describes the system in the Laplace domain and fully specifies the response of the system to any input. By taking the Laplace transform, denoted by $\mathcal{L}$, of Eq.~\eqref{syquation} we obtain:
\begin{equation}
\mathcal{L}(\vecb{J}_t) = \hat H_D(s) \mathcal{L}(\vecb{W}_t),
\label{laplacecont}
\end{equation}
where the transfer matrix is 
$\hat H_D(s)=(s \mathbb{1} - \hat A)^{-1}$. 

Restricting the Laplace variable to the imaginary axis, $s = i\omega$, yields the Fourier transform, $\mathcal{F}$, of the system:
\begin{equation}
\mathcal{F}(\vecb{J}_t) = \hat H_D(i \omega) \mathcal{F}(\vecb{W}_t),
\label{fouriercont}
\end{equation}
This expression allows us to understand the process as a simple input-output relation in frequency domain in the  long-time limit.
The output inherits the essential statistical features of the input, shaped by the filter’s frequency response.
In particular, when $\vecb{W}_t$ is a stationary Gaussian process\footnote{For continuous-time stochastic processes (e.g., Wiener noise appearing here) that are non-square-integrable, the Fourier transform $\mathcal{F}(X_t)$ must be understood as a finite-time, normalized limit:
$\mathcal{F}(J_t)(\omega) := \lim_{T \to \infty} \frac{1}{\sqrt{T}} \int_0^T J_t e^{-i \omega t} \, dt$
which ensures the transform is well-defined and properly captures the long-time spectral content of the process~\cite{trees2001detection}.}, $\vecb{J}_t$ remains stationary and Gaussian.
By multiplying Eq.~\eqref{fouriercont} by its conjugate transpose,  one can get a connection between their asymptotic multivariate periodograms,
\begin{equation}
\vecb{S_{J}}(\omega)=\big |\hat H_D(i \omega)\big |^2\vecb{S_{W}}(\omega),
\end{equation} 
Evaluating $|\hat H_D(i \omega) |^2$ and taking the white noise input $\expect{[{\boldsymbol{S_W}}(\omega)]}=Q \mathbb{1}_2$ one recovers the well-known Lorentzian line-shape centered at the Larmor frequency: 
\begin{equation}
    \expect{[S_{J_z}(\omega)]}=\frac{Q/2}{\gamma^2+ (\omega-\omega_L)^2}+\frac{Q/2}{\gamma^2+ (\omega+\omega_L)^2}
    \label{lorentzian2}
\end{equation}
defined over $\omega\in (-\infty,\infty)$. 
Note that, since the signal is real-valued, the Fourier components at positive and negative frequencies are always complex conjugates of each other. In addition, in the regime where $\omega_L \gg \gamma$, the Lorentzian peak centered at $-\omega_L$ contributes negligibly to the spectrum in the positive-frequency domain. It is therefore sufficient to consider only the positive-frequency branch:
\begin{equation}
\mathbb{E}[S_{J_z}(\omega)] \approx \frac{Q/2}{\gamma^2 + (\omega - \omega_L)^2}, \quad\quad \omega \in (0, \infty).
\end{equation}
To obtain the output signal from Eq.~\eqref{signal} we need to take into account that $J_z$ enters in the photocurrent description sampled at periods $\Delta$ and therefore an analogous analysis in the discrete case must be carried out. In this scenario, the equivalence of the $s$-domain representation is the so-called $z$-domain. The discrete version of the transfer function  for our system is \begin{equation}
    \hat H_D^d(z)=(\mathbb{1}_2-z^{-1}(\mathbb{1}_2+\hat A \Delta))^{-1},
\end{equation}
and the frequency response, i.e. restricting $z=\mathrm{e}^{-i \omega \Delta}$,
\begin{equation}
    \hat H_D^d(\mathrm{e}^{i \omega \Delta})=(\mathbb{1}_2-\mathrm{e}^{-i \omega \Delta}(\mathbb{1}_2+\hat A \Delta))^{-1},  
\end{equation}
with $\omega\in\left(-\frac{\pi}{\Delta},\frac{\pi}{\Delta}\right)$. 
Notice that now the frequency resolution is bounded by the sampling frequency $\frac{1}{\Delta}$. Finally, for a small $\Delta$, one can relate the continuous and discrete versions:
\begin{equation}
  |\hat H_D^d(\mathrm{e}^{i \omega \Delta})|^2 = \frac{|\hat H_D(i \omega)\big |^2 }{ \Delta} + O(1)\,.
\end{equation}

Equipped with these results, and noting that the detected signal, $I$ in Eq.~\eqref{signal}, is linearly coupled to $\boldsymbol{J}$ plus some added independent background noise, we can derive its
PSD in the limit of high sampling rate, i.e. small $\Delta$, and recover the Lorentizan profile found in Eq.
\eqref{lorentzian}.

Following the structure of the previous two subsections, we now present the asymptotic probability distribution of the signal in the continuum, long-time limit. 

As discussed above, the statistical properties of the output signal $I_t$ are inherited from those of the input white noises. In particular, stationarity implies that for any time shift $\tau$, the process $I_{t+\tau}$ has the same statistics as $I_t$. Consequently, the Fourier components $\mathcal{F}_I(\omega)$ and $e^{-i\omega \tau} \mathcal{F}_I(\omega)$ must be identically distributed, which implies that the phase of $\mathcal{F}_I(\omega)$ is uniformly distributed over $[0, 2\pi)$. Since $\mathcal{F}_I(\omega)$ is a linear functional of a Gaussian process, it is itself a complex Gaussian variable. The uniformity of the phase then ensures that the real and imaginary parts of $\mathcal{F}_I(\omega)$ are independent, identically distributed Gaussian variables. 

Moreover, stationarity also implies that the Fourier components $\mathcal{F}_I(\omega)$ and $\mathcal{F}_I(\omega')$ are uncorrelated for $\omega \neq \omega'$ (excluding the case $\omega' = -\omega$, which is perfectly correlated for real-valued signals, as noted above),
and since they are jointly Gaussian, they are statistically independent. Therefore, in the frequency domain, $\mathcal{F}_I(\omega)$ consists of independent complex Gaussian variables across frequencies, each with independent real and imaginary parts.

As a result  the power spectral density $S_I(\omega) = |\mathcal{F}_I(\omega)|^2$ is the sum of the squares of two independent, identically distributed Gaussian variables, i.e. it follows a $\chi^2$ distribution with two degrees of freedom, or equivalently, an exponential distribution:
\begin{equation}
    p(S_I(\omega)) = \frac{1}{\bar S_I(\omega)}\, \mathrm{e}^{- S_I(\omega)/\bar S_I(\omega)},
    \label{exponentialdist}
\end{equation}
where $\bar S_I(\omega) = \expect{[S_I(\omega)]}$ is the mean power spectrum at frequency $\omega$ from Eq.~\eqref{lorentzian} (see also \cite{LuciveroPRA2016, LuciveroPRA2017}). 

The joint probability distribution of the signal across all frequencies factorizes into a product of exponentials given by Eq.~\eqref{exponentialdist} at each frequency. In the continuum limit, this leads to the following expression for the asymptotic log-likelihood ratio (LLR):
\begin{equation}
\begin{aligned}
    &  L_{S_I} =\frac{t}{2\pi}\int_0^\infty \log\left(\frac{\bar S_{I,0}(\omega)}{\bar S_{I,1}(\omega)} \right)d\omega + \\
    &+\frac{t}{2\pi}\int_0^\infty \left(\frac{1}{\bar S_{I,0}(\omega)}-\frac{1}{\bar S_{I,1}(\omega)} \right) S_I(\omega) d\omega\\
\end{aligned}
\label{LIf}
\end{equation}
That is, the LLR is given by a weighted sum of independent exponential random variables,
and yields cumulants matching the leading-order terms in Eq.~\eqref{cumulants}, as expected in the large-$n$ limit. Indeed, these results are fully consistent with the asymptotic diagonalization of Toeplitz covariance matrices discussed in Sec.~\ref{sec:probdist}, where the Fourier basis asymptotically diagonalizes such matrices and the PSDs appear as their eigenvalue spectra \cite{graypaper,grayreview}. There exist a well-established body of results (see Sec.~\ref{sec:moments}),  which provides precise asymptotic expressions for functionals of Toeplitz matrices together with rigorous results on their convergence rates. This framework will allow us to systematically connect the infinite-size limit with the corrections that arise at finite~$n$.

\section{\label{sec:experimentaldata}Experimental data, high-pass filter and log-likelihood ratio}
\subsection{Experimental data and  notch filter}

\begin{figure}[t]
    \centering
    \includegraphics[width=.487\linewidth]{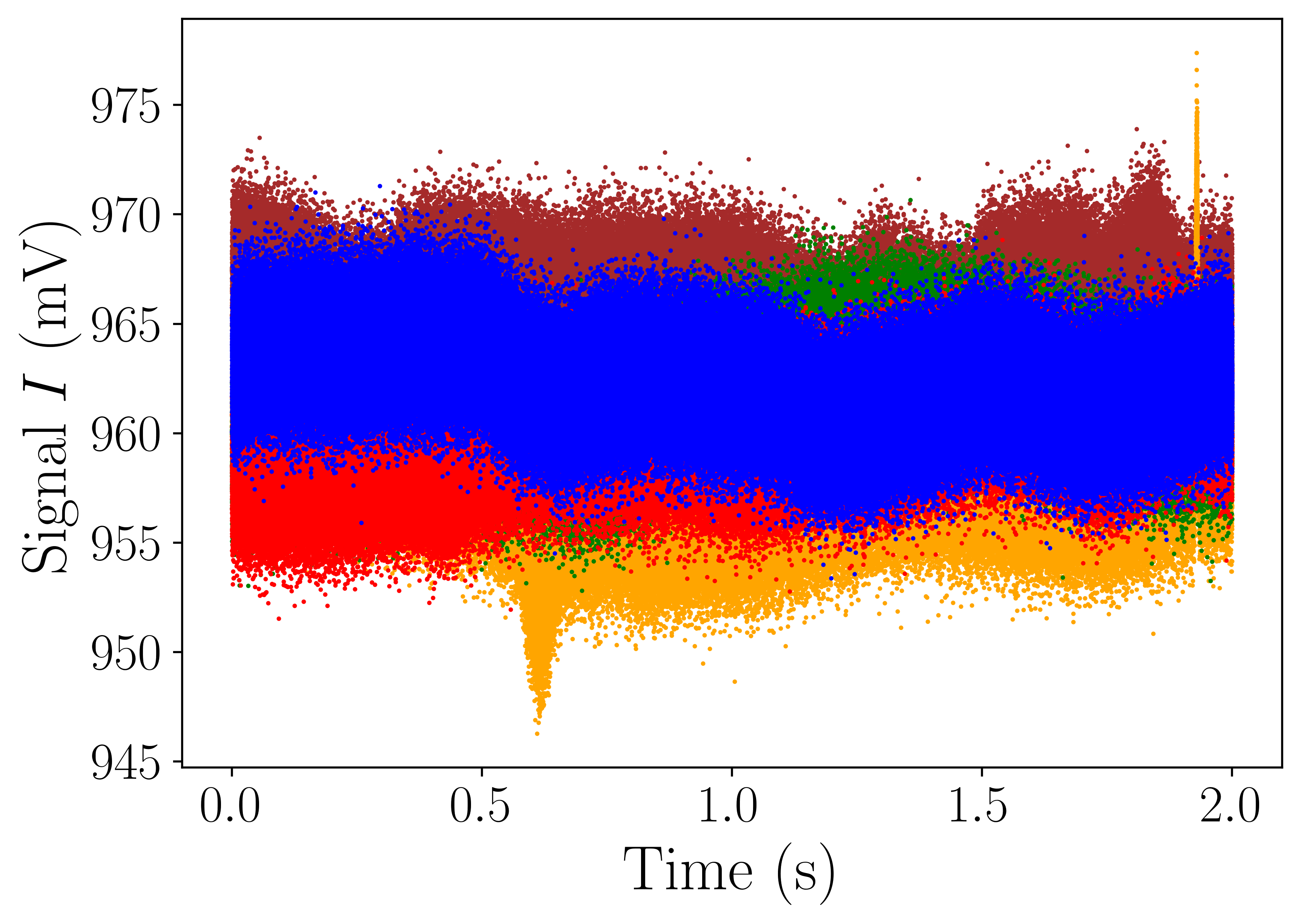}  
    \hfill
    \includegraphics[width=.501\linewidth]{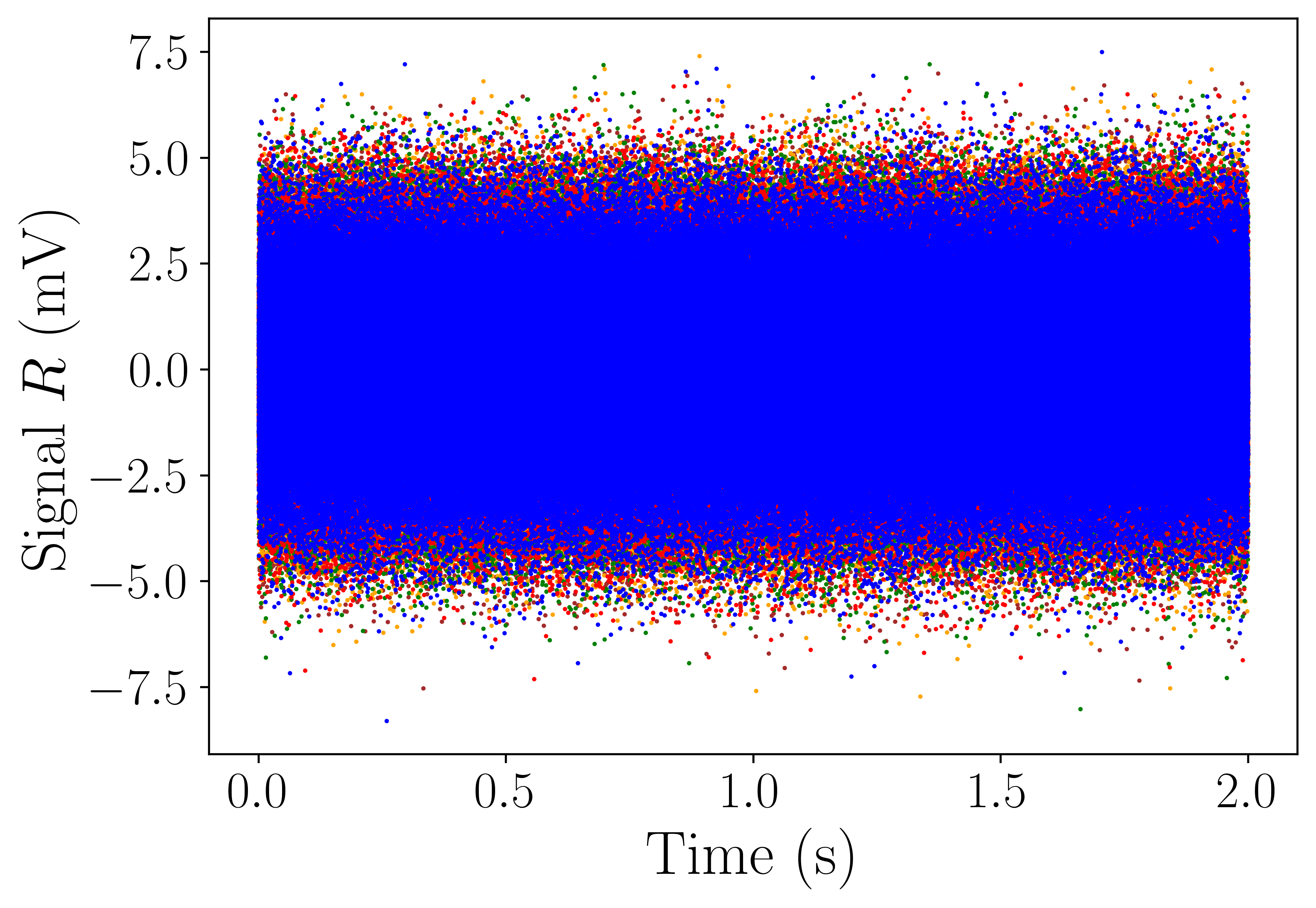}
        
    \includegraphics[width=.995\linewidth]{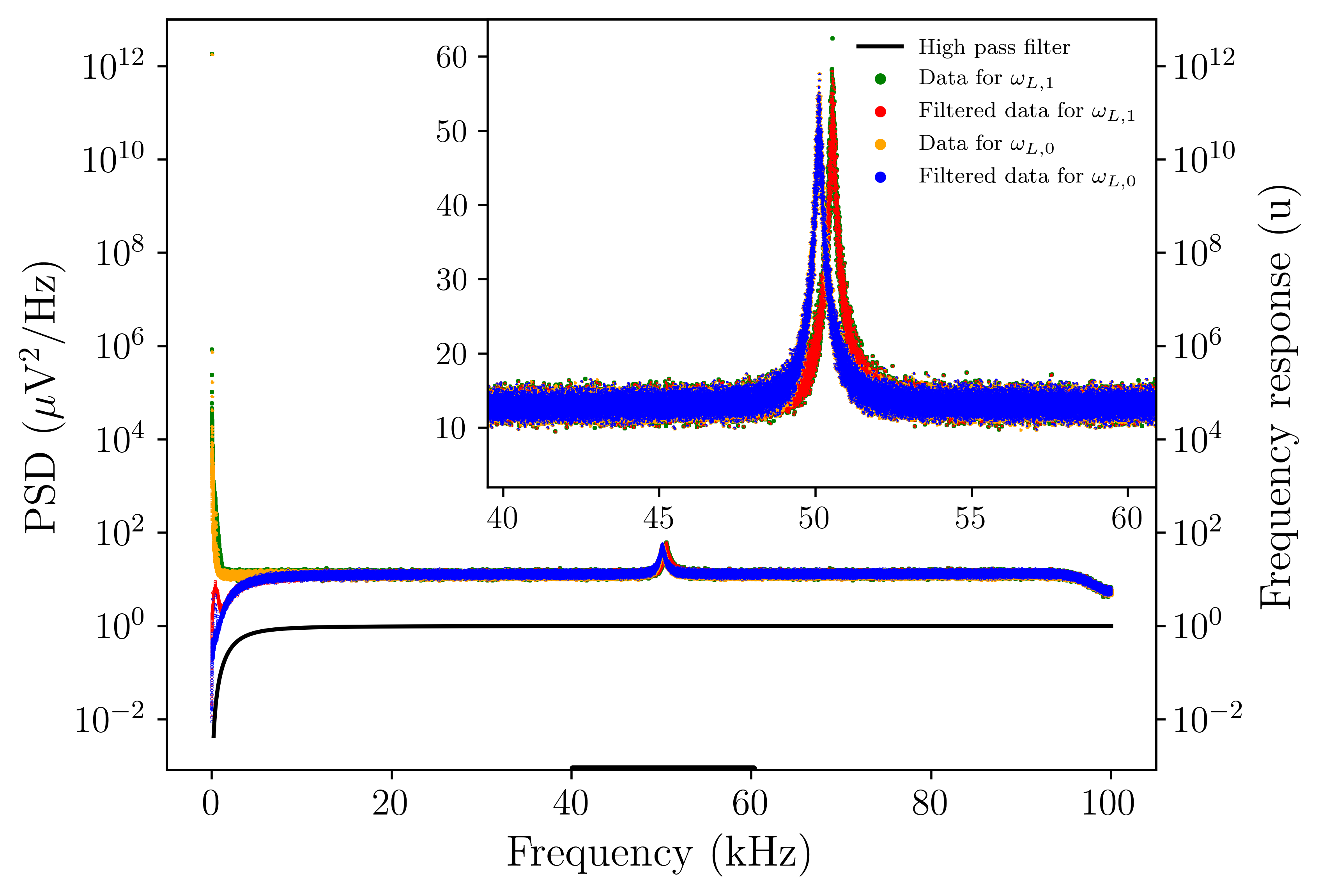}  
    
    \caption{
    Representative time- and frequency-domain traces from two datasets, corresponding to hypotheses differing in the magnitude of the magnetic field, i.e., differing in our model, Eqs.~\eqref{syquation}-\eqref{signal}, only through the Larmor frequencies, labeled $\omega_{L,0}$ and $\omega_{L,1}$. The values of $\omega_L$ and other model parameters are estimated in Sec.~\ref{sec:calibration}. Top: example time traces before (left) and after (right) high-pass filtering. Bottom: PSDs averaged over $200\times 2 \mathrm{s}$ traces, before (green, orange) and after (red, blue) filtering. The frequency response of the filter is indicated on the right-hand y-axis, and the inset (same units as main) zooms into $40–60\mathrm{kHz}$ near the Larmor peaks. 
    }
    \label{rawdata}
\end{figure}
The detection strategies considered in this work rely on the assumption that the observed data align with the underlying theoretical model. However, the current physical model does not account for low-frequency spurious noise, which is particularly prominent near zero frequency and is ubiquitous in laboratory environments. Figure~\ref{rawdata} illustrates examples of random traces corresponding to two different high Larmor frequencies, where the presence of such low-frequency noise can be clearly observed.

A complete characterization of this spurious noise that could be incorporated into our model is futile, due to its diverse and unpredictable origins. Instead, we address the issue by applying a Linear Time Invariant (LTI) causal filter to the observations.
The filter must satisfy three essential properties: first, it must be implementable in real time; second, it must effectively suppress the unwanted noise; and third, its impact on the relevant information in the data, specifically around the Lorentzian peak,  must be minimal. This last point will be verified by quantifying the influence of the filter on our primary statistic, $L_t$, as will be shown in the next subsection.

For this purpose, we  apply to the signal $\boldsymbol{I}_n$ the following infinite impulse response high-pass filter, designed to eliminate the zero-frequency component and attenuate low-frequency noise.  Denoting the output by  $\boldsymbol{R}_n=\left(R_1,..,R_{n}\right)^T$, the resulting signal is: 
\begin{equation}
        R_k= \alpha R_{k-1} + \sqrt{\alpha} (I_k-I_{k-1}),
        \label{notcheq}
\end{equation}
where the parameter $0 < \alpha < 1$  regulates  the strength of the filter.

Figure \ref{rawdata} shows the result of applying the high-pass filter to the time traces. The filtered signals are now centered around zero, with low-frequency components effectively suppressed, yielding data that better matches the model predictions of Eqs.~\eqref{syquation}-\eqref{signal}.

To understand the impact of the high-pass filter, we start by noting that the transformation from $\boldsymbol{I}_n$ to $\boldsymbol{R}_n$ is affine, and therefore the Gaussianity of the observations is preserved. Now   $\boldsymbol{R}_n\sim \mathcal{N} (0,\hat \Sigma_n)$, with 
$\boldsymbol{R}_n=\hat B_n \boldsymbol{I}_n$ and $\hat \Sigma_n=\hat B_n\hat T_n \hat B_n^T$,
where $\hat{T}_n$ and $\hat B_n$ are $n\times n$ Toeplitz matrices with matrix elements respectively given by Eq.\eqref{eq:CovarianceME} and $B_{i,j}=\sqrt{\alpha}(\alpha-1)\alpha^{|i-j|}+\sqrt{\alpha}(2-\alpha)\delta_{i,j}$. After the filter stabilizes, $\hat \Sigma_n$ also becomes a Toeplitz matrix with diagonal entries $\Sigma_k$ that for high Larmor frequencies, i.e., far from spurious noises,
\begin{equation}
\begin{aligned}
        \Sigma_k&\approx\left(1-\frac{b^2}{\omega_L^2}\right) T_k-\\
        &-b\left(\frac{S_{\mathrm{ph}}}{2}+\frac{g_D^2 Q \gamma}{2\omega_L^2}\right)\mathrm{e}^{-b |k|\Delta}+\mathcal{O}\left(\frac{1}{\omega_L^3}\right),
\end{aligned}
\label{corrnotch}
\end{equation}
where  $b$ is implicitly defined via $\alpha = e^{-b\Delta}$ to facilitate taking the continuous-time limit.
Notice that $ \Sigma_k$ is essentially proportional to $ T_k$ ---the correlations that will give rise to the Lorentzian centered at $\omega_L$--- since the term ${b^2}/{\omega_L^2}$ is small in this regime. The other term in Eq.~\eqref{corrnotch} is a contribution at zero-frequency that will be negligible by the functional in the LLR expression. 

\begin{figure}[t]
    \centering   
    \includegraphics[width=0.48\textwidth]{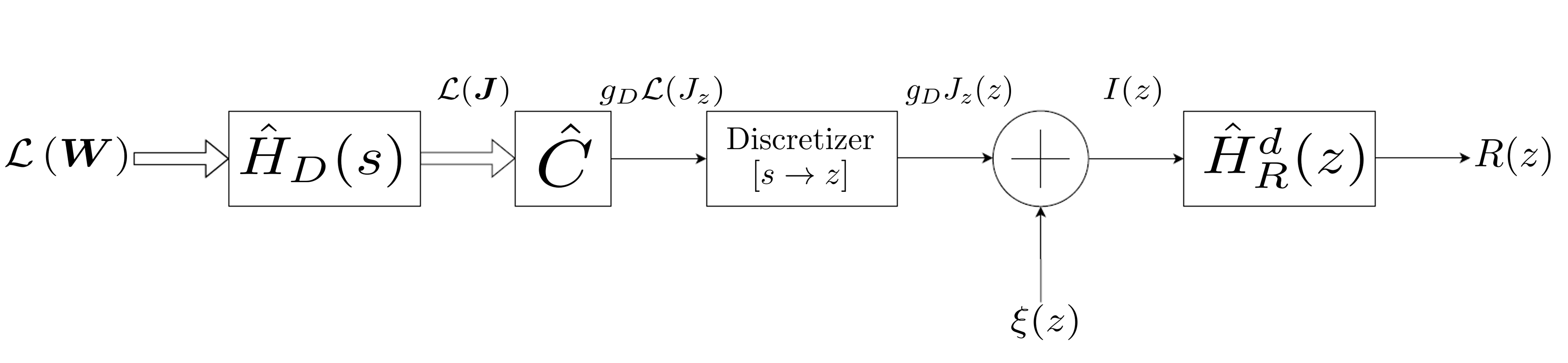}  
 \caption{Block diagram of the process represented with the transfer matrices}
    \label{processasfilter}
\end{figure}

In the frequency domain, the asymptotic transformation from $S_I(\omega)$ to $S_R(\omega)$ is given by the transfer function $H_R^d (z)  =  \sqrt{\alpha} \frac{1-z^{-1}}{1- \alpha z^{-1}}$ evaluated at $z=\mathrm{e}^{-i\omega\Delta}$. Again, taking into account $\alpha=\mathrm{e}^{-b \Delta}$, the square modulus of the frequency response satisfies:
\begin{equation}
\begin{aligned}
        \rvert H_R^d (\mathrm{e}^{i \omega \Delta}) \rvert^2 &= 2 \alpha \frac{1-\cos{(\omega\Delta)}}{1 + \alpha^2 - 2 \alpha \cos{( \omega\Delta)}} =\\
        &=\frac{\omega^2}{b^2+\omega^2} +\mathcal{O}(\Delta), \quad \omega \in \left(-\frac{\pi}{ \Delta},\frac{\pi}{ \Delta}\right)
\end{aligned}
    \label{filtrer}
\end{equation}
where $ \rvert H_R (i \omega ) \rvert^2=\frac{\omega^2}
{b^2+\omega^2}$ is the continuous version. Now one can see how this filter has the effect of eliminating the zero frequency component, reducing the low frequency noise, and at the same time leaving the high-frequency components largely untouched. Other filters, such as the moving average filter or the Kalman filter, could reduce zero-frequency spurious noise, but it would not have the desired effect of annihilating it. Indeed, for large frequencies:
\begin{equation}
\begin{aligned}
       S_R(\omega)&= \frac{\omega^2}{b^2+\omega^2} S_I(\omega)=\\
       &=\left(1-\frac{b^2}{\omega^2}\right)S_I(\omega) + \mathcal{O}\left(\frac{1}{\omega^3}\right).
\end{aligned}
\label{eq:SI1b2}
\end{equation}
For high $\omega_L$, the high-pass filter has negligible effect in the region of interest, i.e. near the Lorentzian peak. 
 Figure~\ref{rawdata} shows the PSD averaged over 200 experimental traces, with and without filtering. The spurious low-frequency noise exceeds the Lorentzian peak by nearly ten orders of magnitude, justifying the use of the high-pass filter. 
Figure~\ref{processasfilter} presents a diagram of the complete process, interpreted as an effective filter.  

\subsection{\label{sec:moments}
Effect of the high-pass filter on the log-likelihood ratio
}
In this subsection, we assess how the test statistic $L_I(\boldsymbol{I}_n)$ is affected when the available data is the filtered signal $\boldsymbol{R}_n = \hat{B} \boldsymbol{I}_n$ rather than the modeled input $\boldsymbol{I}_n$. 
While it is by now clear that in the long-time limit the filter has negligible impact on the relevant parts of the spectrum, our sequential methodology, based on applying a Kalman filter at each step, requires careful assessment of subleading corrections at moderate times. In particular,  it is important to verify that these corrections remain small and that the low-frequency components suppressed by the filter do not contribute significantly to the test statistic.

Since the log-likelihood ratio depends only on the observed data, we define
\[
L_R(\boldsymbol{I}_n) := L_I(\boldsymbol{R}_n),
\]
which has the same functional form as $L_I$ given in Eq.~\eqref{LIToeplitz}, but evaluated on the filtered observations.
Explicitly, this corresponds to
\begin{equation}
\begin{aligned}
    L_{R,n}
=\tfrac{1}{2}\log\frac{\det\hat T_{0,n}}{\det\hat  T_{1,n}}+\tfrac{1}{2}\boldsymbol{I}_n^T \hat B^T\left(\hat T_{0,n}^{-1} -  \hat T_{1,n}^{-1}\right)\hat B\boldsymbol{I}_n.
\end{aligned}
\label{LRToeplitzR}
\end{equation}
 
We quantify the error induced by the filter by
\begin{equation}
 \epsilon_n:=L_{I,n}-L_{R,n}= \frac{1}{2}\boldsymbol{I}_n^T\hat M\boldsymbol{I}_n, 
 \label{error}
\end{equation}
where we combined Eqs.\eqref{LIToeplitz} and \eqref{LRToeplitzR} and defined  $\hat M:=\hat D-\hat B^T\hat D \hat B$ with $\hat D=\hat T_{0,n}^{-1} -  \hat T_{1,n}^{-1}$.

 We expect that the log-likelihood ratio  is primarily dominated by spectral components near the Larmor frequencies. Since the high-pass filter acts trivially at these high frequencies (see Eq.~\eqref{eq:SI1b2}), the difference introduced by filtering should be small. Specifically, we anticipate that the error term $\epsilon_n \sim \frac{b^2}{\omega_c^2} L_{S_I}$, where $ \omega_c:=\tfrac{1}{2}(\omega_0+\omega_1)$, is the high frequency around which both hypotheses $h=0,1$ are centered.

To verify that the effect of the low-frequency filtering can be safely neglected and that the finite-time corrections are small, we compute the mean and variance of $\epsilon_n$. Note that Eq.~\eqref{error} defines $\epsilon_n$ as a quadratic functional of the signal, and as in Sec.~\ref{sec:probdist}, a closed-form expression for the $m$-th cumulant under hypothesis $h$ can be written as:
\begin{align}
   & \tilde\kappa_{m,n}^h=\tfrac{(m-1)!}{2}\textrm{Tr}\left((\hat M \hat T_{h,n})^m\right)=\mathcal{O}(1)+\\
    &+\frac{t (m-1)!}{2\pi}\int_0^\infty \left(\frac{b^2}{\omega^2+b^2}\left(\frac{\bar S_{I,h}(\omega)}{\bar S_{I,0}(\omega)}-\frac{\bar S_{I,h}(\omega)}{\bar S_{I,1}(\omega)} \right)\right)^m d\omega\, . \nonumber
\end{align}  
In particular, the mean and variance can be explicitly computed, yielding:
\begin{equation}
\begin{aligned}
    &  \expect_h{[\epsilon_n]} =\frac{b^2}{\omega_c^2}\expect_h{[ L_{S_I}]}+\mathcal{O}\left(\frac{1}{\omega_c^3}\right)+\mathcal{O}(1)\\
    & \textrm{Var}_h(\epsilon_n)=  \frac{b^4}{\omega_c^4}\textrm{Var}_h( L_{S_I})+\order{\left(\frac{1}{\omega_c^5}\right)}+\mathcal{O}(1)\, ,
\end{aligned}
\label{filtreerror}
\end{equation}
which formally shows that indeed the filter has a very limited effect on the LLR, and hence on the performance of our protocols.

Finally to validate the procedure, we simulate $10^4$ measurement trajectories using the parameters listed in Table~\ref{tab:table1} under hypothesis $h = 0$, with a sampling interval $\Delta = 5~\mu$s, matching our experimental conditions. The simulated data are processed identically to the experimental traces: we apply a high-pass filter with parameter $\alpha = 0.91$, and compute the LLR sequentially using the Kalman filter formalism described in Sec.~\ref{sec:kalmanfilter}, which is formally equivalent to Eq.~\eqref{LRToeplitzR}. We then compare the resulting LLR to that obtained from the unfiltered data to compute the error $\epsilon_n$. 

\begin{figure}[t]
    \centering   \includegraphics[width=0.492\textwidth]{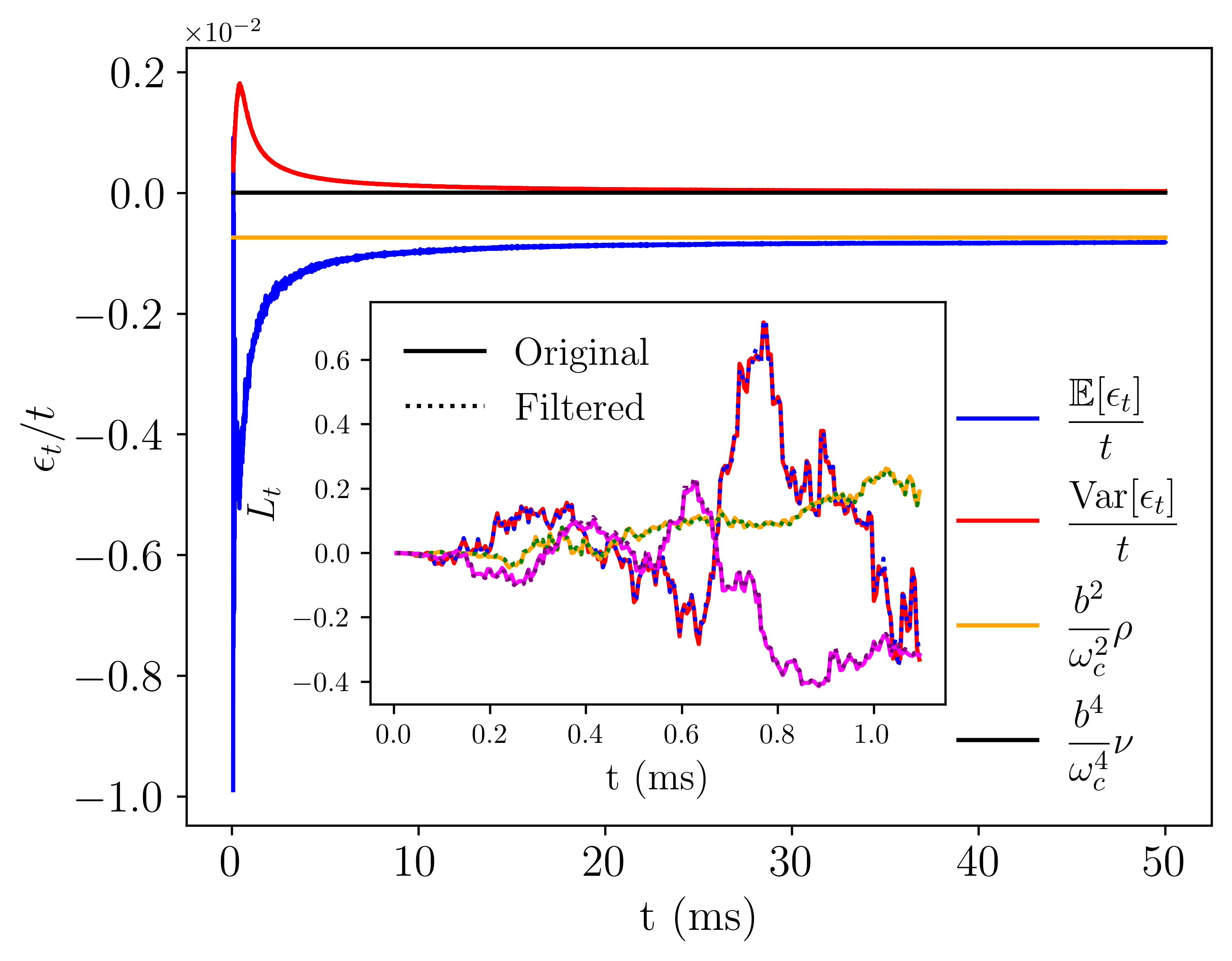}
 \caption{Comparison between analytical and simulated results for the time-normalized error \(\epsilon_t/t\) introduced by the high-pass filter. The mean (blue) and variance (red) are estimated from $10^4$ simulated trajectories and are compared with the theoretical predictions. The asymptotic values are given in terms of the regularized mean and variance of $L_{S_I}$, i.e. of $\rho: =\lim_{t\to\infty}\tfrac{1}{t}\expect_0[L_{S_I}] =-0.207~\mathrm{ms}^{-1}$ and $v: =\lim_{t\to\infty}\tfrac{1}{t}\mathrm{Var}_0[L_{S_I}]= 0.477~\mathrm{ms}^{-1}$ (see Eq.\eqref{filtreerror}).}
 \label{errormeanandvariance}
\end{figure}

Figure~\ref{errormeanandvariance} shows the mean and variance of the time-normalized error $\epsilon_t / t$ alongside theoretical asymptotic predictions from Eq.~\eqref{filtreerror}. With $b^2 / \omega_c^2 = 3.6 \times 10^{-3}$ in our setting, the filter’s relative contribution is minimal. The inset displays representative simulated traces, demonstrating negligible filter effects even at short times.

\section{Sequential strategies on the spin-noise atomic magnetometer}
\label{sec:experimentalimplementation}

We are now in position to demonstrate sequential strategies in a spin-noise quantum system. Specifically, we consider a scenario where all system parameters are fixed except for the Larmor frequency, $\omega_L$. We set $\omega_{L,h} = \omega_c + (-1)^{h} \tfrac{1}{2}{\Delta\omega}$, with $\omega_c$ the frequency around which both hypotheses $h=0,1$ are centered. We have in mind applications such as detecting biomagnetic signals, magnetic anomalies in geophysical surveys, or concealed metallic objects, where a small magnetic field perturbs a baseline field, inducing a shift $\Delta \omega$ in $\omega_L$. Details of $\Delta\omega$ values and the rest of parameters are given below. 

While our setting assumes fully characterized models for both hypotheses, real devices typically require calibration before each use. This is done by operating the sensor under controlled conditions where only the baseline field is present. Therefore, before proceeding, we briefly outline how we estimate model parameters from finite datasets.

\subsection{\label{sec:calibration} Calibrating the Sensor: Estimation of Model Parameters}

We employ the maximum likelihood estimator (MLE) to infer the system parameters, building on the thorough analysis of the likelihood statistic’s properties presented above. Indeed, for long observation times, the PSD components become independent and follow exponential distributions, making the MLE analytically tractable and statistically efficient. That is in the sense that it achieves the Cramér-Rao bound, ensuring optimal precision in parameter estimation given sufficiently large datasets \cite{levin2003power}. 

For a given time trace $\boldsymbol{I}_n$, we estimate the parameters $\boldsymbol{\Theta_h} = \left(S_{\mathrm{ph}}, S_{\mathrm{at}}, \omega_h, \gamma\right)^T$ via maximum likelihood. Recall that the DFT of $\boldsymbol{I}_n$ is unitary and preserves the full information content of the signal, that for stationary processes the Fourier phases carry no information, and that for sufficiently long traces the different frequency components become independent. As a result, given a time trace $\boldsymbol{I}_n$ or equivalently its periodogram $\boldsymbol{S}_I$, the MLE can be obtained by maximizing
 \begin{equation}
  \begin{aligned}
        &-\log p(\boldsymbol{S}_I|\boldsymbol{\Theta_h})=\\
        &=\sum_{k} \log \left(\expect_{\boldsymbol{\Theta_h}}[S_I(\omega_k)]\right) + \frac{ S_I(\omega_k)}{\expect_{\boldsymbol{\Theta_h}}[S_I(\omega_k)]},
  \end{aligned}
  \label{eq:MLEest}
\end{equation} 
where $k$ runs over the first $n/2$ frequencies, since the signal is real valued and the negative frequencies are fully determined by their positive counterparts.

For the expression of $\expect_{\boldsymbol{\Theta_h}}[(S_I(\omega_k))]$, one has two options.  Employ directly the asymptotic Lorentzian given in Eq.~\eqref{lorentzian}, evaluated at the discrete frequencies $\omega_k=\frac{2\pi k}{n\Delta}$,  yielding  the so-called Whittle log-likelihood \cite{alma991041466979704801}, or use the expected expression of the periodogram, which is known as the debiased.  The regime in which we will perform the estimation is long enough to neglect the finite size effects and therefore we will directly employ the Lorentzian approximation.

Notice that, due to the linearity of the log-likelihood in the periodogram components, $S_I(\omega_k)$, if we have 
 $N$ independent traces collected under identical conditions,
the total log-likelihood  takes the same form as Eq.~\eqref{eq:MLEest}, with $S_I(\omega_k)$ replaced by the ensemble average $\tilde S_I(\omega_k) = \frac{1}{N} \sum_{j=1}^N S_I^j(\omega_k)$.

\medskip

In this work, we aim to illustrate the use of sequential methodologies and empirically verify their performance in terms of mean detection times and error probabilities. To this end, we generate multiple time traces under each hypothesis, forming two datasets that enable us to estimate these quantities and compare the performance of sequential and deterministic strategies.

We use these two labeled datasets, corresponding to hypotheses $h_0$ and $h_1$, to infer the relevant system parameters for this batch of experiments. As the data  are independently  distributed and differ solely in the value of the Larmor frequency, we can perform maximum likelihood estimation (MLE) by optimizing the full joint likelihood over both datasets:
\begin{equation}
\begin{aligned}
   & \sum_{k}\left( \log \left(\expect_{\boldsymbol{\Theta_0}}[S_I(\omega_k)]\right) + \frac{\tilde S_{I,0}(\omega_k)}{\expect_{\boldsymbol{\Theta_0}}[S_I(\omega_k)]}+ \right. \\& \left.+\log \left(\expect_{\boldsymbol{\Theta_1}}[S_I(\omega_k)]\right) + \frac{\tilde S_{I,1}(\omega_k)}{\expect[S_I(\omega_k)]}_{\boldsymbol{\Theta_1}}\right)
    \label{both}
    \end{aligned}
\end{equation}

 Since the MLE is known to saturate the Cramér-Rao bound, we can assess the expected estimation errors by computing the inverse of the Fisher information matrix for our model. 
 
 Again, we validate this prediction by simulating the complete estimation procedure. Specifically, we generate synthetic datasets using the parameter values ${\boldsymbol{\Theta}}$ listed in Table~\ref{tab:table1}, matching the number of time traces used in the experimental dataset. This process is repeated over $N_B = 1000$ independent batches, yielding an estimate $\boldsymbol{\hat\Theta}_i$ for each batch. From these, we compute the empirical covariance matrix of the estimator,
\begin{equation}
\hat{\Gamma} = \frac{1}{N_B} \sum_{i=1}^{N_B} (\boldsymbol{\hat\Theta}_i - \boldsymbol{\Theta})(\boldsymbol{\hat\Theta}_i - \boldsymbol{\Theta})^T \,.
\end{equation}
 The simulated mean square error (MSE) values reported in the last column of Table~\ref{tab:table1} correspond to the diagonal entries of this matrix and show good agreement with the theoretical values predicted by the Cramér-Rao bound.

For the estimation, we cut the region of frequencies where the Lorentzian is centered. The Larmor frequencies of both hypothesis are near the 50 (kHz). Therefore, we consider the region of frequencies between $40-60$ (kHz). The experimental datasets for each hypothesis contain 200 traces of $2$ s. The results are detailed in the following Table~\ref{tab:table1}.

\begin{table}[t]
\caption{\label{tab:table1}%
Results for the five estimated parameters. We present the inferred value, the analytical standard deviation $\sigma$ obtained from the Cramér Rao bound and the empirical MSE of the standard deviation simulating the experiments with a total number of 1000 batches.}
\begin{ruledtabular}
\begin{tabular}{ccS[table-format=5.3]S[table-format=1.4]S[table-format=1.4]}
\textrm{Parameter}&
\textrm{Unit}&
\textrm{Estimated Value}&
\multicolumn{1}{c}{$\sigma$}&
\textrm{MSE} \\
\colrule
$\gamma$ & \SI{}{\hertz} & 330.90 & 0.90  & 0.98 \\   
$\omega_{L,0}$ & \SI{}{\hertz}  &    50114.03           &          0.97           &    0.98           \\ 
$\omega_{L,1}$ & \SI{}{\hertz} &  50550.88 & 0.97 & 1.02 \\

$S_{\mathrm{at}}$  & \SI[per-mode=symbol]{}{\micro\volt\squared\per\hertz}      &      31.768       &   0.087                  &       0.094        \\ 
$S_{\mathrm{ph}}$ & \SI[per-mode=symbol]{}{\micro\volt\squared\per\hertz}        &      13.0457      &   0.0038                 &   0.0040     \\
\end{tabular}
\end{ruledtabular}
\end{table}

Notice that the error in the estimation of all parameters is at least two orders of magnitude smaller than the separation between the two hypotheses, since the difference between the two Larmor frequencies is $\Delta\omega = \omega_{L,0} - \omega_{L,1} = 436.85~\mathrm{Hz}$. Therefore, we are in a regime where the inferred parameters can be reliably used for the implementation of the sequential strategies. While the reported errors correspond to theoretical estimates, either from the Cramér-Rao bound or from Monte Carlo simulations, we will later show that our experimental data sets exhibit  parameter fluctuations beyond these estimates, particularly in the parameters that govern the relaxation processes. In Sec.~\ref{sec:sequentialstrategiesexp} and Appendix \ref{sec:gammaphenomenon}, we provide a qualitative model that explains these fluctuations and quantify its effects.

\subsection{\label{sec:sequentialstrategiesexp}Implementation of sequential strategies}

For real-time implementation, both the high-pass and Kalman filters can be efficiently implemented on microcontrollers or field-programmable gate arrays (FPGAs) \cite{magrini2021real, liao2018fpga, auger2013industrial, ruiz2019field} capable of real-time processing, using simple arithmetic operations (see the sequential block diagrams in Appendix~\ref{sec:diagrams}). At each time step $n$, the device computes the filtered signal $R_n$, evaluates the LLR, and applies 
the SPRT or CUSUM thresholds for sequential hypothesis testing or quickest change-point detection, respectively.
In this work, we implement these strategies on a pre-recorded data set. 
This approach allows us to assess the performance of the protocols, such as average detection times and error probabilities, by repeating the analysis over many independent realizations under controlled conditions.

In the hypothesis testing scenario, we segment the recorded trajectories into $10^4$ independent traces of duration $40~\mathrm{ms}$ for each hypothesis in order to increase statistical sampling. The high-pass filter is applied with strength parameter $\alpha = 0.91$, and the acquisition sampling interval is $\Delta = 5~\mu\mathrm{s}$.

\begin{figure}[t]
    \centering   \includegraphics[width=0.48\textwidth]{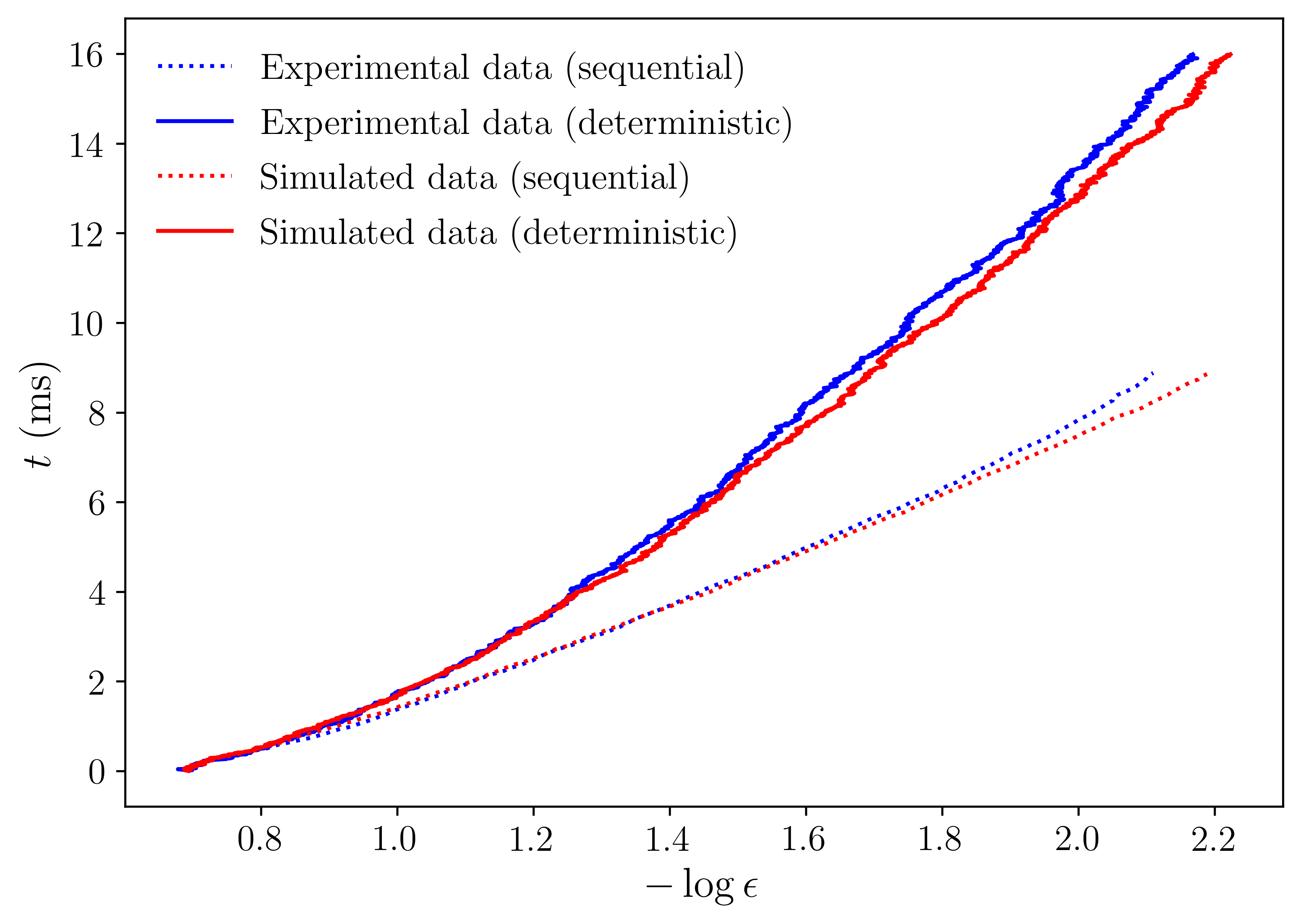}  
 \caption{Relation between error probability and required experiment duration for sequential (dotted) and deterministic (solid) strategies. Each point is obtained by averaging over $10^4$ experimental traces per hypothesis. The horizontal axis shows $\log \epsilon$, where $\epsilon$ is the empirical error probability (i.e. fraction of incorrectly identified hypotheses), and the vertical axis shows the required experiment duration. For the sequential strategy (SPRT), we vary the stopping threshold $a = a_0 = a_1$, compute the error probability and the average stopping time over all runs and plot $(\log \epsilon,  t )$.
For the deterministic strategy, we fix the experiment duration $t$ to various values, and for each value we compute the average error probability $\epsilon$ over all runs, and plot $(\log \epsilon, t)$.
Experimental results (blue) are plotted alongside simulated results (red) for comparison. }
 \label{HT_experiments}
\end{figure}

In Figure~\ref{HT_experiments}, we compare the average time required by the sequential probability ratio test (SPRT) to reach a predefined error level with that of the optimal deterministic strategy achieving the same error. To ensure a fair comparison, we consider a symmetric hypothesis-testing scenario (see Sec.~\ref{sec:hyptesting}), where both types of errors are weighted equally, and define the average error as
$\epsilon = \tfrac{1}{2} \epsilon_0 + \tfrac{1}{2} \epsilon_1$. For the deterministic strategy, we use time traces of fixed length $t$ and apply the optimal decision rule described in Sec.~\ref{sec:hyptestingdet}: decide $h=1$ if $L_t>0$, and $h=0$ otherwise. The empirical error probability is then estimated as the fraction of misidentified traces.   For the SPRT, we set equal error thresholds $\epsilon_0 = \epsilon_1 = \epsilon$. According to Eq.\eqref{ineq}), this determines the upper and lower thresholds, for the SPRT: sampling stops at the first random time $t$ when $L_{R,t} \notin (-a_0, a_1)$, followed by a decision: $h=1$ if $L_t \geq a_1$ and $h=0$ if $L_{R,t} \leq -a_0$, with $a_0=a_1=a$. The empirical error is computed as before, although it is worth emphasizing that, by construction, the SPRT guarantees a conditional error probability $\epsilon$ for every trajectory. 

To validate our model, results are shown for both experimental data (blue) and simulated data (red), generated under identical conditions with matched parameters and the same high-pass filter. In both datasets, the SPRT consistently reaches the target error level in less time than the deterministic strategy, while also providing rigorous error guarantees. It is worth noting that, while sequential strategies are clearly more time- and resource-efficient on average, they come with a tradeoff: the duration of each experiment is inherently random. Although its fluctuations are typically bounded within a characteristic range, the distribution can exhibit long tails, making the exact stopping time unpredictable \cite{gasbarri_sequential_2024}.
 Indeed, in our experimental implementation, we limited traces to $40$~ms to improve statistics; as a result, a tiny fraction of trajectories (55 for $h=0$ and 53 for $h=1$ out of $10^4$ each) did not reach a stopping decision. This negligible portion of the dataset was excluded from the averages shown in Fig.~\ref{HT_experiments}.

When comparing experimental and simulated results in Figure~\ref{HT_experiments}, we observe good agreement at moderate measurement times. As the acquisition time increases, discrepancies appear, with experimental data requiring longer times to reach the same estimation error. This reflects that, at long times, the information gathered about the underlying process becomes increasingly precise, and small system imperfections, such as parameter drifts, stability issues, or unmodeled dynamics, become more significant. In Appendix \ref{sec:gammaphenomenon}, 
we model the effect of small run-to-run variations in the system parameters, particularly those governing spin relaxation, which qualitatively explains the statistics of the  LLR.

To rigorously address scenarios with significant parameter fluctuations, our techniques must be extended to the framework of composite hypothesis testing \cite{trees2001detection, tartakovsky2014sequential}, where each hypothesis $h$ corresponds to a family of possible models defined by parameters in a given set, $\theta_h \in \mathcal{S}_h$. In this setting, different forms of the LLR are employed depending on whether the parameters are partially or completely unknown, and the stopping rules are adapted to select the most likely hypothesis.

\begin{figure}[t]
    \centering   \includegraphics[width=0.48\textwidth]{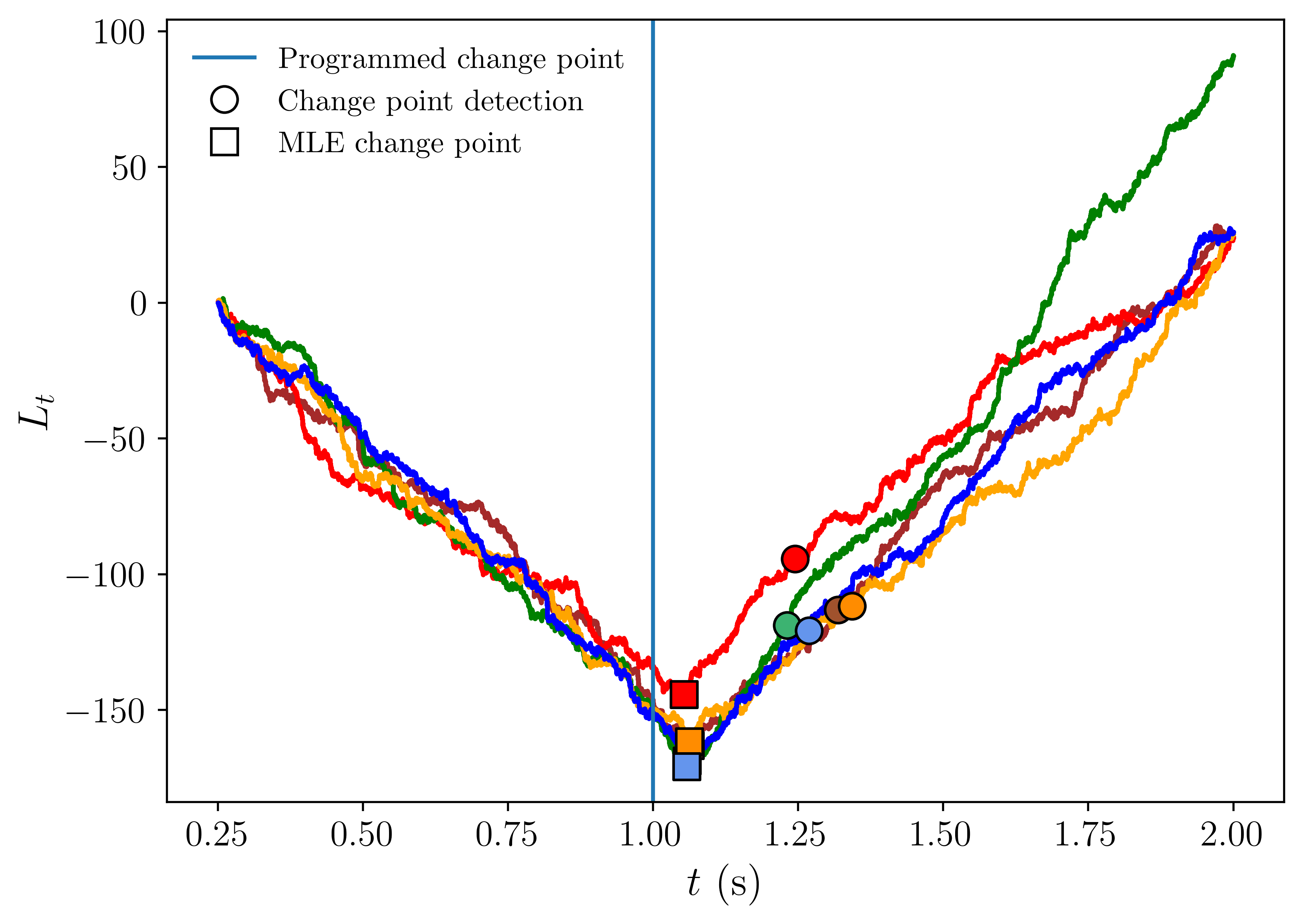}
 \caption{Quickest change-point detection applied to experimental traces with an abrupt magnetic field shift. Each trajectory starts under hypothesis $h_0$ and undergoes a programmed magnetic field change of magnitude $\Delta B = \Delta\omega / \gamma_G \approx 62.3~\mathrm{nT}$ at $t =\SI{1}{\second}$ (indicated by a vertical line), where $\gamma_G \approx 7~\mathrm{Hz/nT}$ is the gyromagnetic ratio. For visualization purposes, the detection threshold is set to the large value $a = 50$. Circles mark the detection times produced by the algorithm for each run, while squares indicate the corresponding maximum-likelihood estimates of the actual change-point times. Note that the programmed change point  does not coincide exactly with the true physical change point, nor with the estimated change points, as discussed in the main text.}
 \label{CP_experiments}
\end{figure}

Finally, we demonstrate the implementation of the quickest change-point detection (QCPD) algorithm using $2$-second measurement traces that initially follow hypothesis $h_0$ and exhibit a sudden transition to hypothesis $h_1$ near $t = \SI{1}{\second}$. For the calibration of the detection model, we employ the parameters previously estimated from the hypothesis testing data sets, as the measurements are acquired under identical conditions, including the application of the same high-pass filter. Representative results for different trajectories are presented in Fig.~\ref{CP_experiments}.

Notice the marked change in the behavior of the LLR after the change-point. We use the CUSUM stopping rule defined in Eq.~\eqref{eq:Tdetection} with a fixed threshold $a = 50$.  

Colored circles along the LLR traces indicate the corresponding stopping times, i.e., the points at which the decision criterion in Eq.~\eqref{eq:Mn} triggers a change-point detection alarm. In addition, as discussed in Sec.~\ref{sec:changepoint}, the minimum of the LLR curve provides the maximum-likelihood estimate of the actual change-point time. Across all runs, this minimum (marked with colored squares) is consistently shifted with respect to the vertical line indicating the nominally programmed change. This systematic offset revealed the presence of an unaccounted technical delay
caused by a timing mismatch between the magnetic field sequence and the data acquisition, resulting in a constant delay of approximately $60~\mathrm{ms}$. 

\begin{figure}[t]
    \centering   \includegraphics[width=0.48\textwidth]{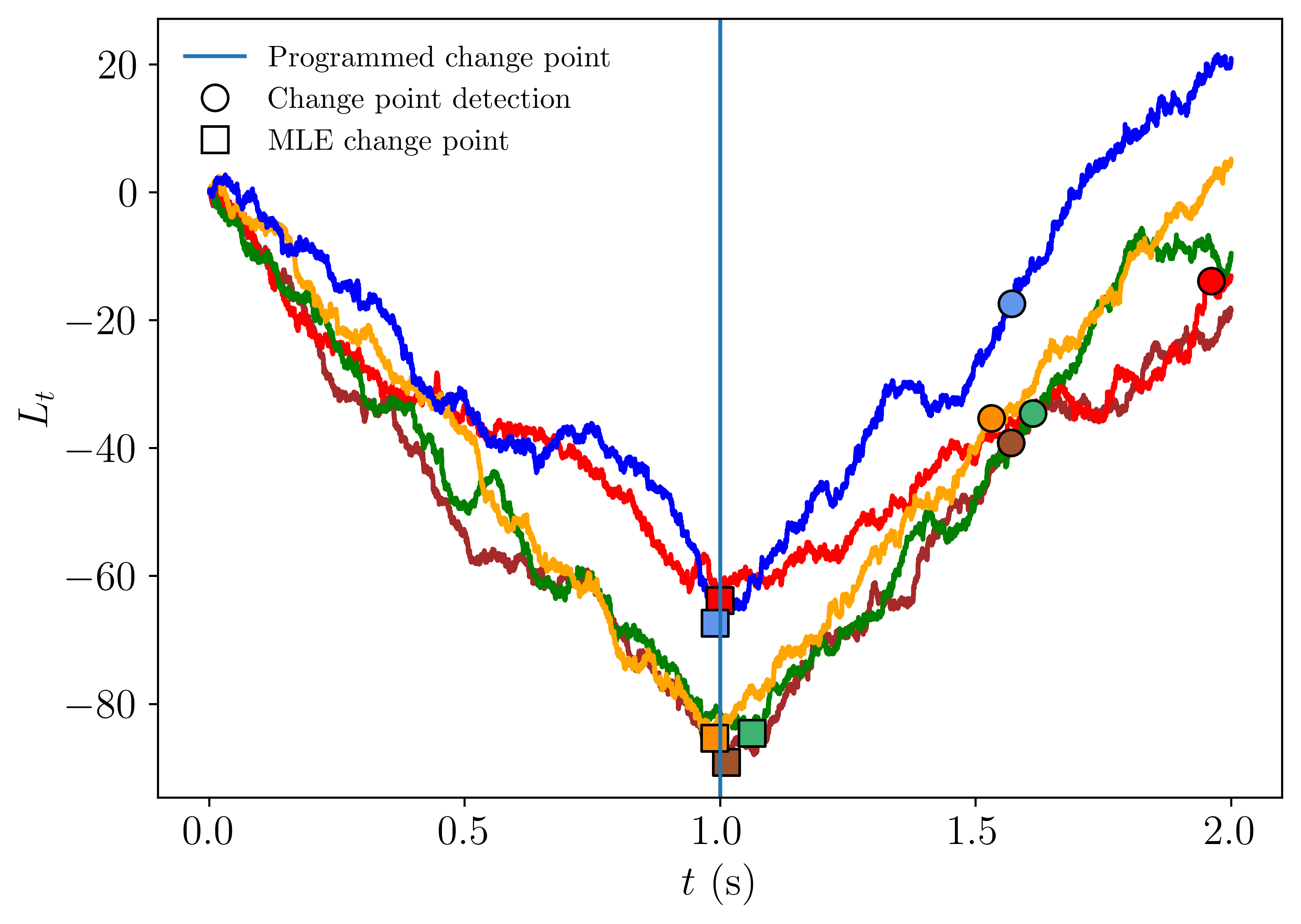}
 \caption{
Same as Fig.~\ref{CP_experiments}, but for a different set of hypotheses with
parameters $\gamma = 418.48~\mathrm{Hz}$, $\omega_{L,0} = 42577.43~\mathrm{Hz}$,
$\omega_{L,1} = 43640.10~\mathrm{Hz}$, $S_{\mathrm{at}} = 0.77090~\mu\mathrm{V}^2/\mathrm{Hz}$,
and $S_{\mathrm{ph}} = 1.4277~\mu\mathrm{V}^2/\mathrm{Hz}$. Unlike Fig.~\ref{CP_experiments},
the magnetic-field sequence and data acquisition are synchronized here, with
$t = 1~\mathrm{s}$ marking the programmed magnetic-field change. Circles mark
detection times and squares the corresponding maximum-likelihood estimates,
which closely match the programmed change-point.
}
 \label{CP_experiments2}
\end{figure}

To address this issue, we performed a second batch of experiments (see Figure \ref{CP_experiments2}), synchronizing the magnetic-field changes and the data acquisition with no delay, and simultaneously recording the coil current and the atomic response on two channels of the data-acquisition system. In this case, the estimated change-points show a good agreement with the programmed times, confirming both the source of the initial offset and the validity of our change-point detection strategy. This new set of hypotheses is less distinguishable than the former set, as indicated by a smaller regularized Kullback--Leibler divergence in Eq.~\eqref{seqslope}. This is also evident in Figure~\ref{CP_experiments2}, which shows that achieving the same detection threshold (false-alarm rate) requires a detection time  two to three times longer, consistent with the theoretical factor of $2.65$ given by the ratio of the slopes in Eq.~\eqref{seqslope}.

\medskip
The threshold $a=50$ considered so far was chosen to be rather large to clearly visualize the detection delay in Figure~\ref{CP_experiments}, and correspond to an astronomical false alarm time of $\sim 10^9$ years. We now turn to a more realistic scenario and consider a more modest bound on mean time of false alarm of $T_{\mathrm{FA}} = 0.5~\mathrm{s}$, which corresponds to a decision threshold of $a= \log\left(T_{\mathrm{FA}} / \Delta\right) \approx 11.5$, and a much quicker detection. We now use a widely adopted version of the CUSUM algorithm:
sequentially compute
\begin{equation}
 L^+_n: = \sum_{i=1}^{n} \Delta L^+_i \; \mbox{ with }\; \Delta L^+_i = \max(0, \Delta L_i),
\end{equation}
and signal a detection alarm as soon as  $L^+_n\geq a$. This algorithm is equivalent to the procedure defined in Eq.~\eqref{eq:Mn}, but accumulates only positive log-likelihood contributions, effectively resetting the reference level to zero whenever the instantaneous evidence for change becomes negative. 

\begin{figure}[t]
\centering
\includegraphics[width=0.48\textwidth]{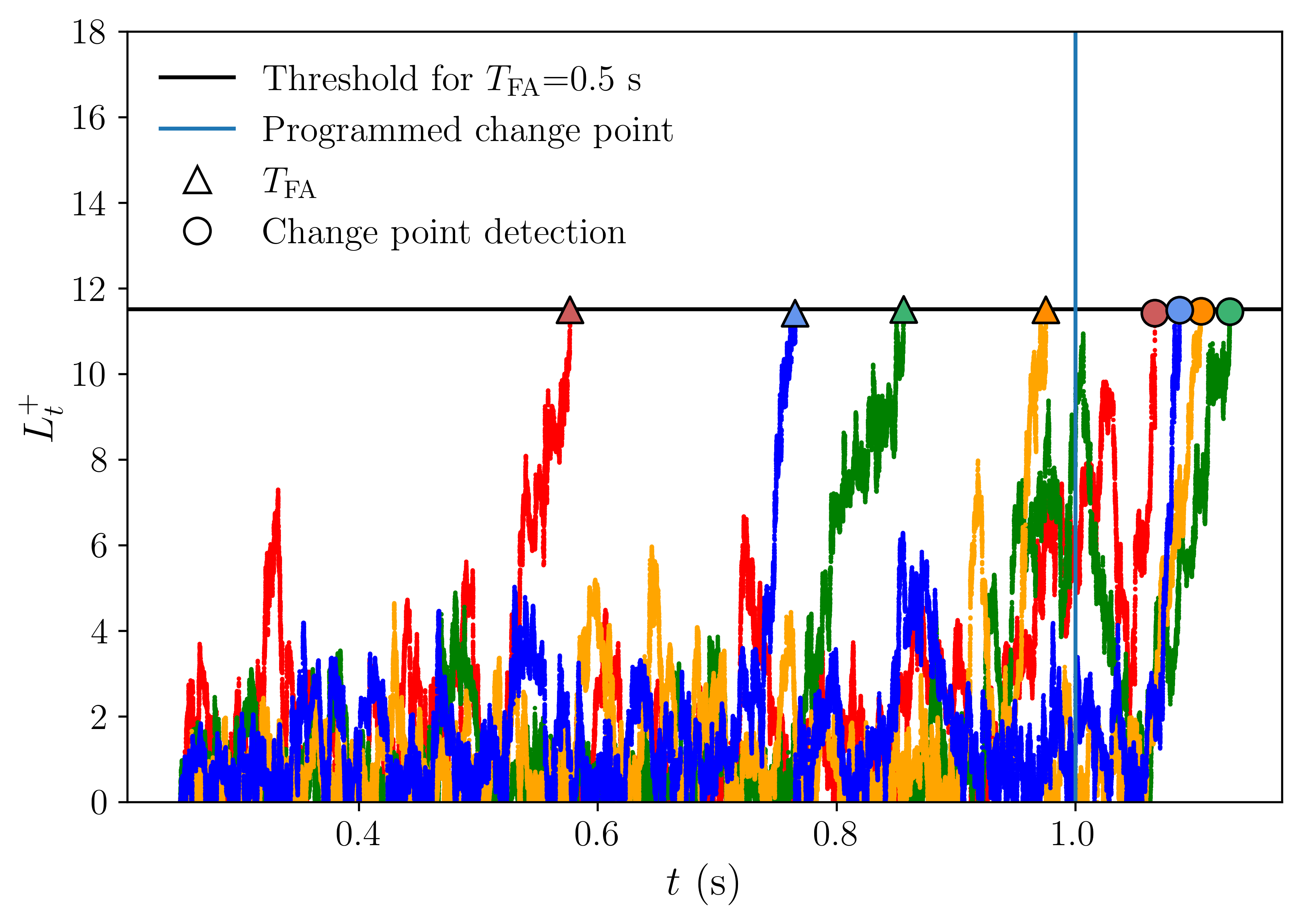}
\caption{Results from 200 experimental runs starting at $t_0 = 0.25~\mathrm{s}$ with a fixed initial magnetic field. A step change of $\Delta B=\SI{62.3}{\nano\tesla}$  is programmed at $t = 1~\mathrm{s}$ (blue vertical line). The plot shows the 4 trajectories that triggered a false alarm, i.e., a detection before the programmed change. Triangles indicate the false alarm times $T_{\mathrm{FA}}$, as determined by the algorithm with a threshold of $a=11.5 $ (black horizontal line) corresponding to $T_{\mathrm{FA}} = 0.5~\mathrm{s}$.}
\label{CP_experiments_validate}
\end{figure}

Figure~\ref{CP_experiments_validate} shows the application of this procedure to a dataset of $200$ experimental trajectories, each lasting $2$ s. We show the 4 traces in which a false alarm was triggered, that is, a change was detected before the programmed nominal change-point at $t = 1 \mathrm{s}$. The sequential analysis begins at $t_0 = 0.25~\mathrm{s}$. In each instance of a false alarm, instead of halting the measurement, we reset the statistic $S_n = 0$ and resumed data acquisition, effectively restarting the detection process from an uninformed state.

Finally, for trajectories that correctly detected the actual change, the average detection delay was found to be approximately $60~\mathrm{ms}$, in agreement with the theoretical value $\bar{\tau} = 56~\mathrm{ms}$ predicted from Eq.~\eqref{delaynoniid} using the mean log-likelihood ratio from Eq.~\eqref{cumulants}.

\section{Optimal asymptotic limits}
\label{sec:ultimatelimits}
In this section, we provide closed-form expressions for key information-theoretic quantities that determine the asymptotic performance bounds at long observation times, both for deterministic and sequential hypothesis testing, as well as for QCPD.

In the \textit{standard deterministic strategy} where the observation time is fixed in advance, for a sequence of $n$ observations $\boldsymbol{I}_n=\left(I_1,...,I_n\right)^T$,  the probability of error is given by
\begin{equation}P_{\textrm{err}}(n)=\int_{\mathbb{R}^n} \textrm{min}\{\pi_0 p_0\left(\boldsymbol{I}_n\right),\pi_1p_1\left(\boldsymbol{I}_n\right)\}d\boldsymbol{I}_n.
\label{perrI}
\end{equation}
Evaluating Eq.~\eqref{perrI} is generally intractable. However, analogously to the i.i.d. case (see Eq.~\eqref{eq:errorrate}), one can derive asymptotic bounds in the long-time limit.

\begin{figure}[t]
    \centering   \includegraphics[width=0.45\textwidth]{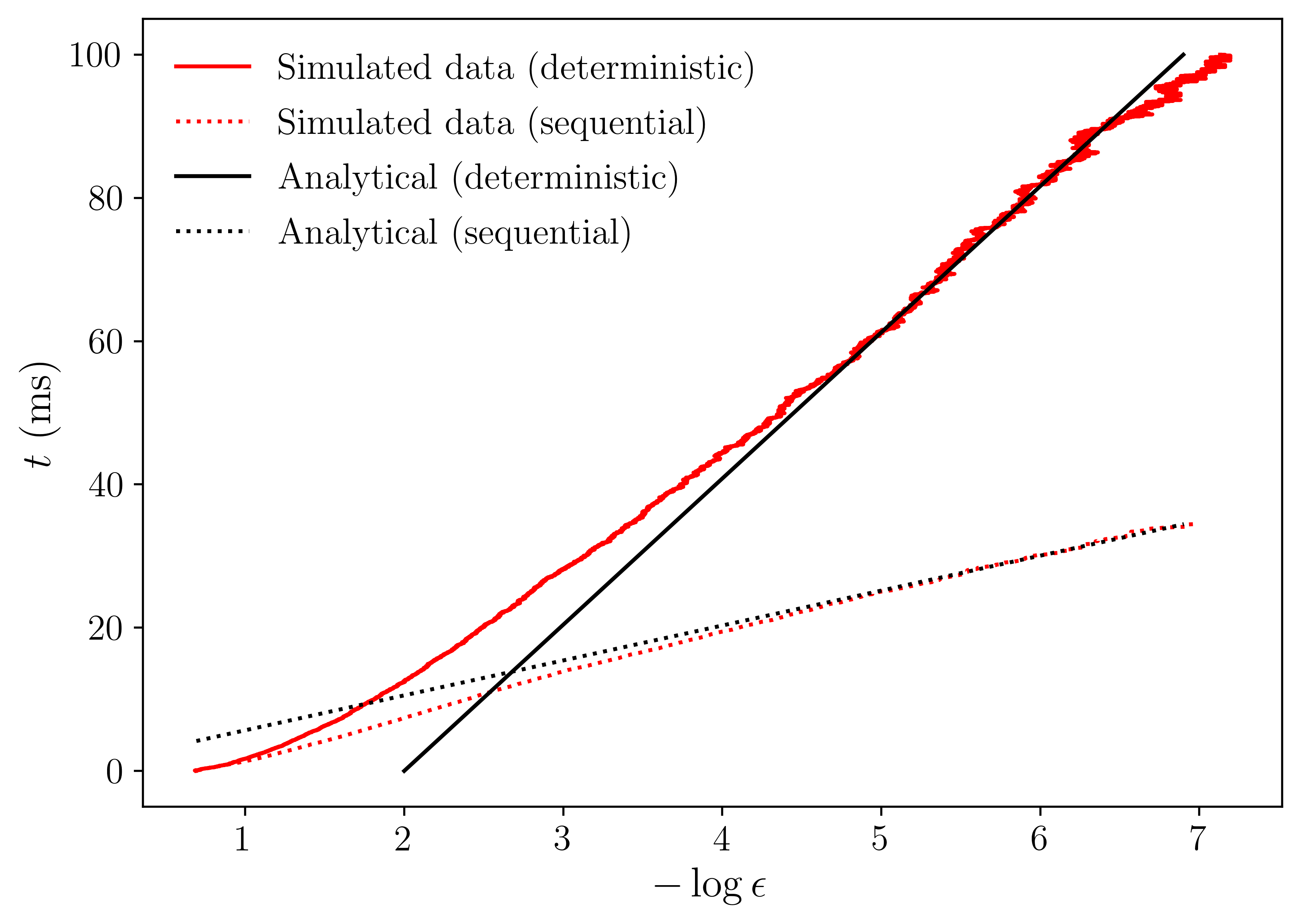} 
 \caption{Asymptotic performance metrics for the sequential probability ratio test  (dotted) and the deterministic probability ratio test (solid), obtained by averaging over $4 \times 10^4$ simulated trajectories per hypothesis. The horizontal axis shows $\log \epsilon$, with $\epsilon$ the error probability, and the vertical axis shows the experiment duration. These quantities are computed as explained in Fig.~\ref{HT_experiments}.   Asymptotic theoretical rates are also shown: the Chernoff information in Eq.~\eqref{chernoffc} (solid black) and the regularized Kullback–Leibler information in Eq.~\eqref{seqslope} (dotted black) give asymptotic linear scaling of time with $-\log \epsilon$, with slopes $0.2055~\textrm{ms}^{-1}$ and $0.0496~\textrm{ms}^{-1}$, respectively. This corresponds to more than a fourfold efficiency gain for the SPRT, $\rho_{\textrm{seq}} / \rho_{\textrm{det}} \simeq 4.14$. 
}    \label{benckmark}
\end{figure}

By applying the inequality $\min(a,b) \leq a^s b^{1-s}$ for all $s \in (0,1)$ to Eq.~\eqref{perrI}, one obtains the Chernoff bound:
\begin{equation}
    P_{\textrm{err}}(n)\leq\underset{s\in (0,1)}{\textrm{min}} \pi_0^s\pi_1^{1-s} \int_{\mathbb{R}^n} p_0\left(\boldsymbol{I}_n\right)^s p_1\left(\boldsymbol{I}_n\right)^{1-s}d\boldsymbol{I}_n.
    \label{eq:Perrn}
\end{equation}
This yields a lower bound on the error exponent:
\begin{align}
-\lim_{t\to\infty}\frac{1}{t}&\log P_{\mathrm{err}}(t)\geq \rho_\mathrm{det}:=\lim_{t\to\infty}\frac{1}{t}\max_s C(s)\nonumber\\
\mbox{with } &
C(s)= - \log\int_{\mathbb{R}^n} p_0\left(\boldsymbol{I}_n\right)^s p_1\left(\boldsymbol{I}_n\right)^{1-s}d\boldsymbol{I}_n 
   \label{chernoffc}
\end{align}
As in the IID  case, this bound is known to be attainable in the asymptotic limit for wide range of stochastic processes \cite{trees_detection_2001} (see also Figure \ref{benckmark}), in which case the regularized Chernoff Information provides a lower bound on the asymptotic rate at which the error probability decays,
\begin{equation}
-\log\epsilon_\mathrm{det}\sim \rho_\mathrm{det} t.
\end{equation}

In the continuum limit, the Chernoff exponent function in Eq.~\eqref{chernoffc} has a closed form in terms of the Kalman estimators, introduced in Sec.~\ref{sec:kalmanfilter} \cite{trees2001detection}
\begin{equation}
\begin{aligned}
    & C(s)=\\
&=\frac{1}{2 G} \int_0^t \left((1-s)\sigma_{1}(t')+s\sigma_{0}(t')-\sigma_{c}(t',s)\right) dt'
\end{aligned}
\label{chernoffkalman}
\end{equation}
where $G$ is the variance of the shot noise and $\sigma_h$ is the continuous version of the innovation variance under hypothesis $h$.  Moreover, $\sigma_{c}(s,t')$ is understood as the variance of the innovations associated to the artificial signal $I_c=\sqrt{s}I_0+\sqrt{1-s}I_1$, {where $I_h$ for $h=0,1$ is the signal under hypothesis $h$}. Equation \eqref{chernoffkalman} provides a useful bound of $P_{\textrm{err}}$ valid for any observation time and predicts a linear scaling of $C(s)$ with $t$, once the variances reach their stationary values. 

Alternatively, the Chernoff function in Eq.~\eqref{chernoffkalman} admits an equivalent representation in the frequency domain. Here, the signal is characterized by its periodogram $\boldsymbol{S}_{n/2}=\left(S_1,...S_{n/2}\right)^T$,  where each $S_j$ follows an independent exponential distribution. This leads to
\begin{equation}
 \begin{aligned}
        C(s)&=\prod_{j=0}^{n/2}\int_0^\infty p_0(S_j)^sp_1(S_j)^{1-s} dS_j=\\
        &=\prod_{j=0}^{n/2}\frac{\bar S_{I,0}(\omega_j)^{1-s}\bar S_{I,1}(\omega_j)^s}{(1-s)\bar S_{I,0}(\omega_j)+s \bar S_{I,1}(\omega_j)},
 \end{aligned}
\end{equation}
which, in the continuum limit  becomes \cite{kazakos_spectral_1980}
\begin{equation}
  C(s)
=\frac{t}{2\pi}\int^{\infty}_0  \log{\frac{\bar S_{I,0}(\omega_j)^{1-s}\bar S_{I,1}(\omega_j)^s}{(1-s)\bar S_{I,0}(\omega_j)+s \bar S_{I,1}(\omega_j)}} d\omega.
\label{eq:chernofffreq}
\end{equation}

The performance metric for  sequential hypothesis testing, and QCPD, is given by the slope of the mean LLR
\begin{equation}
\rho_\mathrm{seq}:=\lim_{t\to\infty}\frac{ |\expect{[ L_{I}]}_h |}{t}
\label{seqslope}
\end{equation}
This quantity corresponds to the time-regularized Kullback–Leibler divergence, and determines the exponential rate of error decay and the average stopping time in sequential strategies. 
In the time domain, using Kalman filter estimators 
the expected LLR is given by \cite{gasbarri_sequential_2024}:
\begin{equation}
\begin{aligned}
     \expect_h{[ L_{I}]} &=\int_0^t \frac{1}{S_{\mathrm{ph}}}\expect_h[dI_t' \left(\mu_1(t')-\mu_0(t')\right)]+\\
& +\int_0^t\frac{dt'}{2 S_{\mathrm{ph}}}\expect_h{[\mu_0(t')^2-\mu_1(t')^2]},
 \end{aligned}
\end{equation}
while in frequency domain, we recall Eq.~\eqref{LIf}

\begin{equation}
\begin{aligned}
     & \expect_h{[ L_{I}]} =\frac{t}{2\pi}\int_0^\infty  \log\frac{\bar S_{I,0}(\omega)} {\bar S_{I,1}(\omega)}d\omega+\\
     &+\frac{t}{2\pi}\int_0^\infty\left(\frac{\bar S_{I,h}(\omega)}{\bar S_{I,0}(\omega)}-\frac{\bar S_{I,h}(\omega)}{\bar S_{I,1}\left(\omega\right)}\right) d\omega+\mathcal{O}(1)
     \label{LLRreq}
     \end{aligned}
\end{equation}
which allows for an analytical computation of $\rho_{\mathrm{seq}}$. Note that the choice of hypothesis $h = 0,1$ only affects the sign of the expected LLR, while its magnitude, and thus $\rho_{\mathrm{seq}}$, remains the same.

The mean detection time in sequential tests then scales as
\begin{equation}
\bar\tau \sim \frac{a}{\rho_\mathrm{seq}},
\label{delaynoniid}
\end{equation}
where $a = -\log \alpha_h$ for a Wald test under hypothesis $h$, and $a = \log (T_\mathrm{FA}/\Delta)$ in the case of QCPD with false alarm constraint $T_\mathrm{FA}$.

In Figure~\ref{benckmark}, we present the results for sequential (dotted) and deterministic (solid) hypothesis testing. To estimate the performance in the asymptotic regime of long observation times, corresponding to small error probabilities, a large number of trials is required. For this reason, we rely on simulated data. Analytical asymptotic error rates, are also plotted for reference. Notably, the numerical results approach the theoretical Chernoff bound, confirming its attainability in this setting.
Furthermore, we note that the sequential strategy outperforms the deterministic one, exhibiting more than a fourfold improvement in time required to achieve a given error probability.

\begin{figure}[t]
    \centering   \includegraphics[width=0.45\textwidth]{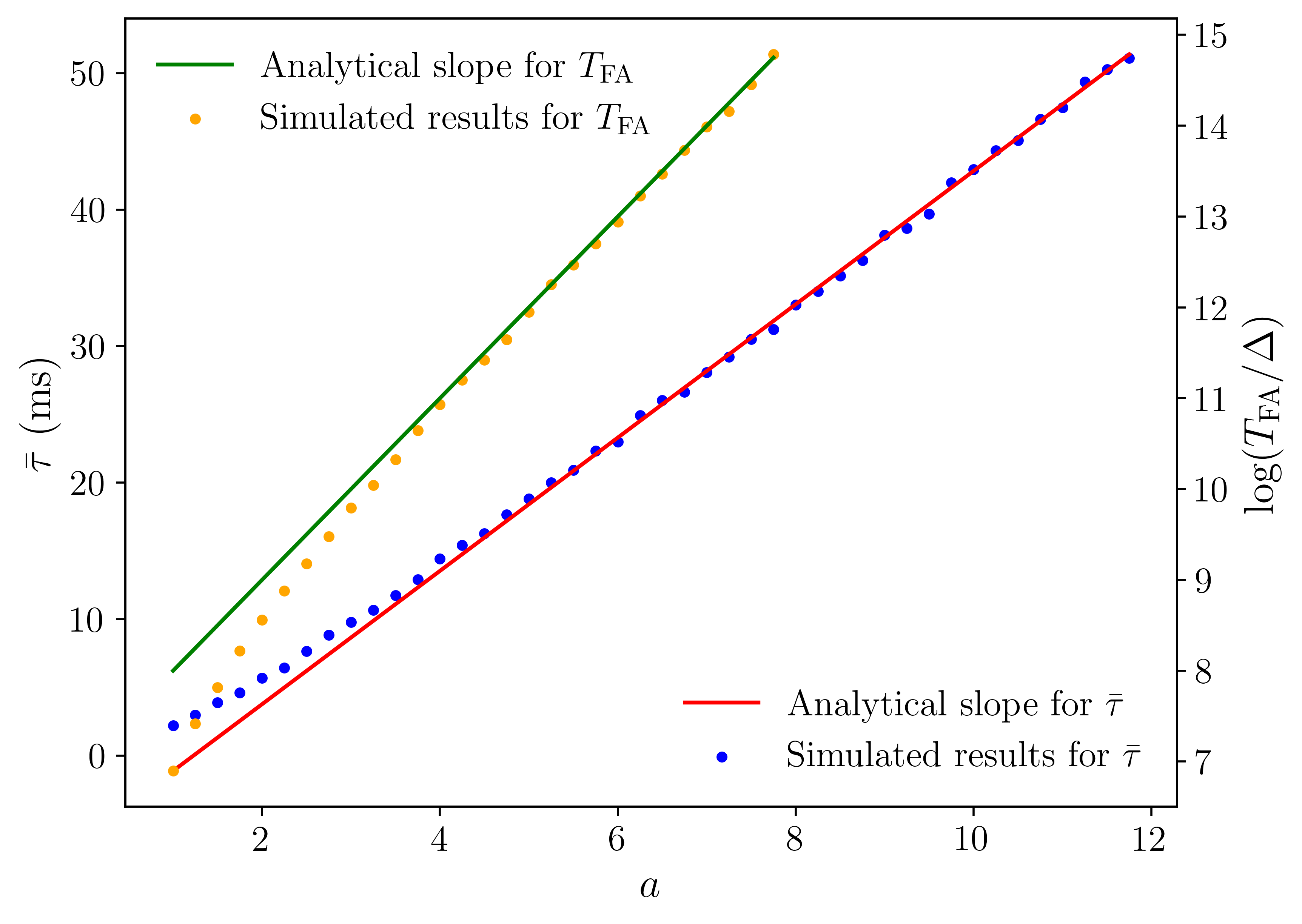} 
 \caption{
 Mean detection time (blue, $5\times 10^3$ traces 
 left axis) and mean false alarm (orange, $10^3$ traces, right axis) for simulated data. The red line shows the asymptotic detection time rate, $\rho_{\textrm{seq}}^{-1}=4.88$ ms from Eq.~\eqref{delaynoniid}, and the green line indicates the expected linear relation between threshold $h$ and $\log(T_{\mathrm{FA}}/\Delta)$, with slope 1. Both slopes match the simulation results.}
     \label{falsealarm}
\end{figure}

Similarly, Figure \ref{falsealarm} illustrates the asymptotic results for the quickest change-point-detection problem. In order to compute these asymptotic results we rely again on simulated data. Notice how the simulated results are in agreement with both slopes validating numerically the theoretical predictions.

In Appendix \ref{sec:ultimateerrorrates} we show that, in our setting where the two models differ only in the Larmor frequency, $\omega_{L,h} = \omega_c + (-1)^h \frac{\Delta \omega}{2}$ and large central frequency $\omega_c$, assuming the Chernoff bound is attainable, the sequential strategies outperform deterministic ones by at least a factor of four.

To conclude this section, we show that when the two hypotheses become arbitrarily close, regardless of the specific way in which they differ, the Chernoff and Kullback-Leibler information, which determine the deterministic and sequential performances respectively,  differ by a factor four.

In order to see this, we can, without loss of generality, choose a parameterization such that the two hypotheses differ only in a single parameter. Specifically, we take $\theta=\theta_0$ under $h_0$ and $\theta_1=\theta_0+\delta\theta$ under $h_1$.  Then, $\bar S_{I,{\theta_0}}(\omega):=\bar S_{I,0}(\omega)$ and $\bar S_{I,1}(\omega)=\bar S_{I,0}(\omega)+\delta\theta \frac{\partial \bar S_{I,\theta}(\omega)}{\partial \theta}|_{\theta=\theta_0} + \mathcal{O}\left( \delta\theta^2\right).$
 
Under these assumptions, and using Eq.~\eqref{eq:chernofffreq}, we readily obtain:
\begin{equation}
   \begin{aligned}
       &\rho_{\textrm{det}}=\underset{s\in (0,1)}{\min} \frac{s(s-1)\delta\theta^2}{4\pi}\int_0^\infty d\omega \left(\partial_{\theta} \ln \bar S_{I,{\theta}}(\omega)|_{\theta=\theta_0}\right)^2  +\\
        &+\mathcal{O}(\delta\theta^3)=\frac{\delta\theta^2}{16\pi}\int_0^\infty d\omega \left(\partial_{\theta} \ln \bar S_{I,{\theta}}(\omega)|_{\theta=\theta_0}\right)^2+ \mathcal{O}(\delta\theta^3)\\
        &=\frac{\delta\theta^2}{8}\FisherInfo(\theta_0)+  \mathcal{O}(\delta\theta^3)
   \end{aligned}
   \label{eq:rhodet}
\end{equation}
where in the last equality we have identified the regularized Fisher Information of the process with respect to parameter $\theta$:
\begin{align}
\FisherInfo(\theta)&:=\lim_{t\to\infty}\frac{1}{t} \expect_{\theta} \left[ \left( \partial_\theta \log p_\theta(\boldsymbol{I}_t) \right)^2 \right]=\nonumber\\
&=\frac{1}{2\pi}\int_0^\infty d\omega \left(\partial_{\theta} \ln \bar S_{I,\theta}(\omega)\right)^2.
\label{eq:fisherg}
\end{align}

Similarly, for nearing
hypothesis, the 
regularized relative entropy becomes
\begin{equation}
\rho_{\textrm{seq}}=\frac{\delta\theta^2}{2}\FisherInfo(\theta_0)+ \mathcal{O}(\delta\theta^3)
\label{eq:drelfisher}
\end{equation}
which, by direct comparison with Eq.~\eqref{eq:rhodet}, implies that $\rho_{\textrm{seq}} = 4\rho_{\textrm{det}}$. That is, sequential strategies require up to four times less observation time to achieve the same error performance as deterministic ones.

\section{\label{sec:Conclusions} Discussion and outlook}

In this work, we have presented what is, to our knowledge, the first experimental implementation of sequential data analysis methods in continuously-monitored quantum sensors. We realize two cornerstone primitives, namely sequential hypothesis testing and quickest change-point detection, in an atomic spin-noise magnetometer. This bridges a gap between quantum sensing theory and realistic, real-time sequential protocols. Our methods enable adaptive decision-making, allowing detection of weak magnetic perturbations with minimal delay and a quantifiable confidence threshold. This represents progress toward making quantum sensors more autonomous, efficient in data use, and compatible with real-world constraints.

We have presented the first comprehensive demonstration that sequential decision protocols can be applied to data from continuously monitored quantum sensors, further exploiting their intrinsic sensitivity. Although decisions were computed post-acquisition, all components of the analysis were designed with real-time feasibility in mind.
 Using a linear Gaussian state-space model of spin-noise magnetometry and processing measurement records with a hybrid continuous-discrete Kalman filter, we evaluated the log-likelihood ratio  via the Sequential Probability Ratio Test (SPRT) for binary discrimination and the Cumulative Sum (CUSUM) algorithm for change-point detection. 

Beyond validating the feasibility of these strategies in a state-of-the-art experimental platform, our results open several promising directions for future research. From a methodological standpoint, the sequential framework can be extended to tackle multiple and composite hypothesis testing, signal segmentation (involving the detection of multiple abrupt changes) and anomaly detection, critical in contexts such as rare signal searches. Introducing adaptive feedback, open and closed loop and in both classical and quantum forms, and quantum enhanced resources, such as spin-squeezed states,  could further enhance detection efficiency or sensitivity. This synergy between quantum sensing and sequential decision theory has yet to be fully explored and could unlock new performance frontiers.

These results confirm that sequential decision-making protocols can be efficiently integrated with continuously monitored quantum systems. While we focused on atomic spin‐noise magnetometry, the methodology, combining state‐space modeling, real‐time filtering, and sequential decision rules, is broadly applicable across other quantum sensing platforms such as optomechanical resonators \cite{barzanjeh2022optomechanics}, superconducting circuits \cite{blais21QuantumCircuit}, NV‑centers in diamond \cite{dohertyNitrogenvacancyColourCentre2013}, and cavity-based architectures \cite{bitarafanOnChipHighFinesseFabryPerot2017}, provided suitable measurement models exist.

Finally, an important open direction lies in model-free approaches, where the log-likelihood ratio is inferred directly from measurement data using machine-learning techniques. Such methods could bypass the need for full model knowledge, making sequential analysis more flexible and robust in complex, real-world environments.

\begin{acknowledgments}
We are grateful to Charikleia Troullinou for insightful discussions during the early stages of this project.
This work was supported by the QuantERA grant
C’MON-QSENS! and Spanish Agencia Estatal de Investigación, project PID2022-141283NB-I00 with the
support of FEDER funds, by the Spanish MCIN
with funding from European Union NextGenerationEU (grant PRTR-C17.I1) and the Generalitat de Catalunya, as well as the Spanish MTDFP
through the QUANTUM ENIA project call - Quantum Spain project, and by the European Union through the Recovery, Transformation and Resilience Plan - NextGenerationEU within the framework of the Digital Spain 2026 Agenda.
ERS was supported by FI AGAUR grant Joan Oró (ref 2024 FI-2 00410). JC also acknowledges support from the ICREA Academia award. AS and MWM acknowledge European Commission projects Field-SEER (ERC 101097313), OPMMEG (101099379) and QUANTIFY (101135931); Spanish Ministry of Science MCIN project SAPONARIA (PID2021-123813NB-I00) and SALVIA (PID2024-158479NB-I00), ``NextGenerationEU/PRTR.'' (Grant FJC2021-047840-I) and ``Severo Ochoa'' Center of Excellence CEX2019-000910-S;  Generalitat de Catalunya through the CERCA program,  DURSI grant No. 2021 SGR 01453 and QSENSE (GOV/51/2022).  Fundació Privada Cellex; Fundació Mir-Puig. MS acknowledges support from Ayuda Ramón y Cajal 2021 (RYC2021-032032-I, MICIU/AEI/10.13039/501100011033, ESF+) as well as Project FEDER C-EXP-256-UGR23 Consejería de Universidad, Investigación e Innovación y UE Programa FEDER Andalucía 2021-2027.
GG acknowledges financial support by the DAAD, the Deutsche Forschungsgemeinschaft (DFG, German Research Foundation, project numbers 447948357 and 440958198), the SinoGerman Center for Research Promotion (Project M-0294), the German Ministry of Education and Research (Project QuKuK, BMBF Grant No. 16KIS1618K), and the Stiftung Innovation in der Hochschullehre.
\end{acknowledgments}

\bibliography{library}
\clearpage
\newpage

\onecolumngrid
\setlength{\baselineskip}{20pt}
\appendix

\begin{center}  
\section*{Supplementary Material}
\end{center}

\section{\label{sec:probmodel} Probability distribution of the measured signal in the steady state}
\label{sec:probdiststeadystate}
The formal solution of the Ornstein–Uhlenbeck process given in Eq.~\eqref{syquation} is well known \cite{jacobs2010stochastic} and can be readily obtained to be  
\begin{equation}
    \vecb{J}_t=\mathrm{e}^{\hat A t}\vecb{J}_0 + \int_0^t \mathrm{e}^{\hat A \left(t-s\right)}\sqrt{Q} \vecb{dW}=\mathrm{e}^{\hat A t}\vecb{J}_0 + \vecb{N}_t,
\label{ss}
\end{equation}
with $\vecb{J}_0$ the initial condition and $\vecb{J}_t$ being a Markovian process. The first term in Eq.~\eqref{ss} is the deterministic solution whereas the second is the stochastic contribution due to Wiener noise, defined by $\vecb{N}_t$. Since Eq.~\eqref{syquation} is a linear system with additive Gaussian noise,  $\vecb{J}_t$ is a stochastic Gaussian variable, and is therefore fully  determined by its mean, $ \expect{[\vecb{J}_t]}$, and covariance matrix  $\mathrm{Cov} \left(\vecb{J}_t,\vecb{J}_{t'}\right)$. Averaging Eq.~\eqref{syquation}, and recalling that $\expect{[dW_i]}=0$, we obtain the differential equation $ d\expect{[\vecb{J}_t]} = \hat A \expect{[\vecb{J}_t]} dt$ whose solution can be verified to be 
\begin{equation}
   \expect{[\vecb{J}_t]}=\mathrm{e}^{\hat A t}\vecb{J}_0, \quad \textrm{ with } \quad \mathrm{e}^{\hat A t}=\mathrm{e}^{-\gamma t} \begin{pmatrix}
          \cos{\left(\omega_L t\right)} &  \sin{\left(\omega_L t\right)} \\
        -\sin{\left(\omega_L t\right)} &  \cos{\left(\omega_L t\right)} 
    \end{pmatrix}.
\end{equation}
Notice that the asymptotic expected value achieved in the steady state is $\displaystyle \expect{[\vecb{J}_t]}_A:=\lim_{t\to\infty}\expect{[\vecb{J}_t]}=0$ regardless of $\vecb{J}_0$.

From Eq.~\eqref{ss}, it is straightforward to check that the noise properties of $\vecb{J}_t$ are inherited by those of  its stochastic component $\vecb{N}_t$. Using that  $\expect{[\vecb{N}_t]}=0$ since $\vecb{N}_t$ is the infinite sum of Wiener noises:
\begin{equation}
    \begin{aligned}
        \textrm{Cov}(\boldsymbol{J}_t,\boldsymbol{J}_{t'})&=\expect{[\boldsymbol{J}_t \cdot \boldsymbol{J}_{t'}^T]}-\expect{[\boldsymbol{J}_t ]}\expect{[\boldsymbol{J}_{t'}^T ]}=   \expect{[\left(\mathrm{e}^{\hat{A}t}\boldsymbol{J}_0+\boldsymbol{N}_{t}\right)\left(\mathrm{e}^{\hat{A}t'}\boldsymbol{J}_0+\boldsymbol{N}_{t'}\right)^T]}-\expect{[\mathrm{e}^{\hat{A}t}\boldsymbol{J}_0+\boldsymbol{N}_{t}]}\expect{[\left(\mathrm{e}^{\hat{A}t'}\expect{\boldsymbol{J}_0}+\boldsymbol{N}_{t'}\right)^T]}=\\
        &=\expect{[\boldsymbol{N}_{t}\boldsymbol{N}_{t'}^T]}= \textrm{Cov}(\boldsymbol{N}_t,\boldsymbol{N}_{t'}).
    \end{aligned}
\end{equation} 
In fact, due to the properties of the independent Wiener noise processes i.e., $\expect{[dW_{s}^idW_{s'}^j]}=\delta_{ij}\delta(s-s')dsds'$ for $i=y,z$ and $j=y,z$, one can see that when $t'<t$, 
\begin{equation}
    \begin{aligned}
      \expect{[\boldsymbol{N}_{t}\boldsymbol{N}_{t'}^T]}&=Q\int_0^t\int_0^{t'} \mathrm{e}^{\hat{A}(t-s)}\expect{[d\boldsymbol{W}_s d\boldsymbol{W}_{s'}^T]} \left(\mathrm{e}^{\hat{A}(t'-s')}\right)^T=Q\int_0^t\int_0^{t'} \mathrm{e}^{\hat{A}(t-s)}\delta(s-s')ds ds'\mathbb{1} \left(\mathrm{e}^{\hat{A}(t'-s')}\right)^T=\\
      &=Q\int_0^{t'}\mathrm{e}^{\hat{A}(t-s')} \left(\mathrm{e}^{\hat{A}(t'-s')}\right)^T ds'=\frac{Q}{2\gamma}\left(\mathrm{e}^{-\gamma(t-t')}-\mathrm{e}^{-\gamma(t+t')}\right)\begin{pmatrix}
       \cos\left(\omega_L\left(t-t'\right)\right) & \sin \left(\omega_L\left(t-t'\right)\right)\\
       -\sin \left(\omega_L\left(t-t'\right)\right) & \cos \left(\omega_L\left(t-t'\right)\right)
    \end{pmatrix}.
    \end{aligned}
\end{equation}
Repeating the same procedure for $t'>t$, we conclude:
\begin{equation}
  \textrm{Cov}(\boldsymbol{J}_t,\boldsymbol{J}_{t'})= \frac{Q}{2\gamma}\left(\mathrm{e}^{-\gamma |t-t'|}-\mathrm{e}^{-\gamma(t+t')}\right) \begin{pmatrix}
       \cos\left(\omega_L \left(t-t'\right) \right) & \sin \left(\omega_L \left(t-t'\right) \right)\\
       -\sin \left(\omega_L  \left(t-t'\right) \right) & \cos \left(\omega_L  \left(t-t'\right) \right)
    \end{pmatrix}.
\end{equation}

Note that for $t=t'$, $\textrm{Cov}(\boldsymbol{J}_t,\boldsymbol{J}_{t})=\frac{Q}{2\gamma}\left(1-\mathrm{e}^{-2\gamma t}\right)\mathbb{1}$, and at $t=t'=0$, the variance is zero since $\boldsymbol{J}_0$ is a fixed initial condition. Moreover, the asymptotic variance is $\textrm{Cov}(\boldsymbol{J}_t,\boldsymbol{J}_{t})_A:=\displaystyle\lim_{t\to \infty} \textrm{Cov}(\boldsymbol{J}_t,\boldsymbol{J}_{t}) =\frac{Q}{2\gamma}\mathbb{1}$. For a process with unknown $\boldsymbol{J}_0$ but that starts in the steady state, i.e.  $\expect{[\vecb{J}_0]}=0$ and $\textrm{Var}, (\boldsymbol{J}_0,\boldsymbol{J}_{0})=\frac{Q}{2\gamma}\mathbb{1}$, the first two moments are:
\begin{equation}
  \begin{aligned}
      \expect{[\vecb{J}_t]}_{\textrm{steady}}&=0 \\
      \textrm{Cov}(\boldsymbol{J}_t,\boldsymbol{J}_{t'})_{\textrm{steady}}&= \frac{Q}{2\gamma} \mathrm{e}^{-\gamma |t-t'|} \begin{pmatrix}
       \cos\left(\omega_L  \left(t-t'\right)  \right) & \sin \left(\omega_L  \left(t-t'\right) \right)\\
       -\sin \left(\omega_L  \left(t-t'\right) \right) & \cos \left(\omega_L\  \left(t-t'\right) \right)
    \end{pmatrix}.
  \end{aligned}
  \label{ssmoments}
\end{equation}
Observe that $\textrm{Cov}(\boldsymbol{J}_t,\boldsymbol{J}_{t'})_{\textrm{steady}}$ is translational invariant since it can be expressed as a function of  the difference $t-t'$.

Regarding the properties of the photocurrent $I$, since $\vecb{J}$ and $\xi$ are independent normal variables, $I$ is also a Gaussian. At the discretized harvesting times $t_m=m \Delta$, the mean and covariance matrix elements of $I$ in the steady state are:
\begin{equation}
\begin{aligned}
\expect{[I_m]}&=g_D\expect{[J_z(t_n)]}=0\\
      \textrm{Cov}(I_m,I_{m-k})&=g_D^2\textrm{Cov}(J_{z}(t_m),J_{z}(t_{m-k}))+\textrm{Cov}(\xi(t_m),\xi(t_{m-k}))=g_D^2 \frac{Q}{2\gamma}\mathrm{e}^{-\gamma |k|\Delta} \cos\left(\omega_L k\Delta\right)+\delta_{0k}\frac{R}{\Delta}:=T_k, 
   \end{aligned}
\end{equation}
where again the second moments of $I$ are translational invariant. With this at hand, we present the normal probability distribution of a sequence of $n$ photocurrent observations in the steady state, $\boldsymbol{I}_n=\{I_1,..,I_{n}\}$, i.e. $\boldsymbol{I}_n\sim \mathcal{N} (\boldsymbol{\mu}_n,\hat T_n )$ with mean and Toeplitz covariance matrix:
\begin{equation}
\begin{aligned}
        \boldsymbol{\mu}_n&=\expect{[\boldsymbol{I}_n]}=\vecb{0},\\
        \hat T_n &=  \begin{pmatrix}
     T_0   & T_1 & T_2 &\cdots & T_n \\
       T_1   & T_0 & T_1  & &T_{n-1} \\
       T_2 & T_1 & T_0 & \ddots & \vdots \\
       \vdots &  & \ddots & \ddots & T_1\\
       T_n &  T_{n-1} & \cdots &T_1 & T_0
    \end{pmatrix}. 
\end{aligned}
\end{equation}

\section{\label{sec:Toeplitz} Toeplitz matrices}

In this Appendix we overview the main definitions and theorems that we use in this work regarding Toeplitz operators, first introduced in Sec.~\ref{sec:probdist}.

A {Toeplitz matrix} $T_n(f)$ of size $n \times n$ is defined by a \textit{symbol} $f(\theta)$, a $2\pi$-periodic function $f\in L^\infty ([0,2 \pi))$, through its entries \cite{bottcher2012introduction}:
\begin{equation}
(T_n(f))_{ij} = C_{i-j}, \quad \text{where} \quad C_k = \frac{1}{2\pi} \int_0^{2\pi} f(\theta) e^{-ik\theta}  d\theta, \quad 1 \leq i,j \leq n.
\end{equation}
Here $L^\infty ([0,2 \pi))$ is the space of all essentially bounded, Lebesgue-measurable functions on the interval $[0,2\pi)$.

The symbol generates the matrix via Fourier coefficients, with $f(\theta) = \sum_{k=-\infty}^{\infty} C_k e^{ik\theta}$. For large $n$, the spectral properties of $T_n(f)$ are governed by harmonic analysis on the unit circle.

For the process under study, recalling Eq.~\eqref{eq:CovarianceME},
\begin{equation}
C_k=T_{i,i+k}=\gamma S_\mathrm{at}\mathrm{e}^{-\gamma |k| \Delta} \cos\left(k \omega_L \Delta  \right)+\delta_{k,0}\frac{S_{\mathrm{ph}}}{\Delta},
\label{eq:CovarianceME2}
\end{equation}
with $S_\mathrm{at}=g_D^2 \frac{Q}{2\gamma^2}$. 
The corresponding symbol \( f(\theta) \) is:
\begin{equation}
f(\theta) = \frac{1}{2}\left(\frac{\gamma S_\mathrm{at} \sinh(\gamma\Delta)}{\cosh (\gamma \Delta) - \cos(\theta -\Delta\omega_L)}
+
\frac{\gamma S_\mathrm{at} \sinh(\gamma\Delta)}{\cosh (\gamma \Delta) - \cos(\theta+\Delta\omega_L)}\right)+ \frac{S_{ph}}{\Delta}
\end{equation}
which, by taking $\theta=\omega \Delta$, at leading order in $\Delta^{-1} \ll 1 $ leads to Lorentzian profile found in Eq.~\eqref{lorentzian}.

\medskip
\noindent \emph{Szegő's Theorems and Spectral Properties}

\begin{theorem}{\bf First Szegő Theorem:}
For $f > 0$ with $\log f \in L^1$, the determinant satisfies:
\begin{equation}
\lim_{n \to \infty} \frac{1}{n} \log \det T_n(f) = \frac{1}{2\pi} \int_0^{2\pi} \log f(\theta)  d\theta \equiv \log G(f),
\end{equation}
where $G(f)$ is the geometric mean of $f$.
\end{theorem}

\begin{theorem}{\bf Strong Szegő Theorem}
If $f$ is sufficiently smooth (e.g., $f \in C^{\infty}$) and positive:
\begin{equation}
\det T_n(f) \sim G(f)^n \cdot E(f), \quad E(f) = \exp\left( \sum_{k=1}^{\infty} k (\log f)_k (\log f)_{-k} \right),
\end{equation}
where $(\cdot)_k$ denote Fourier coefficients. The eigenvalues $\{\lambda_j\}_{j=1}^n$ satisfy $\lambda_j \to f(2\pi j/n)$ in distribution.
\end{theorem}

\emph{Traces and Inverses}
For smooth $f > 0$:
\begin{align}
\text{\underline{Traces of powers:}} \quad & \Tr(T_n(f)^k) = \frac{n}{2\pi} \int_0^{2\pi} f(\theta)^k  d\theta + C_k(f) + o(1) \\
\text{\underline{Inverses:}} \quad & T_n(f)^{-1} \approx T_n(1/f), \quad \Tr(T_n(f)^{-1}) = \frac{n}{2\pi} \int_0^{2\pi} \frac{d\theta}{f(\theta)} + O(1)
\end{align}
Here $C_k(f) = O(1)$ depends on the symbol's smoothness. The Fourier basis $\{e^{ij\theta}\}_{j=1}^n$ asymptotically diagonalizes $T_n(f)$.

\emph{Generalized Traces:}
For products involving inverses of Toeplitz matrices with smooth positive symbols:
\begin{equation}
\Tr\left(\prod_{j=1}^mT_n(f_j(\theta))^{\epsilon_j}\right) = \frac{n}{2\pi} \int_0^{2\pi} \prod_{j=1}^mf_j(\theta)^{\epsilon_j}  d\theta 
+ O(1),
\label{gentraces}
\end{equation}
where $\epsilon_j=\pm 1$.

\begin{proof}
We provide a sketch of the proof using some of the results in \cite{bottcher2012introduction}. A Toeplitz matrix $T_n(f)$ can be understood as $T_n(f)=P_n T(f) P_n$, where $T(f)$ is the  infinite-dimensional Toeplitz matrix (bounded operator) and $P_n: \{x_1,x_2,x_3...\}\to \{x_1,x_2,...x_n,0,0,...\}$ is the projector on the $n$-th dimensional subspace. The main ingredients for the proof are the following: 
\begin{enumerate}
    \item By Proposition 2.12 (Widom), page 40, if $a,b\in L^\infty$, then 
\begin{equation}
   T_n(a b)= T_n(a)T_n(b)+P_n H(a) H(\tilde b) P_n + W_n H(\tilde a)H(b) W_n,
\end{equation}
where $H$ denotes the so called Hankel matrix, $W_n:\{x_1,x_2,x_3...\}\to \{x_n,x_{n-1},...x_1,0,0,...\}$ is a projection and $\tilde a:= a(1/t)$ with $t=e^{i \theta}$. 
\item By Theorem 2.14 (Widom) page 42, if $T(a)$ is invertible, for all sufficiently large $n$
\begin{equation}
    T_n(a)^{-1}=T_n(a^{-1})+P_n K(a) P_n + W_n K(\tilde a)W_n + C_n,
\end{equation}
where $K(a)=H(a^{-1}) H(\tilde a)T^{-1}(a)$, $K(\tilde a)=H(\tilde a^{-1}) H( a)T^{-1}(\tilde a)$ and $\|C_n\|\to 0$ ad $n\to \infty$. Here $\|\cdot \|$ is the operator norm.
\item By Theorem 2.15 (page 43), if $T(a)$ is invertible and $\sum_{n\in \mathbb{Z}}|n|^\alpha|a_n|<\infty$ for some $\alpha>0$, then $\|C_n\|=o(n^{-\alpha}).$
\item Let $W$ be the Wiener algebra and $W \cap B^{1/2}=\{a\in W: \sum_{n\in \mathbb{Z}}(|n|+1)|a_n|^2<\infty\}$. From pages 122-123, if $a\in W \cap B^{1/2}$ and has no zeros, then $a^{-1}\in W \cap B^{1/2}$. Moreover, if $a,b\in W \cap B^{1/2}$ ( i.e. $a,b$ sufficiently smooth), then $H(a)H(\tilde b)$ is a trace class operator, i.e $\|H(a)H(\tilde b)\|_1< \infty$ where $\|\cdot\|_1$ is the trace norm.
\end{enumerate}
Notice that from 1. and 4.,
\begin{equation}
  \begin{aligned}
        |\mathrm{Tr}(T_n(ab)-T_n(a)T_n(b))|&=|\mathrm{Tr}(P_n H(a) H(\tilde b) P_n + W_n H(\tilde a)H(b) W_n)|\leq\\
        &\leq\|P_n H(a) H(\tilde b) P_n + W_n H(\tilde a)H(b) W_n\|_1\leq\\
        &\leq\|P_n H(a) H(\tilde b) P_n\|_1 + \|W_n H(\tilde a)H(b) W_n\|_1<c
  \end{aligned}
    \label{eq1}
\end{equation}
where $c$ is a constant since $\|(P_n H(a) H(\tilde b) P_n\|_1\leq \|P_n\|^2\| H(a) H(\tilde b) \|_1=\| H(a) H(\tilde b) \|_1<\infty$ and the same for $\|W_n H(\tilde a)H(b) W_n)\|_1$ (a trace class operator multiplied by a bounded operator is still trace class). 

For traces involving inverses and $\alpha\geq 1$, from 2., 3. and 4.
\begin{equation}
  \begin{aligned}
        |\mathrm{Tr}(T_n(a)^{-1}-T_n(a^{-1}))|&=|\mathrm{Tr}(P_n H(a^{-1}) H(\tilde a)T^{-1}(a) P_n + W_n H(\tilde a^{-1}) H( a)T^{-1}(\tilde a) W_n)+ C_n|\leq\\
        &\leq\|P_n H(a^{-1}) H(\tilde a)T^{-1}(a) P_n\|_1 + \|W_n H(\tilde a^{-1}) H( a)T^{-1}(\tilde a) W_n)\|_1+||C_n||_1<c'
  \end{aligned}
  \label{eq2}
\end{equation}
since $P_n H(a^{-1}) H(\tilde a)T^{-1}(a) P_n$ and $W_n H(\tilde a^{-1}) H( a)T^{-1}(\tilde a) W_n)$ are trace class due to the bounded property of  $T^{-1}(a)$ and $T^{-1}(\tilde a)$. Moreover,
$||C_n||_1<n\|C_n\|=o(n^{1-\alpha})<o(1)$ for $\alpha\geq 1$.

From Eq.~\eqref{eq2}, 
\begin{equation}
   |\mathrm{Tr}(T_n(b)T_n(a)^{-1}-T_n(b)T_n(a^{-1}))| < \|T_n(b)\|\|T_n(a)^{-1}-T_n(a^{-1})\|_1<c''
   \label{eq3}
\end{equation}
and therefore combining  Eq.~\eqref{eq1} and Eq.~\eqref{eq3},
\begin{equation}
    \mathrm{Tr}(T_n(b)T_n(a)^{-1})=\mathrm{Tr}(T_n(b a^{-1}))+O(1)
\end{equation}
These process can be repeated recursively, when increasing the number of multiplying Toeplitz inside the trace to obtain 
\begin{equation}
    \Tr\left(\prod_{j=1}^mT_n(f_j(\theta))^{\epsilon_j}\right)=\Tr\left(\prod_{j=1}^mT_n(f_j(\theta)^{\epsilon_j})\right)+ O(1)
\end{equation}
and by the Szegő theorems arrive to the result in Eq.~\eqref{gentraces}. 
\end{proof}

The Generalized traces property in Eq.~\eqref{gentraces} is defined for the symbol $f(\theta)$ with $\theta$ in the unit circle, i.e. $\theta\in (0,2\pi)$. When we move to the continuous limit, i.e. $\Delta\to0$, we have that $\omega$ is defined in the interval $(-\infty,\infty)$. Moreover,  $f_j(\omega \Delta) = g_j(\omega) + \order(\Delta)$ and for sufficiently smooth symbols: 
\begin{align}
\Tr\left(\prod_{j=1}^mT_n(f_j(\theta))^{\epsilon_j}\right)= 
\frac{n\Delta}{2\pi} \int_0^{2\pi/\Delta} \prod_{j=1}^mf_j(\omega \Delta)^{\epsilon_j}  d\omega+ O(1)= \frac{t}{2\pi} \int_{-\infty}^{\infty} \prod_{j=1}^mg_j(\omega)^{\epsilon_j}  d\omega + t\order(\Delta)
+ O(1),
\end{align}

\section{Hybrid Kalman filter algorithm}
\label{sec:Kalmanfilter}
In the two step procedure of the  KF algorithm,  we  denote by ${\boldsymbol{j}}_{k | k-1}$ the predicted state of the system at time $t_{k}$, and  the corresponding estimators that construct the mean and covariance of the normal distribution as $\tilde{\boldsymbol{j}}_{k | k-1}$ and $\hat{\Sigma}_{k | k-1}$. In addition, for the computation of the next predictions, we denote the updated estimators by  ${\boldsymbol{\tilde j}}_{k | k}$ and $\hat\Sigma_{k | k}$. Then, the explicit procedure is \cite{kalman2,trees2001detection}:

\textit{Prediction} $\left(\tilde{\boldsymbol{j}}_{k-1 \mid k-1} \rightarrow \tilde{\boldsymbol{j}}_{k \mid k-1}, \textrm{ }\hat\Sigma_{k-1 \mid k-1} \rightarrow \hat\Sigma_{k \mid k-1}\right)$ We define the transition matrix $\hat\Phi$, as the solution of the deterministic part of the state-vector. When $ \hat A$ is time-independent, for any $\Delta$-interval, $\hat\Phi:=\hat\Phi_{\Delta}={e}^{{\hat A} \Delta}$. 
Then, we can construct $\tilde{\boldsymbol{j}}_{k \mid k-1}$ and $\hat\Sigma_{k \mid k-1}$ over the interval $\left[t_{k-1}, t_{k}\right]$, as 
\begin{equation}
\begin{aligned}
\tilde{\boldsymbol{j}}_{k \mid k-1}&=\hat\Phi\tilde{\boldsymbol{j}}_{k-1 \mid k-1}\\
     \hat\Sigma_{k \mid k-1}&=\hat \Phi \hat \Sigma_{k-1 \mid k-1}\hat \Phi +\hat{Q}_{k}^{\Delta},
\end{aligned}
\end{equation}
where $\hat{Q}_{k}^{\Delta}:=\int_{t_{k-1}}^{t_{k}} \hat\Phi_{t_{k}, \tau}  \hat{Q}_{\tau}  \hat \Phi_{t_{k}, \tau}^{T} {~d} \tau$  is the effective covariance matrix of the system noise.

  \textit{Update}   $\left(\tilde{\boldsymbol{j}}_{k \mid k-1} \rightarrow \tilde{\boldsymbol{j}}_{k \mid k}, \textrm{ } \hat\Sigma_{k \mid k-1} \rightarrow \hat\Sigma_{k \mid k}\right)$ 
  To take into account the $k$-th outcome $\boldsymbol{I}_{k}$, 
  one computes 
\begin{equation}
\begin{aligned}
    \tilde{\boldsymbol{j}}_{k \mid k}&=\tilde{\boldsymbol{j}}_{k \mid k-1}+\hat{K}_{k} \tilde{y}_{k} \\
    \hat{\Sigma}_{k \mid k}&=\left(\mathbb{1}-\hat{K}_{k} \hat{C}\right) \hat{\Sigma}_{k \mid k-1}  
\end{aligned}
\end{equation}
where the so-called innovation and the Kalman gain are respectively $  \tilde{y}_{k}=I_{k}-\mu_{k}$ and $\hat{K}_{k}=\hat{\Sigma}_{k \mid k-1} \hat{C}^{T} \hat{S}_{k}^{-1}$. Here $\mu_{k}=\hat{C} \tilde{\boldsymbol{j}}_{k \mid k-1}$ is the prediction of the $k$-th outcome value and $ \sigma_k^2=\expect{[\tilde{y}_{k} \tilde{y}_{k}^{T}]}=\hat{G}_{k}^{\Delta}+\hat{C} \hat{\Sigma}_{k \mid k-1} \hat{C}^{T}$  is the covariance matrix for the $k$-th innovation $\tilde{y}_{k}$ 

 Finally, to initialize the filter at time $t_{0}$ one chooses an appropriate initial distribution, $p \left(\boldsymbol{J}_{0|0}\right) \sim \mathcal{N}\left(\boldsymbol{\tilde J}_{0|0}, \hat{\Sigma}_{0|0}\right)$, representing the knowledge about the state vector. In our case, we start with the steady state solution derived in Sec.~\ref{sec:probmodel}, $\boldsymbol{\tilde J}_{0|0}=\boldsymbol{0}$ and $\hat{\Sigma}_{\boldsymbol{0|0}}=\frac{Q}{2\gamma}\mathbb{1}$.

\section{\label{sec:diagrams} Block diagrams for sequential strategies}

\begin{figure}[H]
    \centering   \includegraphics[width=0.45\textwidth]{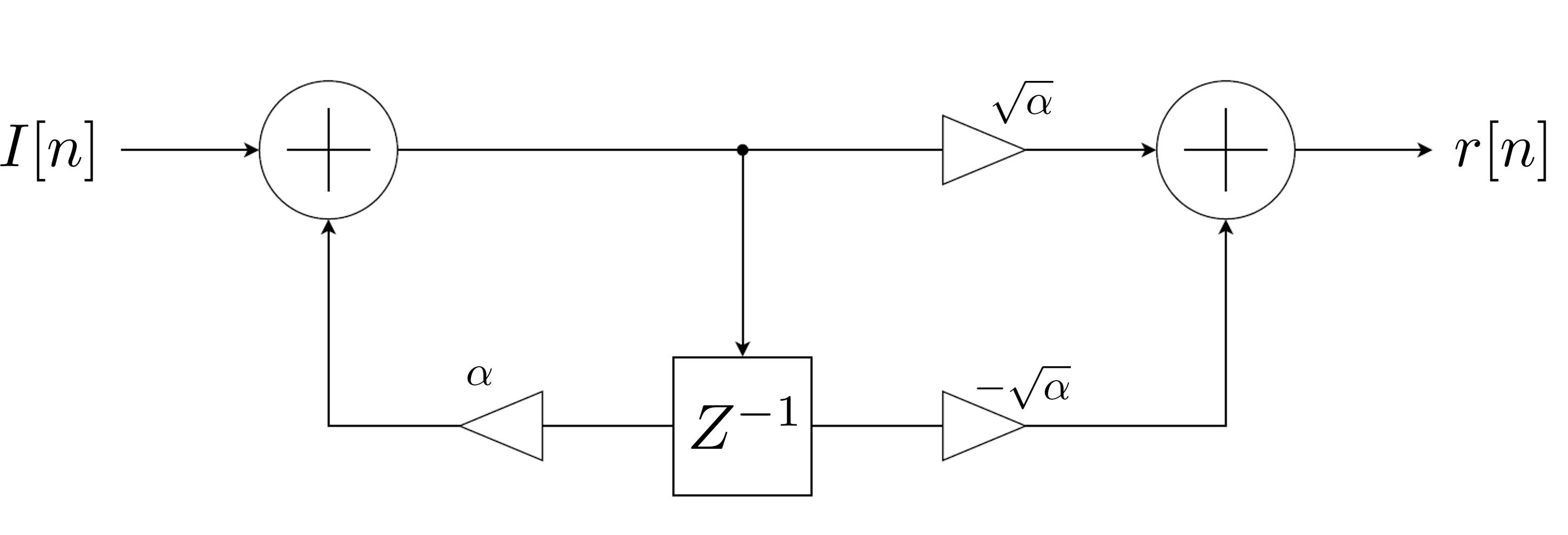}  
 \caption{Block diagram of Eq.~\eqref{notcheq} (see  main text) which describes the  high-pass filter.  Here $n$ represents the time step and $Z^{-1}$ the time delay. Recall that $\alpha$ is the strength of the filter.}
    \label{highpassfilter}
\end{figure}

\begin{figure}[H]
\includegraphics[width=0.95\textwidth]{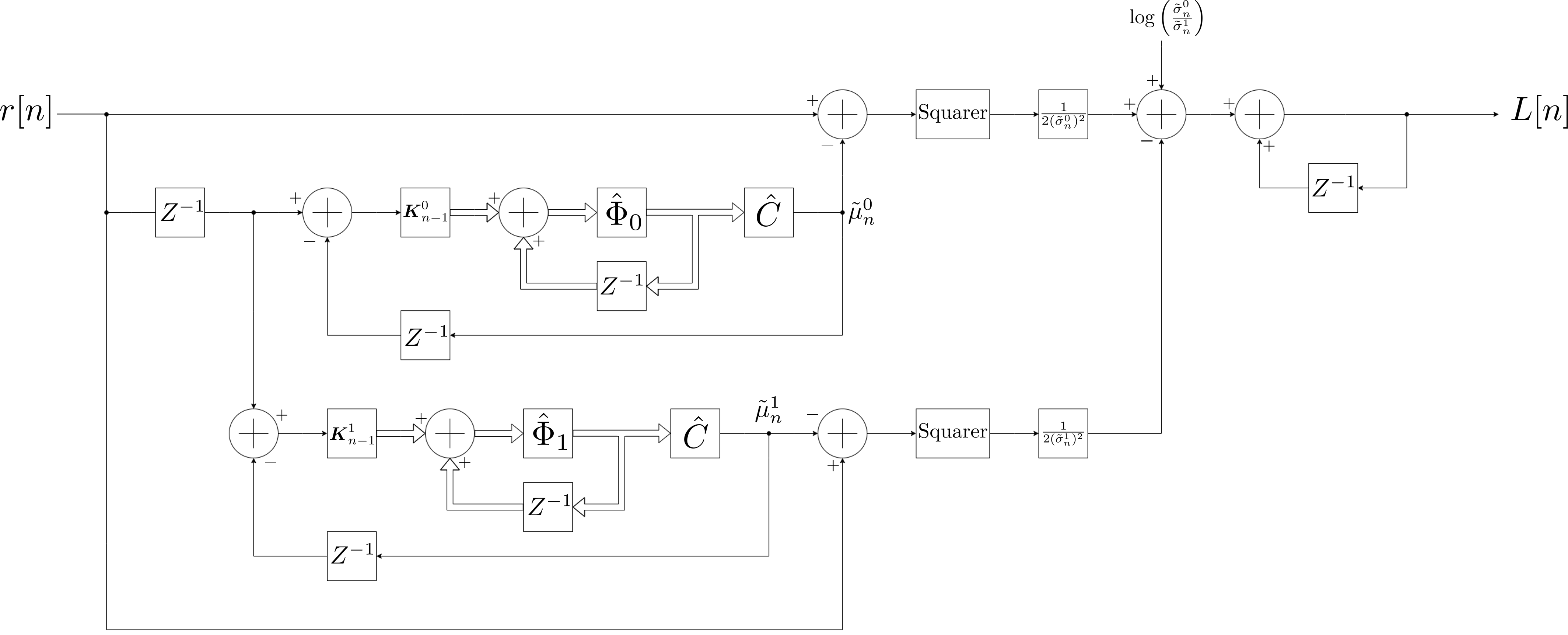}
\caption{\label{fig:llr_realtime} Block diagram of Eq.~\eqref{lrkalman} (see  main text) which describes the  log-likelihood ratio in terms of the Kalman estimators.  Here $n$ represents the time step and $Z^{-1}$ the time delay. The rest of the elements are part of the hybrid Kalman filter algorithm (see Appendix \ref{sec:Kalmanfilter}).
}
\end{figure}

In this section, we present the block diagrams of the high-pass filter and the log-likelihood ratio that can be deployed on an FPGA board for real-time implementation.  As explained in the main text, these two diagrams represent a chain process, i.e. at each time step $n$, the filtered signal $R_n$ is first calculated (Figure \ref{highpassfilter}), then the log-likelihood ratio is evaluated  (Figure \ref{fig:llr_realtime}), and finally, either the SPRT or CUSUM thresholds are applied for sequential hypothesis testing or quickest change-point detection, respectively. 

 \section{Model for stochastic spin relaxation parameter and experimental data behavior}
 
\label{sec:gammaphenomenon}

\begin{figure}[t]
\centering
\includegraphics[width=0.45\textwidth]{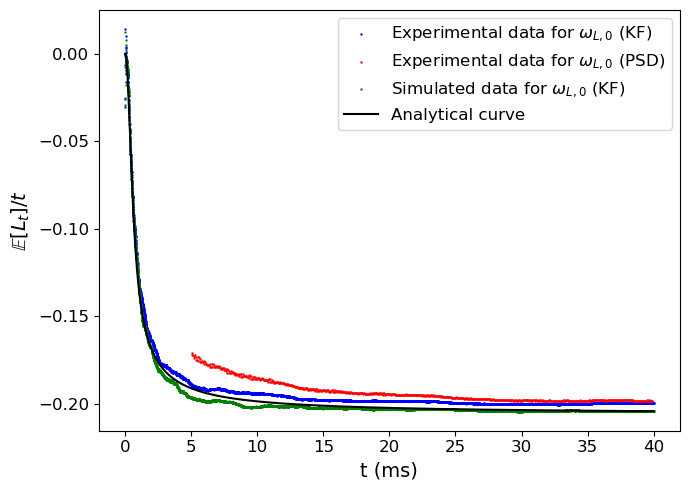} 
\includegraphics[width=0.45\textwidth]{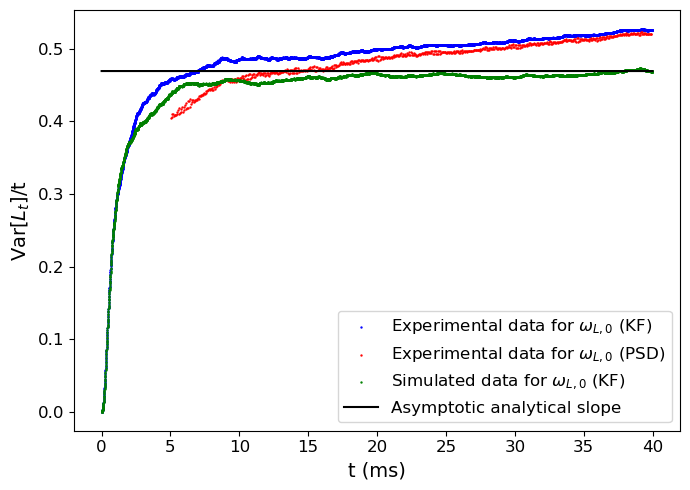} 
\includegraphics[width=0.45\textwidth]{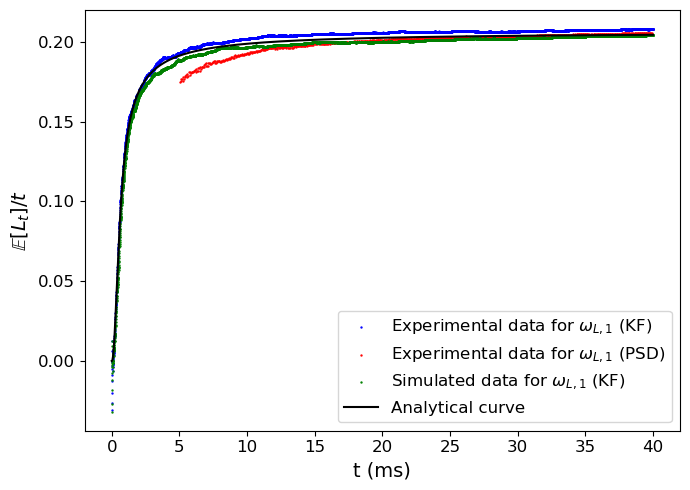} 
\includegraphics[width=0.45\textwidth]{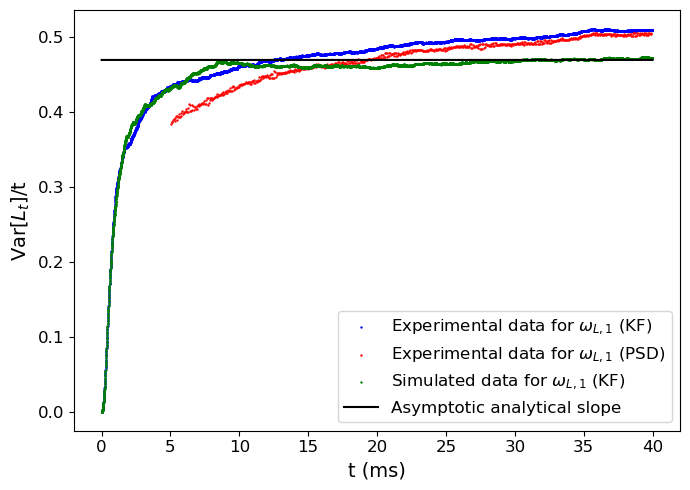} 
\caption{Regularized mean and variance results of the LLR. Top plots for $h=0$ and bottom plots for $h=1$. Left and right show respectively the first and second moments. In all the cases, in blue the computation of the experimental LLR through the Kalman filter is presented and in red the Whittle version. The simulated results are shown in green and the black curves represent the theoretical predictions for constant parameters. }
\label{meanandvariance2}
\end{figure}

As discussed in Section \ref{sec:sequentialstrategiesexp}, at long observation times, we observe discrepancies between the simulated and experimental results in the hypothesis testing scenario. To investigate the origin of these differences, we analyze the LLR as a function of time. In Figure \ref{meanandvariance2}, we present the mean and variance of the statistic obtained from the $10^4$ experimental traces, each lasting $t=40$ (ms) per hypothesis, and compare them with the simulated results and theoretical predictions.  The exact result for the mean is obtained in a recursive way via the Kalman estimators and the analytical slope for the variance is directly the asymptotic solution.

The LLR is computed using both the Kalman filter (shown in blue) and the Whittle log-likelihood method introduced in Sec.~\ref{sec:experimentalimplementation} (shown in red). For the latter, we compute the raw periodogram (no high pass filter applied) and perform a frequency-domain cut, considering only the contributions from the range near the Lorentzian peaks—specifically from $40-60$ (kHz). This range is chosen to be broad enough to retain relevant information while excluding less informative regions and the spurious low frequency noise. Then, the Whittle LLR is
\begin{equation}
  \sum_{k=m}^{l}\log\frac{\bar{S}_1(\omega_k)}{\bar{S}_0(\omega_k)} +\left(\frac{1}{\bar{S}_0(\omega_k)}-\frac{1}{\bar{S}_1(\omega_k)}\right)  S_I(\omega_k)
\end{equation}
where $\omega_k=2\pi k/ t$, and $m$ and $l$ are the integers corresponding to the edges of the region of interest. Observe that both approaches converge as time $t$ increases, and for both hypothesis, the variance grows faster than linearly with time. In contrast, the simulated traces (shown in green) closely follow the expected behavior of the first two moments of the LLR, as indicated by the theoretical curves (shown in black).

\begin{figure}[t]
\centering
\includegraphics[width=0.45\textwidth]{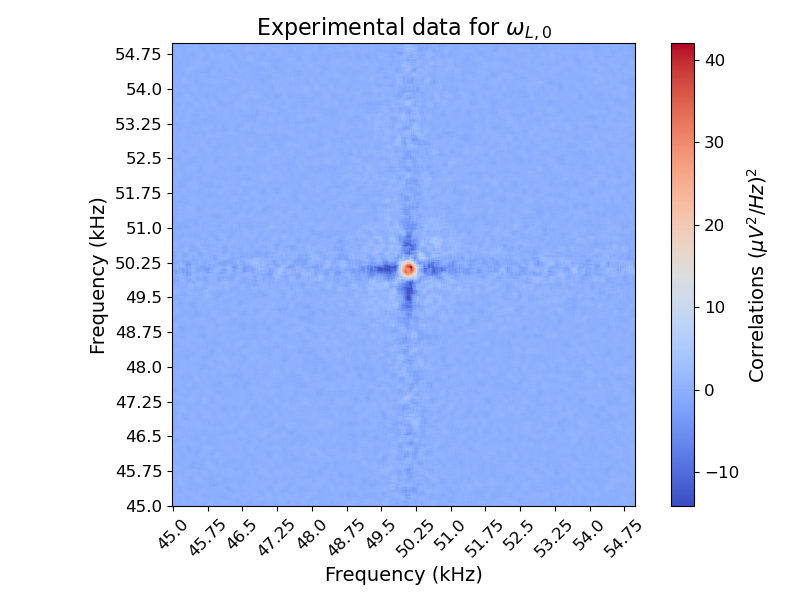} 	
\includegraphics[width=0.45\textwidth]{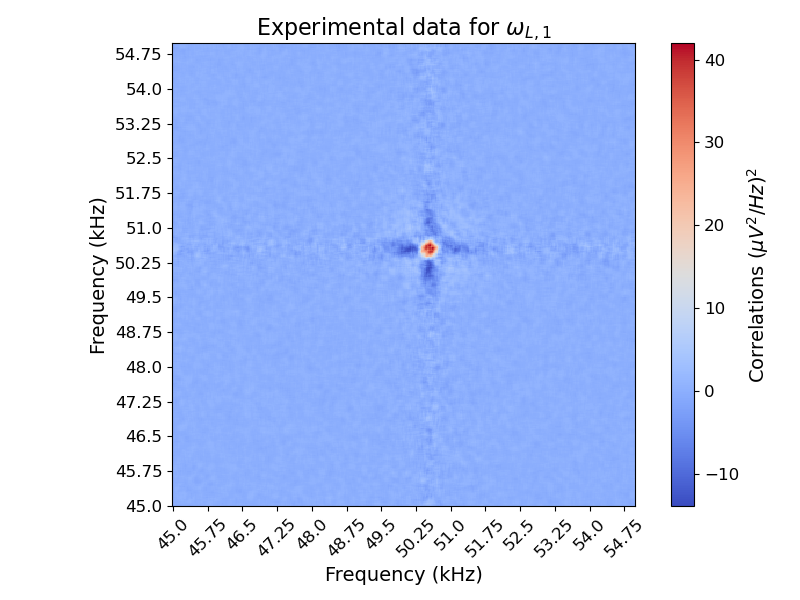} 
\caption{PSD covariance matrix at different frequencies near the Larmor frequency where the diagonal elements are removed. Left and right: results for the experimental dataset under $h=0$ and $h=1$ respectively. }
\label{heatmap}
\end{figure}

To understand the behavior of the LLR, especially with respect to the variance, we computed the covariance matrix across different frequencies of the experimental periodogram.  Figure \ref{heatmap} displays a 2D plot of these frequency correlations near the Larmor frequency, $\omega_{L,h}$ and shows an unexpected structure emerging around this frequency, taking both positive and negative values. To better visualize this structure, we exclude the diagonal elements (i.e., the variance), which is orders of magnitude larger and obscures the finer details. Furthermore, we have applied a $5\times5$ 2D binning to reduce noise and enhance the clarity of the correlation patterns.

\begin{figure}[t]
\centering
\includegraphics[width=0.45\textwidth]{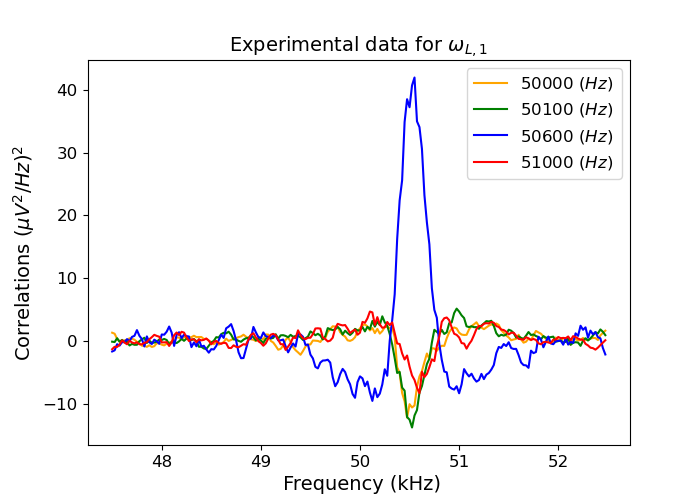} 
\includegraphics[width=0.45\textwidth]{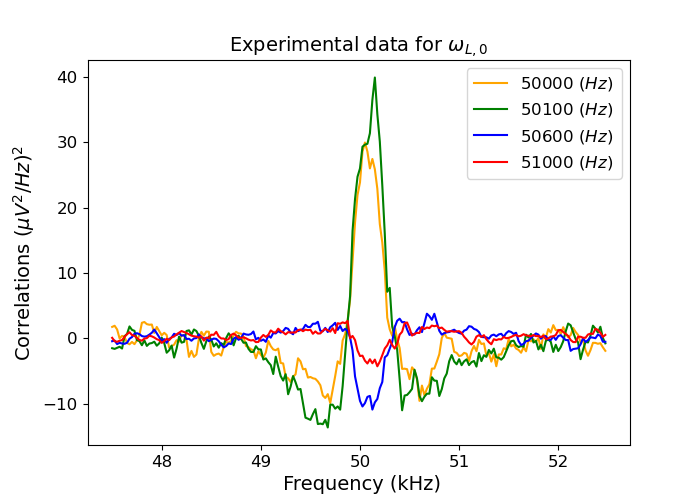}
\caption{Horizontal cut of the PSD covariance matrix at different frequencies. Each color represents a fixed frequency. Again the correlation at same frequencies are removed. Left and right: results for experimental dataset under $h=0$ and $h=1$ respectively.}
\label{heatmaprows}
\end{figure}

For more details about its shape, we present some horizontal cuts of the 2D plots in Figure \ref{heatmaprows} and observe that near the Larmor frequency, the correlations show a sinc function structure.

\begin{figure}[t]
\centering
\includegraphics[width=0.45\textwidth]{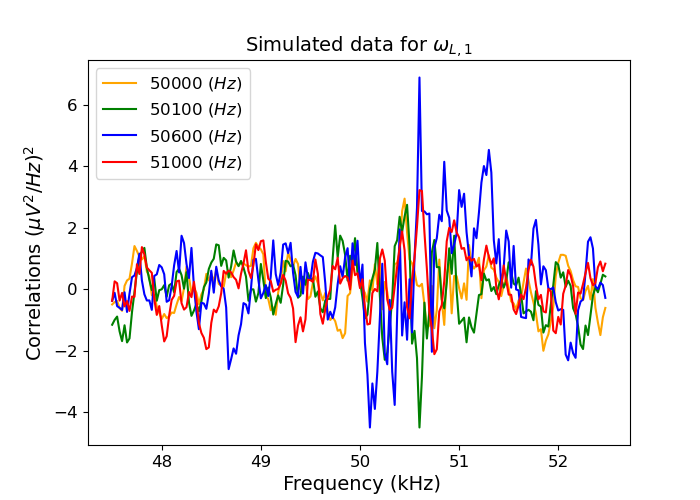} 
\includegraphics[width=0.45\textwidth]{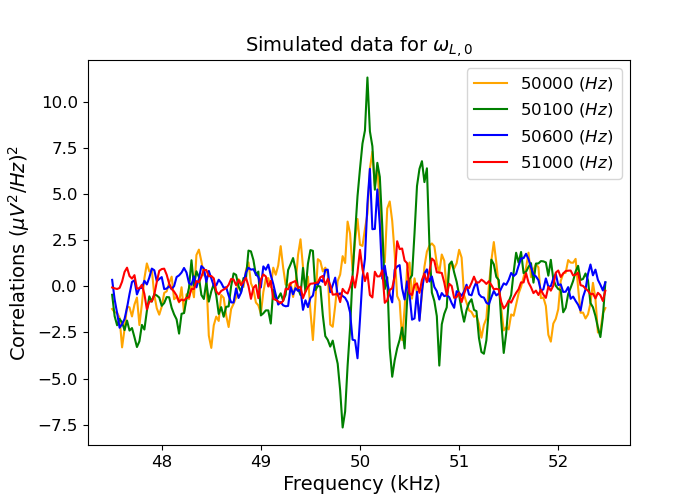}
\caption{Horizontal cut of the PSD covariance matrix at different frequencies. Each color represents a fixed frequency. Again the correlation at same frequencies are removed. Left and right: results for simulated dataset under $h=0$ and $h=1$ respectively.}
\label{heatmaprowssimulated}
\end{figure}

Notice that this profile is not predicted by the physical model in Eqs.~\eqref{syquation}-\eqref{signal}.   At long timescales, is expected to follow an exponential distribution and exhibit no correlations across different frequencies. This theoretical behavior is consistent with the corresponding plots for the simulated data, as shown in Figure 
 \ref{heatmaprowssimulated}. In these simulations, only noise near zero is observed,  and no well-defined pattern emerges.
 
The structure of correlations followed by the experimental datasets, can be explained qualitatively by an stochastic relaxation time. In Eq.~\eqref{syquation}, $\gamma$ is 
 assumed to be constant but it seems to fluctuate run-to-run, during the experimental acquisition time. Let us now briefly explain how run-to-run fluctuations in the relaxation constant can give rise to artificial correlations between different frequency components
of the measured signal. In essence, when a parameter like $\gamma$  fluctuates slightly from run to run, each realization of the signal is effectively drawn from a different process, each with a slightly different frequency response. 
So even though the frequency components remain uncorrelated in any single realization,
the relative amplitudes of all spectral components are jointly influenced by the overall shape of the spectrum 
giving rise to apparent correlations when computing ensemble averages.

We model the fluctuating parameter as
$\tilde\gamma=\gamma+\delta\gamma$, where $\gamma$ is a constant central value and $\delta\gamma$ is a random variable with $\expect{[\delta\gamma]}=0$ and standard deviation $\sigma_{\delta\gamma}$. Moreover, we assume that the total energy of the system is constant, and do not variate for different values of $\gamma$.  The PSD function used is \begin{equation}
    \tilde S_{h,\gamma}(\omega)=\tilde S_{\mathrm{at}} \frac{\gamma}{\gamma^2+(\omega-\omega_{L,h}
    )^2}+S_{\mathrm{ph}}
\end{equation}
where $\tilde S_{at}= S_{at}\gamma= \frac{g_D^2 Q}{2\gamma}$. Notice that $\int_0^\infty  \tilde S_{h,\gamma}(\omega)d\omega=\frac{g_D^2\pi Q}{2\gamma} +\order(\frac{1}{\omega_{L}})$ and for the total area to be independent of $\gamma$, the strength of the Wiener noise has to be proportional to $\gamma$., i.e. $Q:=Q'\gamma$. Then, $\tilde{S}_{at}$ is independent of $\gamma$ and expanding around small $\delta\omega$, we obtain $\tilde S_{h,\gamma+\delta\gamma}(\omega)=\tilde S_{h,\gamma}(\omega)+\tilde g(\omega)\delta\gamma +\mathcal{O}(\delta\gamma^2)$, with
\begin{equation}
    \tilde g(\omega)=\frac{\tilde S_{at} (\gamma+\omega-\omega_L)(\gamma-\omega+\omega_L) }{(\gamma^2+(\omega-\omega_L)^2)^2}.
\end{equation}
It can easily be seen that the correlations under $\delta\gamma$ between two independent PSD at different frequencies is:
\begin{equation}
    \textrm{Cov}((S(\omega),S(\omega'))_{\delta\gamma}=\tilde g(\omega)\tilde g(\omega')\textrm{Var}(\delta\gamma).
    \label{corrgamma}
\end{equation}

\begin{figure}[t]
\centering
\includegraphics[width=0.45\textwidth]{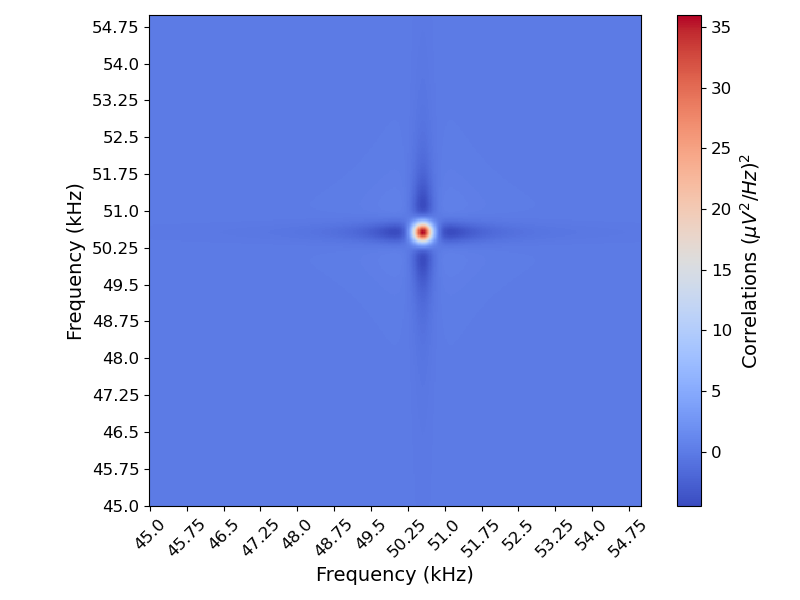} 
\includegraphics[width=0.45\textwidth]{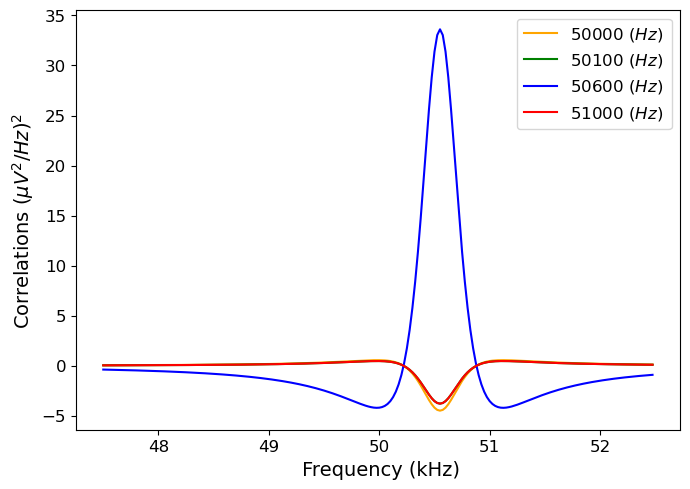}
\caption{Horizontal cut of the PSD covariance matrix at different frequencies. Each color represents a fixed frequency. Again the correlation at same frequencies are removed. Left and right: results for the theoretical predictions under $h=0$ and $h=1$ respectively. The value for $\sigma_{\delta\gamma}$ used is $60$ (Hz). }
\label{heatmaprowstheory}
\end{figure}

In Figure \ref{heatmaprowstheory}, we show the predictions of this stochastic spin relaxation model, and see that they qualitatively reproduce the structures observed in Figures \ref{heatmap} and \ref{heatmaprows}.

 Observe that it is crucial that the total energy of the PSD is independent of the fluctuations. If we do not take into account this fact and use 
\begin{equation}
     S_{h,\gamma}(\omega)=S_{\mathrm{at}} \frac{\gamma^2}{\gamma^2+(\omega-\omega_{L,h}
    )^2}+S_{\mathrm{ph}}
\end{equation}
where $ S_{at}=\frac{g_D^2 Q}{2\gamma^2}$ and $Q$ is independent of $\gamma$, the results after expanding around small $\delta\omega$ are $ S_{h,\gamma+\delta\gamma}(\omega)= S_{h,\gamma}(\omega)+ g(\omega)\delta\gamma +\mathcal{O}(\delta\gamma^2)$, with
\begin{equation}
     g(\omega)=\frac{2 S_{at} \gamma (\omega-\omega_L)^2 }{(\gamma^2+(\omega-\omega_L)^2)^2}>0 \quad \forall \omega,
\end{equation}
and the negative valleys in $\textrm{Cov}((S(\omega),S(\omega'))_{\delta\gamma}$ cannot be theoretically explained.

With all this information, we are in a position to study the effect on the log-likelihood ratio $\tilde{L}_t$ under the proposed model for stochastic fluctuations.  Combining Eq.~\eqref{LIf} and Eq.~\eqref{corrgamma}, the statistic asymptotically presents these first two moments:

\begin{equation}
\begin{aligned}
\lim_{t\to\infty}\frac{\expect{[\tilde L_t]}_{h,\delta\gamma}}{t}&=\frac{1}{2 \pi} \int_{0}^{\infty} \left(\frac{1}{\bar{S}_0(\omega)}-\frac{1}{\bar{S}_1(\omega)}\right)  {\bar{S}_h(\omega')} d\omega \\
\lim_{t\to\infty}\frac{\textrm{Var}(\tilde L_{t})_{h,\delta \gamma}}{t^2}  &=\frac{1}{4 \pi^2}\int_{0}^{\infty} \int_{0}^{\infty} \left(\frac{1}{\bar{S}_0(\omega)}-\frac{1}{\bar{S}_1(\omega)}\right)  \left(\frac{1}{\bar{S}_0(\omega')}-\frac{1}{\bar{S}_1(\omega')}\right) \textrm{Cov}(S(\omega),S(\omega'))_{\delta\gamma} d\omega d\omega'=\\
&=\textrm{Var}(\delta\gamma)\frac{1}{4 \pi^2}\int_{0}^{\infty} \int_{0}^{\infty} \left(\frac{1}{\bar{S}_0(\omega)}-\frac{1}{\bar{S}_1(\omega)}\right)  \left(\frac{1}{\bar{S}_0(\omega')}-\frac{1}{\bar{S}_1(\omega')}\right) \tilde g(\omega)\tilde g(\omega') d\omega d\omega'
\end{aligned}
\label{varall2}
\end{equation}

Notice that the leading term of $\textrm{Var}(L_{t})_{h,\delta \gamma}$  scales quadratically with time $t$, unlike   $\textrm{Var}(L_{t})_{h}$, which corresponds to the case with constant parameters and grows linearly with time. This linear behavior can be derived using the same reasoning as in Sec.~\ref{sec:moments}. 

\begin{figure}[t]
\centering
\includegraphics[width=0.45\textwidth]{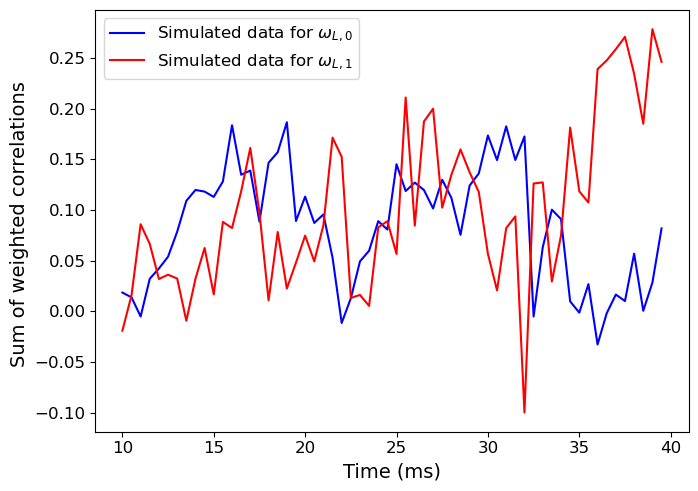} 
\includegraphics[width=0.45\textwidth]{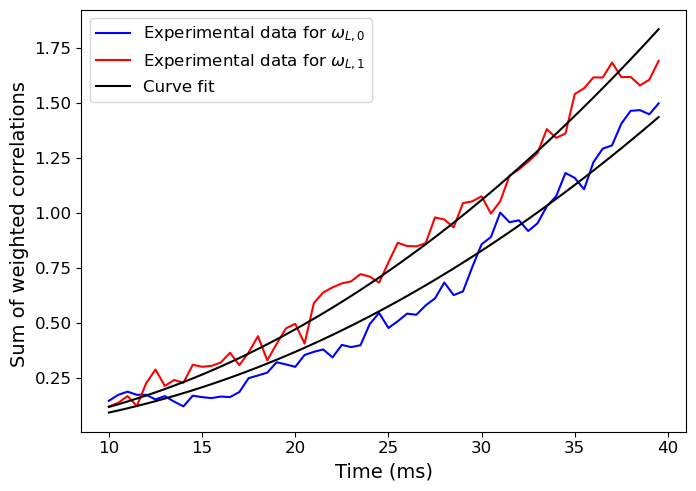} 
\caption{Sum of correlations at different frequencies weighed with the LLR functional in dotted curves. In stripped curves, the fitting outcomes are represented. Red and blue show the results for $h=0,1$ datasets respectively. In the plots and fits, the correlations at same frequencies are excluded. }
\label{corrLt}
\end{figure}

Finally, we can use the expression of $\textrm{Var}(L_{t})_{h,\delta \gamma}$ 
 in Eq.~\eqref{varall2} to estimate $\textrm{Var}(\delta\gamma)$ and automatically obtain the standard deviation $\sigma_{\delta\omega}$. In Figure \ref{corrLt}, we show the  contribution of the correlations between different frequencies in the LLR, for both experimental and simulated datasets,  computed via the Whittle log-likelihood ratio for the range of $40-60$ (kHz). The left plot shows the results for the simulated traces, where, as expected, the contribution remains constant over time. In contrast, the right plot displays the experimental data, where artificial fluctuations increase with time. The fitted curves are shown in black, from which we extract standard deviations of  $\sigma_{0,\delta\gamma}=57.3$ (Hz)  and $\sigma_{1,\delta\gamma}=50.7$ (Hz) for the datasets of $\omega_{L,0}$ and $\omega_{L,1}$ respectively. 

\begin{figure}[t]
\centering
\includegraphics[width=0.45\textwidth]{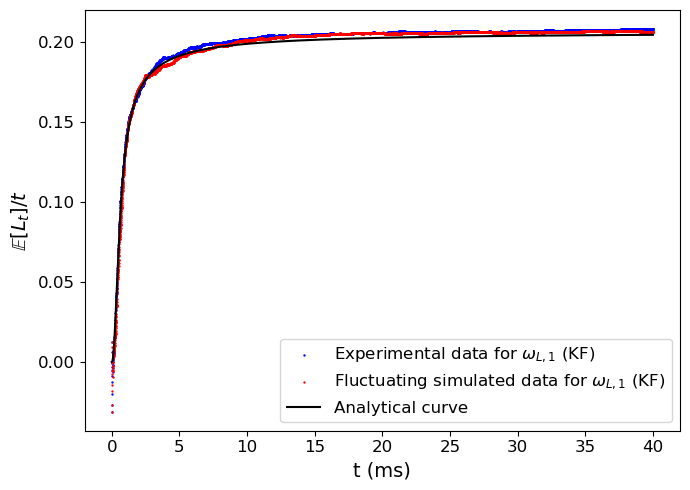} 
\includegraphics[width=0.45\textwidth]{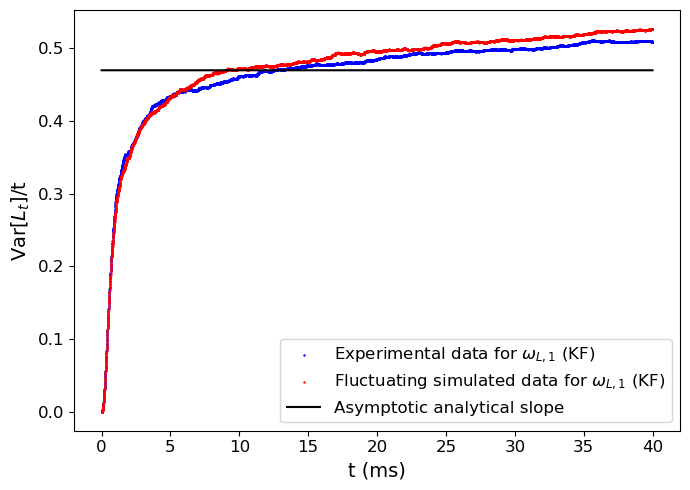} 
\caption{LLR results as a function of time for the simulated data with fluctuating spin relaxation 
(red). The experimental results are shown in  (blue).}
\label{likelihoodstoch}
\end{figure}

Finally, we take the largest value for $\sigma_{\delta\gamma}$, i.e. $\sigma_{0,\delta\gamma}$ and generate time traces under the conditions that this model predicts.  
In Figure \ref{likelihoodstoch}, we present the results of the LLR for simulated data with fluctuating spin relaxation (red), and compare them with the experimental results (blue). 

\section{\label{sec:ultimateerrorrates} Ultimate error rates}
In this section we provide analytical asymptotic expressions for the sequential and Chernoff rates. Recall that we are in a scenario where the choice of hypothesis $h=0,1$ only affects the sign of the LLR and the sequential rate can be obtained from Eq.~\eqref{LLRreq}. For high $\omega_c$,
\begin{equation}
\begin{aligned}
\rho_{\textrm{seq}}&=\frac{S_{\mathrm{at}}\gamma\Delta\omega^2\left(S_{\mathrm{at}}\left(-3 \sqrt{S_{\mathrm{ph}}}+\sqrt{S_{\mathrm{at}}+S_{\mathrm{ph}}}\right)\gamma^2+S_{\mathrm{ph}}\left(-\sqrt{S_{\mathrm{ph}}}+\sqrt{S_{\mathrm{at}}+S_{\mathrm{ph}}}\right)\left(4\gamma^2+\Delta\omega^2\right)\right)}{2\sqrt{S_{\mathrm{at}}+S_{\mathrm{ph}}}\left(S_{\mathrm{at}}^2\gamma^4+2S_{\mathrm{at}}S_{\mathrm{ph}}\gamma^2\Delta\omega^2+S_{\mathrm{ph}}^2\Delta\omega^2\left(4\gamma^2+\Delta\omega^2\right)\right)}-\\
&-\frac{S_{\mathrm{at}}^2\gamma^4\Delta\omega^2}{5\pi S_{\mathrm{ph}}²\omega_c^5} +\mathcal{O}\left(\frac{1}{\omega_c^6}\right)\\
\end{aligned}
\label{seq_an}
\end{equation}

On the other hand, for the  Chernoff rate computation we rely on  Eq.~\eqref{eq:chernofffreq}. To obtain the value of $s$ that provides the tightest bound in our experimental scenario, i.e. $C(s^*):=\max_s C(s)$, assuming one can compute the following quantity,
\begin{equation}
   \begin{aligned}
        \frac{\partial }{\partial s}\lim_{t\to\infty}\left(\frac{C(s)}{t}\right)&= \int^{\infty}_0  \frac{d\omega}{2\pi} \frac{\partial }{\partial s} \log{\frac{\bar S_{I,0}(\omega_j)^{1-s}\bar S_{I,1}(\omega_j)^s}{(1-s)\bar S_{I,0}(\omega_j)+s \bar S_{I,1}(\omega_j)}}
        =\frac{(1 - 2s) S_{\mathrm{at}}^2 \gamma^4 \Delta\omega^2}{5\pi S_{\mathrm{ph}}^2 \omega_c^5}   +    \\
        &+\frac{((-1 + 2s) S_{\mathrm{at}}^2 \gamma^4 \Delta\omega^2 (2(S_{\mathrm{at}} + 2S_{\mathrm{ph}}) \gamma^2 - S_{\mathrm{ph}} \Delta\omega^2))}{7\pi S_{\mathrm{ph}}^3 \omega_c^7} +\frac{(1 + 6 (-1 + s) s)S_{\mathrm{at}}^3\gamma^6\Delta\omega^3}{12 \pi S_{\mathrm{ph}}^3\omega_c^8} + \mathcal{O}\left(\frac{1}{\omega_c^9}\right),
   \end{aligned}
\end{equation}
 and set it equal to zero to find the relative extreme. Observe that for large $\omega_c$, the first two terms of the expansion are zero at $s^*=1/2$. Then, evaluating Eq.~\eqref{eq:chernofffreq} at $s^*=1/2$, we obtain the following closed form at large $\omega_c$
 \begin{equation}
\begin{aligned}
\rho_\textrm{c}:&=\lim_{t\to\infty} \frac{C(s^*)}{t}=\frac{S_{\mathrm{at}}^2\gamma^4\Delta\omega^2}{20 \pi S_{\mathrm{ph}}^2\omega_c^5}+\left(1+\sqrt{\frac{S_{\mathrm{at}}+S_{\mathrm{ph}}}{S_{\mathrm{ph}}}}\right)\gamma+\\
    &-\frac{1}{2 \sqrt{S_{\mathrm{ph}}}}\left(\sqrt{2 S_{\mathrm{at}} \gamma^2+4 S_{\mathrm{ph}}\gamma^2-S_{\mathrm{ph}}\Delta\omega^2-2\gamma \sqrt{S_{\mathrm{at}}^2\gamma^2-2S_{\mathrm{ph}}\left(S_{\mathrm{at}}+2S_{\mathrm{ph}}\right)\Delta\omega^2}}\right)\\
    &-\frac{1}{2 \sqrt{S_{\mathrm{ph}}}}\left(\sqrt{2 S_{\mathrm{at}} \gamma^2+4 S_{\mathrm{ph}}\gamma^2-S_{\mathrm{ph}}\Delta\omega^2+2\gamma \sqrt{S_{\mathrm{at}}^2\gamma^2-2S_{\mathrm{ph}}\left(S_{\mathrm{at}}+2S_{\mathrm{ph}}\right)\Delta\omega^2}}\right)+\mathcal{O}\left(\frac{1}{\omega_c^6}\right).
\end{aligned}
\label{maxchernoff}
\end{equation}

The leading terms of the sequential and Chernoff rates in Eq.~\eqref{seq_an} and Eq.~\eqref{maxchernoff} are 4-parameter dependent, i.e. $\Delta\omega,\gamma,S_{at}$ and $S_{ph}$ ($\omega_c$ no longer plays a role). Therefore, one can express these two rates in terms of the following two ratios that provide  an idea of the shape and relative distance between the two hypothesis under testing
\begin{equation}
   c_a:= \frac{\Delta\omega}{\gamma}>0, \quad  c_b:= \frac{S_{\textrm{at}}}{S_{\textrm{ph}}}>0.
\end{equation}
 In the first case, $c_a$ is an indicator of the overlap between the two Lorentzian since it takes into account their width and distance in between.  In the second case,  $c_b$ stands for the ratio between the Lorentzian peak, i.e. and the strength of the spin signal,  and the background noise. With this change of variable, it turns out that both slopes become a 2-parameter function. Then, with the aim to compare both the sequential and Chernoff rates we define the quantity,
\begin{equation}
    r(c_a,c_b):=\frac{\rho_{\textrm{seq}}}{\rho_\textrm{c}},
\end{equation}
and study its behavior at different parameter regimes, i.e. $\forall c_a,c_b>0$. It can be symbolically checked with computational software, like, for instance, Wolfram Mathematica, that  $r(c_a,c_b)\geq 4$  $\forall c_a,c_b>0$. Therefore, for processes where the Chernoff bound is asymptotically attainable, the sequential strategy outperforms the deterministic one by at least a factor of 4 in terms of less observation time to achieve the same error performance.

\begin{figure}[t]
\centering
\includegraphics[width=0.45\textwidth]{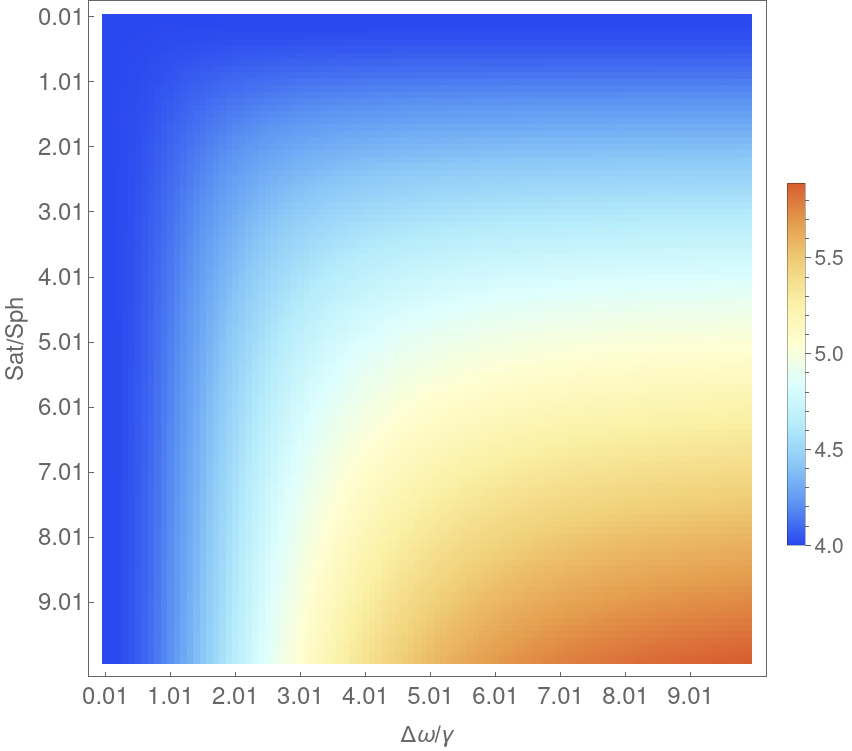} 
\caption{Heatmap of $r(c_a,c_b)$  for different values of $c_a:=\Delta\omega/\gamma$ and $c_b:=S_{\textrm{at}}/S_{\textrm{ph}}$ that range up to 10.  }
\label{heatmap_chernoffvsseq}
\end{figure}

In Figure \ref{heatmap_chernoffvsseq}, we show a heatmap of $r(c_a,c_b)$ for different values of $c_a,c_b$ that range up to 10. Notice that for small $c_a,c_b$, i.e. in the regimes where the two hypothesis are more difficult to distinguish, $r(c_a,c_b)$ saturates the minimum value of 4. Moreover, in scenarios where the two Lorentzians are widely separated or the shot noise is very small compared to the spin signal, $r(c_a,c_b)$ increases up to values bigger than 5.5. 

\end{document}